\documentstyle[aps,eqsecnum,epsf]{revtex}

\ifx\epsffile\undefined
\def\insertfig#1{}
\else
\def\insertfig#1{{\baselineskip=4pt
\centerline{\epsfxsize=\hsize\epsffile{#1}}}}\fi

\def\ssqr#1#2{{\vbox{\hrule height #2pt
      \hbox{\vrule width #2pt height#1pt \kern#1pt\vrule width #2pt}
      \hrule height #2pt}\kern- #2pt}}
\def\sqr{\mathchoice\ssqr8{.4}\ssqr8{.4}\ssqr{5}{.3}\ssqr{4}{.3}}

\def\bsqr{\ssqr{10}{.1}}

\def\nbox{\hbox{$\bsqr\bsqr\bsqr\bsqr\raise2.7pt\hbox{$\,\cdot\cdot\cdot
\cdot\cdot\,$}\bsqr\bsqr\bsqr$}}

\def\nboxconj{\vbox{\hbox{$\sqr$}
\nointerlineskip\kern-.3pt\hbox{$\sqr$}
\nointerlineskip\kern-.3pt\hbox{$\sqr$}
\nointerlineskip\kern-.3pt\moveright2.9pt\hbox{$\cdot$}
\nointerlineskip\kern-.3pt\moveright2.9pt\hbox{$\cdot$}
\nointerlineskip\kern-.3pt\moveright2.9pt\hbox{$\cdot$}
\nointerlineskip\kern-.3pt\hbox{$\sqr$}
\nointerlineskip\kern-.3pt\hbox{$\sqr$}}}

\def\nboxadj{\vbox{\hbox{$\sqr\sqr\,$}
\nointerlineskip\kern-.3pt\hbox{$\sqr$}
\nointerlineskip\kern-.3pt\hbox{$\sqr$}
\nointerlineskip\kern-.3pt\moveright2.9pt\hbox{$\cdot$}
\nointerlineskip\kern-.3pt\moveright2.9pt\hbox{$\cdot$}
\nointerlineskip\kern-.3pt\moveright2.9pt\hbox{$\cdot$}
\nointerlineskip\kern-.3pt\hbox{$\sqr$}
\nointerlineskip\kern-.3pt\hbox{$\sqr$}}}

\def\nboxss{\vbox{\hbox{$\sqr\sqr\sqr\sqr$}
\nointerlineskip\kern-.3pt\hbox{$\sqr\sqr$}
\nointerlineskip\kern-.3pt\hbox{$\sqr\sqr$}
\nointerlineskip\kern-.3pt\moveright2.9pt\hbox{\hbox{$\cdot$}
\hskip2.5pt\hbox{$\cdot$}}
\nointerlineskip\kern-.3pt\moveright2.9pt\hbox{\hbox{$\cdot$}
\hskip2.5pt\hbox{$\cdot$}}
\nointerlineskip\kern-.3pt\moveright2.9pt\hbox{\hbox{$\cdot$}
\hskip2.5pt\hbox{$\cdot$}}
\nointerlineskip\kern-.3pt\hbox{$\sqr\sqr$}
\nointerlineskip\kern-.3pt\hbox{$\sqr\sqr$}}}

\def\nboxas{\vbox{\hbox{$\sqr\sqr\sqr$}
\nointerlineskip\kern-.3pt\hbox{$\sqr$}
\nointerlineskip\kern-.3pt\hbox{$\sqr$}
\nointerlineskip\kern-.3pt\moveright2.9pt\hbox{$\cdot$}
\nointerlineskip\kern-.3pt\moveright2.9pt\hbox{$\cdot$}
\nointerlineskip\kern-.3pt\moveright2.9pt\hbox{$\cdot$}
\nointerlineskip\kern-.3pt\hbox{$\sqr$}}}

\def\nboxsa{\vbox{\hbox{$\sqr\sqr\sqr$}
\nointerlineskip\kern-.3pt\hbox{$\sqr\sqr\sqr$}
\nointerlineskip\kern-.3pt\hbox{$\sqr\sqr$}
\nointerlineskip\kern-.3pt\moveright2.9pt\hbox{\hbox{$\cdot$}
\hskip2.5pt\hbox{$\cdot$}}
\nointerlineskip\kern-.3pt\moveright2.9pt\hbox{\hbox{$\cdot$}
\hskip2.5pt\hbox{$\cdot$}}
\nointerlineskip\kern-.3pt\moveright2.9pt\hbox{\hbox{$\cdot$}
\hskip2.5pt\hbox{$\cdot$}}
\nointerlineskip\kern-.3pt\hbox{$\sqr\sqr$}
\nointerlineskip\kern-.3pt\hbox{$\sqr\sqr$}}}

\def\nboxaa{\vbox{\hbox{$\sqr\sqr$}
\nointerlineskip\kern-.3pt\hbox{$\sqr\sqr$}
\nointerlineskip\kern-.3pt\hbox{$\sqr$}
\nointerlineskip\kern-.3pt\moveright2.9pt\hbox{$\cdot$}
\nointerlineskip\kern-.3pt\moveright2.9pt\hbox{$\cdot$}
\nointerlineskip\kern-.3pt\moveright2.9pt\hbox{$\cdot$}
\nointerlineskip\kern-.3pt\hbox{$\sqr$}}}

\def\nboxsix{\vbox{\hbox{$\sqr\sqr$}
\nointerlineskip\kern-.3pt\hbox{$\sqr\sqr$}
\nointerlineskip\kern-.3pt\hbox{$\sqr$}
\nointerlineskip\kern-.3pt\hbox{$\sqr$}}}

\def\nboxfour{\vbox{\hbox{$\sqr\sqr$}
\nointerlineskip\kern-.3pt\hbox{$\sqr\sqr$}}}

\def\nboxA{\vbox{\hbox{$\bsqr\bsqr\bsqr\bsqr\raise2.7pt\hbox{$\,\cdot\cdot\cdot
\cdot\cdot\,$}\bsqr\bsqr$}\nointerlineskip
\kern-.3pt\hbox{$\bsqr$}}}

\def\nboxB{\vbox{\hbox{$\bsqr\bsqr\bsqr\bsqr\raise2.7pt\hbox{$\,\cdot\cdot\cdot
\cdot\cdot\,$}\bsqr$}\nointerlineskip
\kern-.3pt\hbox{$\bsqr\bsqr$}}}

\def\nboxD{\vbox{\hbox{$\bsqr\bsqr\raise2.7pt\hbox{$\,\cdot\cdot\cdot
\cdot\cdot\,$}\bsqr\bsqr\bsqr\bsqr\bsqr\bsqr$}\nointerlineskip
\kern-.2pt\hbox{$\bsqr\bsqr\raise2.7pt\hbox{$\,\cdot\cdot\cdot
\cdot\cdot\,$}\bsqr$}}}

\def\nboxE{\vbox{\hbox{$\bsqr\bsqr\bsqr\raise2.7pt\hbox{$\,\cdot\cdot\cdot
\cdot\cdot\,$}\bsqr\bsqr\bsqr\bsqr$}\nointerlineskip
\kern-.2pt\hbox{$\bsqr\bsqr\bsqr\raise2.7pt\hbox{$\,\cdot\cdot\cdot
\cdot\cdot\,$}\bsqr$}}}

\def\nboxF{\vbox{\hbox{$\bsqr\bsqr\bsqr\bsqr\raise2.7pt\hbox{$\,\cdot\cdot\cdot
\cdot\cdot\,$}\bsqr\bsqr$}\nointerlineskip
\kern-.2pt\hbox{$\bsqr\bsqr\bsqr\bsqr\raise2.7pt\hbox{$\,\cdot\cdot\cdot
\cdot\cdot\,$}\bsqr$}}}

\def\nboxG{\vbox{\hbox{$\bsqr\bsqr\bsqr\bsqr\raise2.7pt\hbox{$\,\cdot\cdot\cdot
\cdot\cdot\,$}\bsqr$}\nointerlineskip
\kern-.2pt\hbox{$\bsqr\bsqr\bsqr\bsqr\raise2.7pt\hbox{$\,\cdot\cdot\cdot
\cdot\cdot\,$}\bsqr$}}}

\def\nboxH{\vbox{\hbox{$\bsqr\bsqr\raise2.7pt\hbox{$\,\cdot\cdot\cdot
\cdot\cdot\,$}\bsqr\bsqr\bsqr\raise2.7pt\hbox{$\,\cdot\cdot\cdot
\cdot\cdot\,$}\bsqr$}\nointerlineskip
\kern-.2pt\hbox{$\bsqr\bsqr\raise2.7pt\hbox{$\,\cdot\cdot\cdot
\cdot\cdot\,$}\bsqr$}}}

\begin{document}

\def\Tr{{\rm Tr}}
\def\onebox{{\vbox{\hbox{$\sqr\thinspace$}}}}
\def\twobox{{\vbox{\hbox{$\sqr\sqr\thinspace$}}}}
\def\threebox{{\vbox{\hbox{$\sqr\sqr\sqr\thinspace$}}}}

\def\N{N_c}

\twocolumn[       
{\tighten
\preprint{\vbox{
\hbox{UCSD/PTH 94--18}
\hbox{PUPT--1505}
\hbox{hep-ph/9411234}
}}
\title{Spin-Flavor Structure of Large $\bf \N$ Baryons}
\author{Roger F. Dashen${}^1$, Elizabeth Jenkins\footnotemark${}^{1,2}$ and
Aneesh V.
Manohar${}^{1,2}$}
\address{${}^1$\ Department of Physics, University of California at
San Diego, La Jolla, CA 92093}
\address{${}^2$\ Joseph Henry Laboratories of Physics, Princeton University,
Princeton, NJ 08544}
\bigskip
\date{November 1994}
\maketitle
\widetext
\vskip-1.5in
\rightline{\vbox{
\hbox{PUPT--1505}
\hbox{UCSD/PTH 94--18}
\hbox{hep-ph/9411234}
}}
\vskip1.5in
\begin{abstract}
The spin-flavor structure of large $\N$ baryons is described in
the $1/\N$ expansion of QCD using quark operators. The complete set of quark
operator identities is obtained, and used to derive an operator reduction rule
which
simplifies the $1/\N$ expansion. The operator reduction rule is applied to the
axial
currents,
masses, magnetic moments and hyperon non-leptonic decay amplitudes in the
$SU(3)$
limit, to first order in $SU(3)$ breaking, and without assuming $SU(3)$
symmetry.
The connection between the Skyrme and quark representations is discussed. An
explicit formula is given for the quark model operators in terms of the Skyrme
model
operators to all orders in $1/\N$ for the two flavor case.
\end{abstract}

}

] 

\narrowtext

\footnotetext{${}^*$National Science Foundation Young Investigator.}

\section{Introduction}\label{sec:intro}

The properties of baryons have recently been studied in QCD in a systematic
expansion in $1/\N$, where $\N$ is the number
of colors\cite{dm,j,djm,jm}.
In the limit $\N \rightarrow \infty$, it
has been shown that the baryon  sector of QCD has an exact contracted $SU(2F)$
spin-flavor symmetry, where
$F$ is the number of light quark flavors\cite{dm,gervaissakita}.  This
contracted spin-flavor symmetry follows from consistency conditions on
meson-baryon scattering amplitudes which must be satisfied for the theory
to be unitary.  The spin-flavor structure of baryons for finite $\N$ is
given by studying $1/\N$ corrections to the large $\N$ limit.  The
consistency conditions severely restrict the form of subleading $1/\N$
corrections, so definite predictions can be made at subleading
orders.  The $1/\N$ expansion has been used to obtain results for baryon
axial vector currents and magnetic moments up
to corrections of relative order $1/\N^2$ and for baryon masses up to
relative order $1/\N^3$ for two and three light quark flavors.
Salient results include the vanishing of $1/\N$ corrections to pion-baryon
coupling ratios in a given strangeness sector and an equal
spacing rule for decuplet $\rightarrow$ octet pion couplings in different
strangeness sectors.  For the case of three flavors, additional
results have been obtained which do not assume $SU(3)$ flavor symmetry and
are therefore valid to all orders in $SU(3)$ symmetry
breaking.  These results give insight into the structure of
flavor $SU(3)$ breaking in the baryon sector.  The predictions of the
$1/\N$ expansion for baryons are in good agreement with experiment, and
explain the phenomenological success of spin-flavor symmetry for the baryon
sector of QCD.

There are two natural approaches to the study of the spin-flavor algebra of
baryons for large $\N$.  One can solve the consistency conditions by
constructing irreducible representations of the $\N \rightarrow \infty$
contracted $SU(2F)$ spin-flavor symmetry\cite{djm}.  These irreducible
baryon representations are constructed using standard techniques from the
theory of induced representations, and are very closely related to the
collective coordinate quantization of the Skyrmion in the Skyrme
model\cite{skyrme,anw,mattis}.  One can also construct solutions to the
consistency conditions using quark operators, an approach which is closely
related to the non-relativistic quark model.  The two methods are
equivalent, since the non-relativistic quark model and Skyrme model
are identical in the $\N\rightarrow\infty$ limit \cite{am}.
The quark model approach was discussed in detail in refs.~\cite{cgo,lm},
and used to derive results for large $\N$ baryons.  One nice feature of the
quark approach is that it is closely tied to the intuitive picture of
baryons as quark bound states, and the $1/\N$ counting is simply related to
quark Feynman diagrams.  This connection is obvious when the quarks are
heavy.

The Skyrme and non-relativistic quark model realizations of the large $\N$
spin-flavor algebra for baryons in QCD are identical in the $\N \rightarrow
\infty$ limit.  At finite $\N$, the Skyrme and quark representations differ
in their organization of $1/\N$ corrections, but they give equivalent
results at a given order in $1/\N$.  In the Skyrme representation, the
contracted spin-flavor algebra is realized exactly, which implies that the
irreducible baryon representations of the contracted algebra are infinite
dimensional.  In contrast, the quark representation uses the
non-relativistic quark model algebra for finite $\N$.  The baryon spectrum,
in this case, is finite; it consists of a tower of baryon states that
terminates at spin $N_c/2$.  The Skyrme and quark representations both give
rise to operator identities which eliminate redundant operators at a given
order in the $1/\N$ expansion.  These identities are much simpler in the
Skyrme representation than in the quark representation for two
flavors\cite{djm}.  (In particular, some of the operator identites used in
the original analysis are not obvious in the quark description, and have
not been derived using this method.  The derivation is supplied in this
work.) However, both the Skyrme and quark representations become quite
complicated for more than two flavors.  There has been no derivation of all
the non-trivial operator identities for the case of three flavors using
either method.  These operator identities are required for a systematic
analysis of the $1/\N$ expansion for baryons in the three flavor case.

In this paper, we study the spin-flavor structure of baryons for an
arbitrary number of colors and flavors.  We present a brief review of the
quark model representation in Section~\ref{sec:qrep}, and the $1/\N$ counting
rules for baryon operators in Section~\ref{sec:ncount}.  All the baryon
operators are classified using the $SU(2F)$ spin-flavor symmetry group in
Section~\ref{sec:classify}.  All the non-trivial operator identities among
the baryon operators are derived in Section~\ref{sec:derive}.  The set of
independent operators and the operator identities have an elegant
group-theoretic classification.  The operator analysis of
Sections~\ref{sec:classify}--\ref{sec:derive} is done for the general case
of $F$ quark flavors.  The special cases of two and three light flavors
which are of principal physical interest are considered explicitly in
Section~\ref{sec:2and3}.  The operator identities are used to derive a
simple operator reduction rule in Section~\ref{sec:opanalysis}, which gives
the linearly independent baryon operators at a given order in the $1/\N$
expansion.  The operator analysis for the case of baryons with three flavors
without any assumption of $SU(3)$ symmetry is given in
Section~\ref{sec:broken}.
The operator analysis is then used to study the static baryon
properties (axial currents, masses, magnetic moments, and hyperon non-leptonic
decays) in the $SU(3)$ limit, to first order in perturbative $SU(3)$ breaking,
and for completely broken $SU(3)$ flavor symmetry in
Sections~\ref{sec:axial}--\ref{sec:nonlep}.  Readers not
interested in some of the details can skip
Sections~\ref{sec:classify}--\ref{sec:opanalysis}, and refer only to the
identities
(for three flavors) in Tables~\ref{tab:su6iden},~\ref{tab:broken0}
and~\ref{tab:brokennot0} and the operator reduction rules in
Sections~\ref{sec:opanalysis} and~\ref{sec:broken} before proceeding to
the discussion of the static baryon  properties.  Additional group theory
required
in the analysis is given in the  appendices.  We reproduce some of the earlier
results for the baryon axial currents, masses and magnetic moments
\cite{dm,j,djm,jm,cgo,lm}.  In addition, new results are  presented for the
three flavor
case in the symmetry limit, and to first  order in symmetry breaking.  New
results are
also presented for the hyperon  non-leptonic decay amplitudes.  To make the
results accessible to a  wider audience, we will present detailed comparisons
of the
large $\N$  predictions with the experimental data in another paper
\cite{ddjm}.

The operator analysis in this paper is discussed almost entirely using
the quark representation.  The connection with the Skyrme model is
discussed in Section~\ref{sec:skyrme}.  The quark representation uses the
algebraic structure of the non-relativistic quark model to classify all the
baryon operators.  It is important to stress, however, that the results of
this paper do not assume that the non-relativistic quark model is valid, or
that the quarks in the baryon are non-relativistic.  A more detailed
discussion of the connection between the quark basis and large $\N$ QCD
can be found in refs.~\cite{djm,cgo,lm}.  Finally, we restrict our analysis
principally to the ground-state baryons.  Excited baryons have been
considered in ref.~\cite{cgkm}.

\section{The Quark Representation}\label{sec:qrep}

The quark representation of the spin-flavor symmetry of large $\N$ baryons
is based on the non-relativistic quark model picture.  However, as emphasized
in the
introduction, using the quark model realization of the contracted
spin-flavor symmetry does not mean that we are treating the quarks in the
baryon as non-relativistic.  The non-relativistic quark model algebra
provides a convenient way of writing the results of a $1/\N$ calculation in
QCD, which are valid even for baryons with massless quarks.  We will refer
to the quark representation rather than the non-relativistic quark model,
to emphasize this distinction.

In the quark representation, one defines a set of quark creation and
annihilation operators, $q^\dagger_{ \alpha}$ and $q^{\alpha}$, where
$\alpha=1, \ldots, F$ represents the $F$ quark flavors with spin up, and
$\alpha=F+1,\ldots, 2F$, the $F$ quark flavors with spin down.  The
antisymmetry of the $SU(\N)$ color $\epsilon$-symbol and Fermi statistics
implies that the ground-state baryons contain $\N$ quarks in the completely
symmetric representation of spin $\otimes$ flavor (see
Fig.~\ref{fig:groundstate}), so one can omit the color quantum numbers of
the quark operators for the spin-flavor analysis and treat them as bosonic
objects.  Thus, the quark operators satisfy the bosonic commutation
relation
\begin{equation}\label{IIi}
\left[q^{\alpha},q^\dagger_{\beta}\right] = \delta^\alpha_\beta .
\end{equation}
In this work, we consider the spin-flavor structure of the ground-state
baryons for $\N$ large and finite, and odd.  The completely symmetric
$\N$-quark representation of $SU(2F)$ contains baryons with spin $1/2$,
$3/2$, $\ldots$ , $\N/2$, which transform as the flavor representations
shown in Table~I, respectively.  For two flavors, the baryon states can be
labeled by their spin $J$ and isospin $I$, $(J,I) = (1/2,1/2)$,
$(3/2,3/2)$, $\ldots$, $(\N/2,\N/2)$.  For three flavors, the spin-1/2
baryons have the weight diagram shown in Fig.~\ref{fig:weight1/2}, and the
spin-3/2 baryons have the weight diagram shown in Fig.~\ref{fig:weight3/2}.
Generically, the spin $J$ weight diagram has an edge with $2J+1$
weights, and an edge with $(N_c+2)/2-J$ weights. The multiplicity
starts at one for the outermost weights, and increases by one as one moves
inward,
until one reaches the point at which the weights are triangular. From this
point inwards, the multiplicity remains constant. The dimension of the
representation is $ab(a+b)/2$, where $a$ and $b$ are the number of
weights on the two edges. The weight diagrams of the spin-$1/2$ and spin-$3/2$
baryons reduce to
the baryon octet and  decuplet for $\N=3$.
For $F>2$, the baryon flavor representations grow rapidly with $\N$, and
are not the same as the flavor representations for $\N=3$.  This dependence
of the flavor representations on $\N$ leads to subtleties in obtaining
results for $\N=3$.

Quark operators can be classified according to whether they are zero-body,
one-body, $\ldots$, or $n$-body operators.  A zero-body operator
contains no $q$ or $q^\dagger$.  There is a unique zero-body operator, the
identity operator~$\openone$.  A one-body operator acts on a single quark.  The
one-body operators consist of the quark number operator $q^\dagger q$ and
the spin-flavor adjoint $q^\dagger \Lambda^A q$, $A = 1, \ldots, (2F)^2-1$,
where $\Lambda^A$ is a spin-flavor generator.  Two-body operators involve
two $q$'s and two $q^\dagger$'s, and act upon two quarks.  Two-body
operators can be written either as bilinears in the one-body operators or
in normal ordered form (e.g.  $q^\dagger q^\dagger q q$).  Normal ordered
two-body operators are ``pure'' two-body operators, in the sense that they
have vanishing matrix elements on single quark states.  Similarly, one can
consider $n$-body operators either as polynomials of degree $n$ in the
one-body operators, or in normal ordered form with $n$ $q$'s and $n$
$q^\dagger$'s.  $n$-body operators acting on a $\N$-quark state typically
have matrix elements of order $\N^n$ because of combinatoric factors
associated with inserting the operator in the $\N$-quark state.

Since we eventually will be interested in classifying operators according
to their spin and flavor representations, it is convenient to decompose the
$SU(2F)$ adjoint one-body operator $q^\dagger \Lambda^A q$ into
representations of $SU(2)\times SU(F)$,
\begin{eqnarray}
&&J^i = q^\dagger \left(J^i \otimes \openone \right) q\qquad (1,0),\nonumber \\
&&T^a = q^\dagger \left(\openone \otimes T^a \right) q\qquad
(0,adj),\label{IIii} \\
&&G^{ia} = q^\dagger \left(J^i \otimes T^a \right) q \qquad (1, adj),
\nonumber
\end{eqnarray}
where $J^i$ are the spin generators, $T^a$ are the flavor generators, and
$G^{ia}$ are the spin-flavor generators.  The transformation properties of
these generators under $SU(2) \times SU(F)$ are given in eq.~(\ref{IIii}),
where $adj$ denotes the $F^2-1$ dimensional adjoint representation of
$SU(F)$.  Throughout this work, uppercase letters ($A$, $B$, \ldots) denote
indices transforming according to the adjoint representation of the
$SU(2F)$ spin-flavor group, lowercase letters ($a$, $b$, \ldots) denote
indices transforming according to the adjoint representation of the $SU(F)$
flavor group, and ($i$, $j$, \ldots) denote indices transforming according
to the vector representation of spin.  The matrices $J^i$ and $T^a$ on the
right-hand side of eq.~(\ref{IIii}) are in the fundamental representations
of $SU(2)$ and $SU(F)$, respectively, and are normalized so that
\begin{eqnarray}\label{two}
\Tr\ J^i J^j &=& {1\over 2}\, \delta^{ij},\nonumber \\
\Tr\ T^a T^b &=& {1\over 2}\, \delta^{ab}. \label{IIiii}
\end{eqnarray}
The spin-flavor matrices $\Lambda^A$ normalized to
\begin{equation}
\Tr\ \Lambda^A \Lambda^B = {1\over 2}\, \delta^{AB},\label{IIIiv}
\end{equation}
are $\left(J^i \otimes \openone \right)/\sqrt{F}$, $\left(\openone \otimes T^a
\right)/ \sqrt2$ and $\sqrt2 \left(J^i \otimes T^a \right)$, so that the
properly normalized $SU(2F)$ operators are $J^i/\sqrt{F}$, $T^a/\sqrt2$ and
$\sqrt2\, G^{ia}$.

\section{Large $\bf \N$ Power Counting}\label{sec:ncount}

The baryons in QCD are color singlet states of $\N$ quarks. The
$\N$-dependence of operator matrix elements in baryon states can be obtained
using the double line notation of `t~Hooft~\cite{thooft}. The $\N$ counting
rules were
discussed extensively by Witten~\cite{witten}, and more recently in
refs.~\cite{cgo,lm}.
The basic result can be given very simply using an illustrative example.

Consider the baryon matrix element of a one-quark QCD operator ${\cal
O}_{QCD}=\bar q \Gamma q$, where $\Gamma$ is a Dirac and flavor
matrix.\footnote{Note
that the quark field $q$ in the QCD operator is not the same as the quark
operator $q$ of the quark representation.}\
For example, the operator could be the flavor singlet axial current, with
$\Gamma=\gamma_\mu\gamma_5$, or a flavor octet vector current, with $\Gamma
= T^a \gamma_\mu$, etc.  The baryon matrix element of ${\cal O}_{QCD}$ is
obtained by inserting the operator on any of the $\N$ quark lines, as show
in Fig.~\ref{fig:opmatrix}a.  There are $\N$ insertions, and each graph is
of order one, so that a one-quark QCD operator has a matrix element which
is at most of order $\N$.  The matrix element is not necessarily of order
$\N$, however, since there may be cancellations among the $\N$ insertions
on the various quark lines.  All planar graphs with additional gluon
exchanges (Fig.~\ref{fig:opmatrix}b) are of the same order in $\N$ as
Fig.~\ref{fig:opmatrix}a, whereas graphs with additional exchanges of
non-planar gluons are suppressed
by powers of $1/\N$ relative to Fig.~\ref{fig:opmatrix}a.

The QCD operator is given by an expansion in $1/\N$ in terms of operators
in the quark representation.  At leading order, the QCD operator has an
expansion of the form
\begin{equation}
{\cal O}_{QCD}= \sum_{n,k} c^{(n)}_k {1\over \N^{n-1}} {\cal O}^{(n)}_k ,
\label{IIIi}
\end{equation}
where the sum is over all possible $n$-body operators ${\cal O}^{(n)}_k$,
$n= 0, \ldots, \N$, with the same spin and flavor quantum numbers as ${\cal
O}_{QCD}$, with coefficients $c^{(n)}_k$ of order unity.  Subleading $1/\N$
corrections to the leading order expression~(\ref{IIIi}) can be included by
adding $1/\N$ corrections to the coefficients $c^{(n)}_k$.  The complicated
QCD dynamics is parametrized by the unknown coefficients $c_k^{(n)}$.  A
comparison of the form of this expansion with the Feynman diagrams in
Fig.~\ref{fig:opmatrix} is suggestive.  The one-body operator can be thought
of as arising from the operator insertion graphs depicted in
Fig.~\ref{fig:opmatrix}a.  The single gluon exchange graphs of
Fig.~\ref{fig:opmatrix}b produce two-body operators with an extra factor of
$1/\N$ from the two gauge coupling constants at the gluon vertices, and so
on.  Non-planar gluon exchange graphs
result in $1/\N$ corrections to the operator coefficients.  There are $\N$
quarks in the baryon, so one can terminate the expansion at $\N$-body
operators.  An
$n$-body operator is typically of order $\N^n$, so that all of the terms in
eq.~(\ref{IIIi}) are of the same order in the $1/\N$ expansion as the
leading term.  In the limit $\N\rightarrow\infty$, one obtains an infinite
series of operators which are equally important even at leading order in
$1/\N$.  Since we cannot evaluate the coefficients $c^{(n)}_k$, the entire
$1/\N$ expansion would be intractable, were it not for a series of operator
identities which allows the number of operators to be reduced to a finite
set at a given order in $1/\N$.  In this paper, we derive all these
operator identities, and classify the independent operators at any given
order in $1/\N$ for some quantities of interest, such as the baryon
axial currents, masses, magnetic moments and non-leptonic decay amplitudes.

An important feature of the above $\N$ counting for the $n$-body quark
operators is that the $\N$ counting is preserved under commutation.  The
commutator of an $m$-body operator with an $n$-body operator is an
$\left(m+n-1\right)$-body operator,
\begin{equation}
\left[{\cal O}^{(m)},{\cal O}^{(n)}\right]={\cal O}^{(m+n-1)},\label{IIIii}
\end{equation}
and
$\left(1/\N^{m-1}\right)\left(1/\N^{n-1}\right)=\left(1/\N^{(m+n-1)-1}\right)$.
In contrast, the anticommutator of an $m$-body and $n$-body operator is
typically an $(m+n)$-body operator.  The commutator has one less $q^\dagger
q$ than the anticommutator, because quark operators acting on different
quark lines commute.  The commutativity of quark operators acting on
different quark lines forces one quark in ${\cal O}^{(m)}$ to act on the
same quark line as a quark in ${\cal O}^{(n)}$ to produce a non-zero
commutator, and reduces the $\left(m+n\right)$-body operator to an
$\left(m+n-1 \right)$-body operator.

We have given the quark counting rules for a one-quark QCD operator.
Similarly, it is easy to see that a $m$-quark QCD operator is given as an
expansion in terms of $n$-body operators with coefficients of order
$\N^{m-n}$. It need not be the case that $n\ge m$. For example,
in $\Delta I=1/2$ weak decays, a four-quark (i.e. two-body) QCD operator
can produce a one-body quark operator.\footnote{A similar result was found
in the chiral quark model\cite{amhg}.}

\section{Quark Operator Identities: Classification}\label{sec:classify}

In this section, we classify all independent operator identities among
the $n$-body operators in the quark representation.  These identities have
an elegant group theoretical structure.  Readers not interested in the
details can look at the identities for three flavors in
Table~\ref{tab:su6iden}, and skip to the operator reduction rule at the end
of Section~\ref{sec:opanalysis}.  The general structure of the identities
is that certain $n$-body operators can be reduced to linear combinations of
$m$-body operators, where $m<n$.  Since $n$-body operators acting on an
$\N$-quark baryon state are generically of order $\N^n$, the coefficient of
the $m$-body operator is typically of order $\N^{n-m}$.  For example, some
three-body operators can be reduced to two-body operators with coefficients
of order $\N$, one-body operators with coefficients of order $\N^2$, and
zero-body operators with coefficients of order $\N^3$.  We will show that
the only independent operator identities which are required are those which
reduce two-body operators to linear combinations of one-body and/or
zero-body operators.  All identities for $n$-body operators with $n>2$ can
be obtained by recursively applying two-body identities.  This result leads to
a
tremendous  simplification in the analysis, since there are only a finite
number of
identities which need to be written explicitly for the two-body case.
Explicit expressions for the two-body identities are derived in
Section~\ref{sec:derive}, and are given in Table~\ref{tab:su2fiden} for an
arbitrary number of flavors, and in Tables~\ref{tab:su4iden} and
\ref{tab:su6iden} for two and three flavors, respectively.

\subsection{Zero-Body Operators}

There is a unique zero-body operator, the identity operator $\openone$, which
has matrix elements $\N^0=1$.  The identity operator transforms as a
singlet under the spin-flavor group $SU(2F)$ and as a singlet under
$SU(2)\times SU(F)$.  There are no operator identities at this level.

\subsection{One-Body Operators}

The one-body operators transform under $SU(2F)$ as the tensor product of a
quark and antiquark representation.  A quark is in the fundamental
representation of $SU(2F)$, and transforms as a tensor with one upper
index.  The antiquark transforms as an $SU(2F)$ tensor with one lower
index.  Thus, the one-body operators transform as
\begin{equation}\label{IVi}
{\rm 1-body:}\ \ \left( \overline{\onebox} \otimes \onebox \right) = 1 + adj
= 1 + T^\alpha_\beta ,
\end{equation}
where $T^\alpha_\beta$ is a traceless tensor which transforms as the
adjoint representation of $SU(2F)$.

The independent one-body operators were listed in Section~\ref{sec:qrep}; they
are
the quark number operator $q^\dagger q$ and the spin-flavor operators
$q^\dagger \Lambda^A q$, which consist of $J^i$, $T^a$, and $G^{ia}$.  The
quark number operator $q^\dagger q$ is a singlet under $SU(2F)$ and $J^i$,
$T^a$ and $G^{ia}$ together form the adjoint representation of $SU(2F)$,
which agrees with the analysis of eq.~(\ref{IVi}).  These one-body
operators transform as $(0,0)$, $(1,0)$, $(0,adj)$ and $(1,adj)$,
respectively, under $SU(2)\times SU(F)$.

The only operator identity allowed at
this stage is one relating the one-body and zero-body $SU(2F)$ singlets.
This identity is trivial,
\begin{equation}
q^\dagger q = \N\ \openone . \label{IVii}
\end{equation}
Note that this identity has the general structure stated at the beginning
of this section: a one-body operator is written as $\N$ times a zero-body
operator.

\subsection{Two-Body Operators}

The non-trivial identities occur among two-body operators.  The two-body
operators transform as the tensor product of a two-quark and two-antiquark
state.  Since the quarks in the ground-state baryon representation are in a
completely symmetric state (Fig.~\ref{fig:groundstate}), any two quarks
transform according to the two-index symmetric tensor representation of
$SU(2F)$, and any two antiquarks are in the complex conjugate
representation.  Thus, the two-body operators transform as
\begin{eqnarray}
{\rm 2-body:}\ \left( \overline{\twobox} \otimes \twobox \right) &=&
 1 + adj + {\bar s s} \nonumber \\  &=& 1 + T^\alpha_\beta +
T^{(\alpha_1 \alpha_2)}_{(\beta_1 \beta_2)} ,\label{IViii}
\end{eqnarray}
where $T^{(\alpha_1 \alpha_2)}_{(\beta_1 \beta_2)}$ is a traceless tensor
which is completely symmetric in its upper and lower indices.  This tensor
representation of $SU(2F)$ will be called the ${\bar s s}$ representation.

It is convenient to write the two-body operators as products of two
one-body operators, rather than to write them in normal-ordered form
directly.  The quark number operator $q^\dagger q$ can be eliminated using
the identity eq.~(\ref{IVii}), so we only need to consider bilinears of the
$SU(2F)$ adjoint representation $q^\dagger \Lambda^A q$, which consists of
$J^i$, $T^a$ and $G^{ia}$.  Any product of two operators can always be
written as the symmetric product (an anticommutator), or the antisymmetric
product (a commutator).  The commutator can be eliminated using the
$SU(2F)$ Lie algebra commutation relations listed in
Table~\ref{tab:su2fcomm}.  The anticommutator transforms as the symmetric
product of two $SU(2F)$ adjoints,
\begin{equation}\label{IViv}
\left( adj \otimes adj \right)_S = 1 + adj + {\bar a a} + {\bar s s},
\end{equation}
where ${\bar a a} = T^{[\alpha_1 \alpha_2]}_{[\beta_1 \beta_2]}$ transforms as
a
traceless tensor which is antisymmetric in its upper and lower indices.
The decomposition of the symmetric tensor product of two adjoints for an
arbitrary $SU(Q)$ group, and for the special cases $Q=6$ and $Q=4$ are
listed in Table~\ref{tab:adj2s}, using the Dynkin notation for the
irreducible representations.  Each of the representations in $\left( adj
\otimes adj \right)_S$ occurs in eq.~(\ref{IViii}) except for the ${\bar a a}$
representation.

The structure of all two-body identities can now be determined.  We quote
the results here; a detailed derivation of all the identities is presented
in Sec.~\ref{sec:derive}.  The two-body identities can be divided into
three different sets:
\begin{enumerate}
\item {There is a linear combination of two-body operators which is an
$SU(2F)$ singlet.  This linear combination can be written as a coefficient
of order $\N^2$ times the zero-body unit operator $\openone$.  The $SU(2F)$
singlet in $\left( adj \otimes adj \right)_S$ is the Casimir operator,
which equals
\begin{equation} \left\{q^\dagger\Lambda^A q,q^\dagger
\Lambda^A q\right\} = \N\left(\N+2F\right)\left(1-{1\over 2F}\right) \openone,
\end{equation}
where the coefficient of the $\openone$ operator is the $SU(2F)$ Casimir for
the completely symmetric baryon representation Fig.~\ref{fig:groundstate}.}
\item {There is a linear combination of
two-body operators which transforms as an $SU(2F)$ adjoint.  This linear
combination can be written as a coefficient of order $\N$ times the
one-body adjoint operator.  The $SU(2F)$ adjoint in $\left( adj \otimes adj
\right)_S$ is obtained by contraction with the $SU(2F)$ $d$-symbol
$d^{ABC}$; it equals
\begin{equation}\hskip-2em\label{blotz}
d^{ABC} \left\{q^\dagger \Lambda^B q ,  q^\dagger \Lambda^C q \right\} =
2\left( \N + F\right)\left( 1 - {1 \over F} \right)\
q^\dagger \Lambda^A q \ ,
\end{equation}
for the completely symmetric baryon representation. The coefficient on the
right-hand
side of eq.~(\ref{blotz}) is the ratio of the cubic and quadratic Casimirs of
the
completely symmetric baryon representation Fig.~\ref{fig:groundstate}.}
\item {Comparison of eqs.~(\ref{IViii}) and~(\ref{IViv}) shows that
there is no ${\bar a a}$ representation for two-body operators acting on the
completely symmetric baryon representation, but there is an ${\bar a a}$
representation in $\left( adj \otimes adj \right)_S$.  Thus, the linear
combination of bilinears in one-body operators which transforms as
an ${\bar a a}$ must vanish for the completely symmetric baryon representation.
This set of identities eliminates certain bilinears in $\{J^i,T^a,G^{ia}\}$
from the set of independent two-body operators.  }
\end{enumerate}

\subsection{Three-Body Operators and Generalization}

Three-body operators act on symmetric tensor products of three-quark
states, and transform as the representations
\begin{eqnarray}
&&{\rm 3-body:}\ \  \left( \overline{\threebox} \otimes \threebox \right)
\nonumber \\
&&=  1 + T^\alpha_\beta + T^{(\alpha_1 \alpha_2)}_{(\beta_1 \beta_2)}
 + T^{(\alpha_1\alpha_2\alpha_3)}_{(\beta_1\beta_2\beta_3)}.\label{IVvii}
\end{eqnarray}
The only new tensor occurring at three-body is the traceless tensor
$T^{(\alpha_1\alpha_2\alpha_3)}_{(\beta_1\beta_2\beta_3)}$.  Any
three-body operator can be written as a trilinear in the one-body
operators.  The one-body quark number operator can be trivially replaced by
$\N \openone$, so only the adjoint one-body operators need to be considered.
Any trilinear product of adjoint one-body operators which is not completely
symmetric can be reduced to two-body operators using the $SU(2F)$
commutation relations given in Table~\ref{tab:su2fcomm}, so one only needs
to consider completely symmetric trilinears in the adjoint one-body
operators.  The decomposition of $\left(adj \otimes adj \otimes adj
\right)_S$ is given in Table~\ref{tab:adj3s} for a general $SU(Q)$ group
and for the special cases $Q=6$ and $Q=4$.

First consider operator identities which relate the three-body singlet,
adjoint and ${\bar s s}$ representations to zero-body, one-body, and two-body
operators times coefficients of order $\N^3$, $\N^2$, and $\N$,
respectively.  In normal ordered form, it is easy to see that these
three-body identities are obtained by contraction of a pair of $q^\dagger$, $q$
indices.  Contraction of pairs of quark indices is already described by the
two-body
$\rightarrow$~one-body and two-body $\rightarrow$~zero-body identities, so
judicious application of these identities yields the required three-body
identities.

Table~\ref{tab:adj3s} shows that there are ten irreducible $SU(2F)$
representations in $\left(adj \otimes adj \otimes adj\right)_S$ for $F \ge
2$, of which only four are present in eq.~(\ref{IVvii}).  Thus, in
principle, there are six sets of identities which vanish identically for
the three-body case.  (For $F =2$, there are eight irreducible $SU(2F)$
representations in $\left(adj \otimes adj \otimes adj\right)_S$, and thus,
in principle, four sets of identities which vanish identically for the
three-body case.) It is clear, however, that not all of these identities
are really new, since at least some of the three-body identities are simply
products of two-body identities in the ${\bar a a}$ representation and one-body
operators in the adjoint representation, $J^i$, $T^a$, or $G^{ia}$.  The
tensor product of ${\bar a a}$ with the adjoint representation is given in
Table~\ref{tab:aaadj}.  Comparison of Tables~\ref{tab:adj3s}
and~\ref{tab:aaadj} shows that all  the representations in
$\left(adj\otimes adj \otimes adj\right)_S$ which are not present in
eq.~(\ref{IVvii}) occur in $\left({\bar a a}\otimes adj\right)$.  This
observation is not sufficient to conclude that all the three-body
identities are given in terms of two-body identities, however.  The three-body
representations in Table~\ref{tab:adj3s} are contained in the completely
symmetric tensor product of three adjoints, $\left(adj\otimes adj \otimes
adj\right)_S$.  The representations in $\left({\bar a a} \otimes adj\right)$ of
Table~\ref{tab:aaadj} are in the tensor product $\left(adj\otimes adj
\right)_S\otimes adj$, since ${\bar a a}$ is contained in $\left(adj\otimes
adj\right)_S$.  To find the representations which reduce to the two-body
${\bar a a}$ identities, one has to impose the additional constraint that the
three adjoints in $\left({\bar a a}\otimes adj\right)$ are completely
symmetric.
Not all the irreducible representations of Table~\ref{tab:aaadj} survive
when this constraint is imposed, but all the irreducible representations of
Table~\ref{tab:adj3s} which are not present in eq.~(\ref{IVvii}) do
survive, as can be checked explicitly.  This observation leads to the
conclusion that there are no new vanishing three-body identities which are
not simply products of the one-body and
two-body identities which have already been determined.

This conclusion can be generalized to $n$-body operators.  There are no new
vanishing identities for $n$-body operators, $n\ge3$, which are not
products of the original ${\bar a a}$ two-body identities and one-body
operators.
The ${\bar a a}$ two-body identities result because any product of two adjoint
one-body operators which is antisymmetric in creation or annihilation
operators must vanish when it acts on the completely symmetric baryon
representation.  Similarly, the vanishing $n$-body identities result from
products of $n$ one-body operators which are not completely symmetric in
the $n$ creation and $n$ annihilation operators.  Any representation of the
permutation group which is not completely symmetric in all the quark
creation (or annihilation) operators must be antisymmetric in at least one
pair.  Two one-body operators containing an antisymmetric quark pair vanish
by the two-body identities derived earlier, so $n$-body operators
containing an antisymmetric quark pair automatically vanish by the two-body
identities, and there are no new identities which vanish for $n\ge3$.

In summary, we have classified all non-trivial operator identities for
$SU(2F)$ quark operators.  For $n$-body quark operators, the only
representations of the $SU(2F)$ group which are allowed are $1 +
T^{\alpha_1}_{\beta_1} + T^{(\alpha_1\alpha_2)}_{(\beta_1\beta_2)} + \ldots
+ T^{(\alpha_1\alpha_2\ldots \alpha_n)}_{(\beta_1\beta_2\ldots\beta_n)}$.
All other representations can be eliminated using the two-body ${\bar a a}$
operator identities.  Furthermore, the only ``purely'' $n$-body
representation is $T^{(\alpha_1\alpha_2\ldots
\alpha_n)}_{(\beta_1\beta_2\ldots\beta_n)}$.  The non-vanishing two-body
operator identities can be used to write $n$-body operators that transform
as $T^{(\alpha_1\alpha_2\ldots \alpha_m)}_{(\beta_1\beta_2\ldots\beta_m)}$
($m<n$) as $m$-body operators times coefficients of order $\N^{n-m}$.

\section{Two-Body Quark Identities: Derivation}\label{sec:derive}

In this section, all of the non-trivial two-body operator identities
are derived explicitly for an arbitrary number of light quark flavors.
The identities are listed in Table~\ref{tab:su2fiden}. The $SU(2F)$ group
theory which is needed for the computation is given in
appendix~\ref{app:suq}.

\subsection{Two-Body $\bf \rightarrow$ Zero-Body Identity}

The $SU(2F)$ singlet in the symmetric product of two $SU(2F)$ generators is
the Casimir operator $\Lambda^A\Lambda^A$, which is a constant for a given
irreducible representation.  The Casimir for an $SU(Q)$ irreducible
representation $R$ (see \cite{grosstaylor}) is
\begin{equation}\label{VIIi}
C_2(R) = {1 \over 2}\left( NQ - {N^2\over Q} + \sum_i r_i^2 - \sum_i c_i^2
\right),
\end{equation}
where $r_i$ is the number of boxes in the $\imath^{\rm th}$ row of the
Young tableau, $c_i$ is the number of boxes in the $\imath^{\rm th}$
column of the Young tableau, and $N=\sum_i r_i=\sum_i c_i$ is the total
number of boxes.  The Casimir for the completely symmetric $SU(2F)$ baryon
representation with a single row of $\N$ boxes (Fig.~\ref{fig:groundstate})
is
\begin{equation}
C_2 =  {1\over2}\N\left(\N+2F\right)\left(1-{1\over 2F}\right),\label{VIIii}
\end{equation}
so the Casimir identity\footnote{The Casimir operator in the completely
symmetric baryon representation can also be computed directly using the
quark operators and the Fierz identity eq.~(\ref{Aix}).} is
\begin{equation}\label{VIIci}
\left\{ q^\dagger \Lambda^A  q,\ \  q^\dagger \Lambda^A q \right\} =
\N\left(\N+2F\right)\left(1-{1\over 2F}\right) \ \openone.
\end{equation}
The Casimir operator $\Lambda^A\Lambda^A$ equals $J^2/F + T^2/2 + 2\,G^2$
using the properly normalized $SU(2F)$ generators.  Combining this relation
with eq.~(\ref{VIIci}) gives the first identity in
Table~\ref{tab:su2fiden}.  Note that the coefficient of the zero-body
operator is of order $\N^2$, as expected.

\subsection{Two-Body $\bf \rightarrow$ One-body Identity}

The linear combination of two-body operators which transforms as an
$SU(2F)$ adjoint reduces to the adjoint one-body operator
\begin{equation}\label{VIIiii}
d^{ABC} \left\{ q^\dagger \Lambda^A q, q^\dagger \Lambda^B q\right\}
=D(R)\ q^\dagger \Lambda^C q ,
\end{equation}
where $D(R)$ is a constant which must be determined for the completely
symmetric baryon representation.  The two-body operator in
eq.~(\ref{VIIiii}) can be written as
\begin{eqnarray}
&&2\,d^{ABC}\ \sum_{r,s=1}^{\N} q^\dagger_{r \alpha}
\left(\Lambda^A\right)^\alpha_\beta
q^\beta_{r} \ q^\dagger_{s \gamma} \left(\Lambda^B\right)^\gamma_\zeta
q^\zeta_{s} \nonumber\\
&&\ \ =2\,d^{ABC}\ \left(\Lambda^A\right)^\alpha_\beta
\left(\Lambda^B\right)^\gamma_\zeta\sum_{r,s=1}^{\N}\ q^\dagger_{r \alpha}
q^\beta_{r} q^\dagger_{s \gamma}
q^\zeta_{s}, \nonumber
\end{eqnarray}
where $q_{r}$ denotes the quark annihilation operator acting on the $r^{\rm
th}$ quark in the baryon, and the sums on $r$ and $s$ run over the $\N$
quarks in the baryon.  Using the $SU(2F)$ identity
\begin{eqnarray}
d^{ABC}\left(\Lambda^A\right)^\alpha_\beta&&
\left(\Lambda^B\right)^\gamma_\zeta = -{1\over 2F}\left[ \delta^\alpha_\beta
\left(\Lambda^C\right)^\gamma_\zeta + \left(\Lambda^C\right)^\alpha_\beta
\delta^\gamma_\zeta\right]\nonumber\\
&&\  + {1\over2}\left[\delta^\alpha_\zeta
\left(\Lambda^C\right)^\gamma_\beta + \left(\Lambda^C\right)^\alpha_\zeta
\delta^\gamma_\beta\right],\label{VIIiv}
\end{eqnarray}
the two-body operator can be rewritten as
\begin{eqnarray}
&&-{2\N\over F} q^\dagger\Lambda^C q +
\sum_{r,s} \left(\Lambda^C\right)^\gamma_\beta q^\dagger_{r \alpha}
 q^\beta_{r} q^\dagger_{s \gamma} q^\alpha_{s} \nonumber\\
&&\ \ +\sum_{r,s} \left(\Lambda^C\right)^\alpha_\zeta   q^\dagger_{r \alpha}
q^\beta_{r} q^\dagger_{s \beta} q^\zeta_{s} \ .\label{VIIv}
\end{eqnarray}
The first term is in the form required by eq.~(\ref{VIIiii}), but the last
two terms need further simplification.  The summation over $r$ and $s$ for
these terms can be divided into a sum
over $r=s$, and over $r\not=s$. The terms with $r=s$ in eq.~(\ref{VIIv}) are
\begin{equation}\label{VIIvi}
\sum_{r} \left(\Lambda^C\right)^\gamma_\beta q^\dagger_{r \alpha}
 q^\beta_{r} q^\dagger_{r \gamma} q^\alpha_{r}
+\sum_{r} \left(\Lambda^C\right)^\alpha_\zeta   q^\dagger_{r \alpha}
q^\beta_{r} q^\dagger_{r \beta} q^\zeta_{r}.
\end{equation}
The normal ordered version of this operator vanishes, since each quark in
the baryon is only singly occupied (any two annihilation operators
$q_{r}^\alpha q_{r}^\beta$ acting on the same quark line must vanish (even
if $\alpha\not=\beta$)). Normal ordering eq.~(\ref{VIIvi}) using the quark
commutator eq.~(\ref{IIi}) yields
\begin{eqnarray}\label{VIIvii}
&&\sum_{r} \left(\Lambda^C\right)^\gamma_\beta q^\dagger_{r \alpha}
 q^\alpha_{r} \delta^\beta_\gamma
+\sum_{r} \left(\Lambda^C\right)^\alpha_\zeta   q^\dagger_{r \alpha}
 q^\zeta_{r} \delta^\beta_\beta \\
&& = 2F\sum_{r}  q^\dagger_{r \alpha}
\left(\Lambda^C\right)^\alpha_\zeta  q^\zeta_{r} \nonumber
=2F\ q^\dagger \Lambda^C q \nonumber
\end{eqnarray}
where the first term vanishes because $\Lambda^C$ is traceless and
$\delta^\beta_\beta=2F$ for the second term.  Finally, the contribution
of the last two terms in eq.~(\ref{VIIv}) with $r\not=s$ must be evaluated.
Since $q_{r}$ and $q^\dagger_{s}$ act on different quark lines, they can be
treated as commuting operators.  Thus the two sums in eq.~(\ref{VIIv}) with
$r\not=s$ are equal to each other (as can be seen by exchanging the dummy
indices $r$ and $s$), and their sum is equal to
\begin{equation}
2 \sum_{r\not=s} \left(\Lambda^C\right)^\gamma_\beta q^\dagger_{r \alpha}
q^\dagger_{s \gamma} q^\beta_{r}  q^\alpha_{s}.\nonumber
\end{equation}
This operator acts on the $(r,s)$ quark pair in the initial baryon.  The
initial baryon is completely symmetric in flavor, so one can exchange
the flavor labels $\alpha$ and $\beta$ on the initial quark pair.  This
gives the equivalent operator
\begin{eqnarray}
&&2 \sum_{r\not=s} \left(\Lambda^C\right)^\gamma_\beta q^\dagger_{r \alpha}
q^\dagger_{s \gamma} q^\alpha_{r}  q^\beta_{s} = 2 \sum_{r\not=s}
q^\dagger_{s \gamma}\left(\Lambda^C\right)^\gamma_\beta q^\beta_{s}
q^\dagger_{r \alpha}
q^\alpha_{r}\nonumber\\
&&\ \  = 2\left(\N-1\right) q^\dagger
\Lambda^C q ,\label{VIIviii}
\end{eqnarray}
since there are $(\N-1)$ values of $r\not=s$ for a given value of $s$, and
$q^\dagger_{r} q_{r}$ is unity since each quark line is singly occupied.
Combining eqs.~(\ref{VIIv})--(\ref{VIIviii}) with eq.~(\ref{VIIiii}) gives
the final form of the identity
\begin{equation}\label{VIIix}
d^{ABC} \left\{ q^\dagger\Lambda^A q,q^\dagger \Lambda^B q\right\} =
2\left(\N + F\right)\left(1-{1\over F}\right) q^\dagger \Lambda^C q.
\end{equation}

The identity eq.~(\ref{VIIix}) for the $SU(2F)$ adjoint two-body operator
can be decomposed under $SU(2)\times SU(F)$ into the three representations
$(1,0)$, $(0,adj)$ and $(1,adj)$.  The identities for these three
$SU(2)\times SU(F)$ representations can be obtained by substituting in the
properly normalized $SU(2F)$ generators $J^i/\sqrt F$, $T^a/\sqrt 2$, and
$\sqrt2\, G^{ia}$ into eq.~(\ref{VIIix}), and using the decomposition of
the $SU(2F)$ $d$-symbol under $SU(2)\times SU(F)$ given in
eq.~(\ref{Bv}).  Eq.~(\ref{VIIix}) then yields the three identities in the
second block of Table~\ref{tab:su2fiden}.

\subsection{Vanishing Two-Body Operators}

The final set of identities is obtained by combining the two adjoint
one-body operators into the ${\bar a a}$ representation, and setting it equal
to
zero.  The ${\bar a a}$ representation can be obtained by using the $SU(2F)$
projection operators discussed in appendix~\ref{app:suq}.  The
decomposition of the ${\bar a a}$ representation into irreducible $SU(2) \times
SU(F)$ representations is given in eq.~(\ref{Bi}).  Another method for
obtaining the vanishing two-body identities is to simplify anticommutators
of $J^i$, $T^a$ and $G^{ia}$ using the same techniques applied in the
derivation of eq.~(\ref{VIIix}).  One then finds linear combinations of the
anticommutators which vanish.  The ${\bar a a}$ identities are contained in the
third
block of Table~\ref{tab:su2fiden}.

The simplest method for obtaining the vanishing two-body identities uses a
trick.  The baryon $SU(2)\times SU(F)$ representations in the completely
symmetric $SU(2F)$ representation have identical Young tableaux for the spin
and flavor
subgroups (see Table~\ref{tab:su2f->suf}).   Eq.~(\ref{VIIi}) then implies that
the
$SU(F)$ Casimir operator $T^a T^a$ of an  arbitrary baryon flavor
representation is
simply related to its $SU(2)$  Casimir $J^i J^i$,
\begin{equation}\label{VIIx}
T^2 = J^2 + {1\over {4F}}\N\left(\N+2F\right)\left(F-2\right).
\end{equation}
This relation, which transforms as an $SU(2)\times SU(F)$ singlet, is a
linear combination of the Casimir identity eq.~(\ref{VIIci}) and the
$(0,0)$ element of the ${\bar a a}$ representation of $SU(2F)$.  The singlet of
the ${\bar a a}$ representation is obtained by finding the linear combination
of
eq.~(\ref{VIIx}) and the Casimir identity which orthogonal to the Casimir
identity.  This linear combination,
\begin{eqnarray}
&&4F\left(2-F\right)\ \left\{G^{ia},G^{ia}\right\} + 3 F^2\ \left\{T^a,
T^a\right\}\nonumber\\
&&\qquad + 4\left(1-F^2\right)\ \left\{J^i,J^i\right\}=0,\label{VIIxi}
\end{eqnarray}
is the first identity in the third block of Table~V.  All the other
elements of the $SU(2F)$ ${\bar a a}$ irreducible representation can be
obtained
by applying raising and lowering operators to the $(0,0)$ element,
eq.~(\ref{VIIxi}).  In our case, all other elements of the ${\bar a a}$
representation can be obtained by commuting identity eq.~(\ref{VIIxi}) with
the generators $J^i$, $T^a$ and $G^{ia}$.  Since $J^i$ and $T^a$ are spin
and flavor generators, they do not produce identities which are in new
$SU(2)\times SU(F)$ representations.  Thus, only commutators of $G^{ia}$
with eq.~(\ref{VIIxi}) need to be evaluated to obtain the other $SU(2)\times
SU(F)$ identities in the $SU(2F)$ ${\bar a a}$ representation.  Applying
successive commutators, and projecting onto definite $SU(2) \times SU(F)$
channels gives the remaining identities in Table~\ref{tab:su2fiden}.  For
ease of notation, we have not explicitly written some of the spin and
flavor projectors in Table~\ref{tab:su2fiden}, but have simply indicated
that both sides of a given equation are to be projected into the relevant
channel.  For example, $\left\{G^{ia}, G^{ja}\right\} (J=2)$ indicates that
only the spin-two piece is to be retained, and is a shorthand notation for
$\left\{G^{ia}, G^{ja}\right\}- (1/3)\delta^{ij}\left\{G^{ka},
G^{ka}\right\}$.

\section{The Operator Identities for Two and Three
Flavors}\label{sec:2and3}

The group theoretic structure of the independent quark operators and the
complete set
of non-trivial quark operator identities
were derived in the previous two sections for an arbitrary number of flavors.
Specialization of these results to two and
three flavors is useful for application of this formalism to QCD.
There is considerable simplification in the results for two flavors, since many
of the $SU(F)$ representations vanish for $F=2$. There is also some
simplification for three flavors.

\subsection{Two Flavors}

For two light quark flavors, the zero---three-body operators transform
according to
the $SU(4)$ irreducible representations
\begin{eqnarray}
&{\rm 0-body:}&\ \ \left( 0 \times 0 \right)= 1, \nonumber\\
&{\rm 1-body:}&\ \ \left( 4 \times \overline 4 \right)= 1 + 15, \nonumber\\
&{\rm 2-body:}&\ \ \left( \overline{10} \otimes 10 \right) =  1 + 15 + 84 ,
\\
&{\rm 3-body:}&\ \ \left( \overline{20} \otimes 20 \right)
= 1 + 15 + 84 + 300.\nonumber
\end{eqnarray}
The symmetric product of two adjoints (see Table~III) transforms as
\begin{equation}
\left( 15 \otimes 15 \right)_S = 1 + 15 + 20 + 84,
\end{equation}
so the vanishing two-body identities transform in the $20$-dimensional
representation
of $SU(4)$.  The $SU(4)$ singlet, adjoint and ${\bar a a}$ two-body identities
are listed in Table~VII.  The $SU(2) \times SU(2)$ decompositions of the $15$,
$20$
and $84$ are given in appendix~B.

\subsection{Three Flavors}

For three light quark flavors, the zero---three-body operators transform
according to
the $SU(6)$ irreducible representations
\begin{eqnarray}
&{\rm 0-body:}&\ \ \left( 0 \times 0 \right)= 1, \nonumber\\
&{\rm 1-body:}&\ \ \left( 6 \times \overline 6 \right)= 1 + 35, \nonumber\\
&{\rm 2-body:}&\ \ \left( \overline{21} \otimes 21 \right) = 1 + 35 + 405,
\\
&{\rm 3-body:}&\ \ \left( \overline{56} \otimes 56 \right)
=  1 + 35 + 405 + 2695. \nonumber
\end{eqnarray}
The symmetric product of two adjoints (see Table~III) transforms as
\begin{equation}
\left( 35 \otimes 35 \right)_S = 1 + 35 + 189 + 405 ,
\end{equation}
so the vanishing two-body identities transform in the $189$-dimensional
representation
of $SU(6)$.  The $SU(6)$ singlet, adjoint and ${\bar a a}$ two-body identities
are
listed in Table~VIII. The $SU(2) \times SU(3)$ decompositions of the $35$,
$189$
and $405$ are given in appendix~B.
The ${\bar a s}+{\bar s a}$ and ${\bar s s}$ representations of $SU(3)$ are the
$10+\overline{10}$
and $27$, respectively.
The ${\bar a a}$ representation doesn't exist for the $SU(3)$ flavor group.

\section{Operator Analysis for the $\bf 1/\N$
Expansion}\label{sec:opanalysis}

We now analyze the spin-flavor structure of baryons in large-$\N$
QCD for $\N$ finite and odd. Any QCD operator which transforms
as an irreducible representation of $SU(2) \times SU(F)$ can be written
as an expansion in $n$-body quark operators, $n=0, \ldots, \N$,
which transform according to the same irreducible representation
(see eq.~(\ref{IIIi})).
The quark operator identities can be used
to construct a linearly independent and complete operator basis of $n$-body
operators with the correct transformation properties.  The operator
basis for any $SU(2) \times SU(F)$ representation contains a finite
number of operators.  The $1/\N$ expansion for any QCD operator can
be simplified by retaining only those operators in the operator
basis which contribute at a given order in $1/\N$.  In this section,
the generic structure of the $1/\N$ expansion for the ground-state
baryons is discussed.  Sections~\ref{sec:axial} through~\ref{sec:nonlep}
derive $1/\N$
expansions for certain static properties of baryons.  The analysis
for two flavors is straightforward and reproduces earlier results.
The analysis for three (or more) flavors is much more subtle and
leads to many new results, as well as reproducing some old results.

In the $\N\rightarrow\infty$ limit, it has been shown that the baryon
states form degenerate irreducible representations of the $SU(2F)$
spin-flavor algebra generated by spin, flavor, and the space components of
the axial flavor currents\cite{dm,djm}.
Matrix elements of the axial currents within a
given irreducible baryon representation are of order $\N$, whereas matrix
elements of the axial currents between different irreducible representations
are at most of order $\sqrt{\N}$.  The mass of the degenerate baryon multiplet
is of order $\N$.  The degeneracy of the baryon spectrum is broken by $1/\N$
corrections, and it has been shown that the $1/\N$ correction
to the baryon masses is proportional to $J^2$  (in the flavor
symmetry limit)\cite{j}.  This baryon mass spectrum is depicted in
Fig.~\ref{fig:spectrum}.  The degenerate baryon $SU(2F)$ multiplet splits
into a tower of states with spin $1/2, \ldots , \N/2$.
Mass splittings between baryon states at the bottom of the tower
($J$ of order one) are of order $1/\N$,
whereas mass splittings between baryon states at the top of the tower ($J$
of order $\N$) are of order one.  The mass difference between the baryon states
at the bottom and top of the towers is of order $\N$, and is of the same order
in $\N$ as the average mass of the baryon multiplet. The $1/\N$ correction to
the
baryon masses is of order $\N$ near the top of the tower, and is not a
small perturbation. However, it is small near the bottom of the tower,
where the baryons have spins which are of order one.  The $1/\N$ expansion
is therefore valid for the lowest spin states in the $SU(2F)$ baryon
representation.
Thus, our analysis considers baryons in the limit $\N\rightarrow\infty$ with
$J$ fixed.

The $1/\N$ expansion of a QCD operator is in terms of a basis of $n$-body
quark operators, where $n=0, \ldots, \N$.  A generic $n$-body operator
can be written as a polynomial of homogeneous degree $n$ in the one-body
operators $J^i$, $T^a$ and $G^{ia}$,
\begin{equation}\label{on}
{\cal O}^{(n)} =\sum_{\ell,m} \left(J^i\right)^\ell
\left(T^a\right)^m \left(G^{ia}\right)^{n-\ell-m},
\end{equation}
so the expansion of a QCD one-body operator has the form
\begin{equation}\label{ope}
{\cal O}_{QCD} =\sum_{\ell,m,n} c^{(n)} {1\over \N^{n-1}} \left(J^i\right)^\ell
\left(T^a\right)^m \left(G^{ia}\right)^{n-\ell-m},
\end{equation}
where summation over different $n$-body operators ${\cal O}^{(n)}_k$ is
implied.  The generalization of eq.~(\ref{ope}) to an $m$-body QCD operator
is clear from
the discussion in Section~\ref{sec:ncount}.
An important feature of the operator expansion~(\ref{ope}) can now be
explained.
We argued in Sec.~\ref{sec:ncount} that the matrix elements of
one-body operators are typically of order $\N$, though they can be smaller
if there are cancellations between insertions of the operator on the various
quark lines.  There is an important example of such a cancellation:
the baryon states with a valid $1/\N$ expansion are restricted to those
states for which the matrix element of the
one-body operator $J$ is of order one, not of order $\N$. As a result, every
factor of
$J$ on the right hand side of eq.~(\ref{ope}) comes with a $1/\N$ suppression.

The quark operator identities can be used to eliminate redundant operators
from the expansion eq.~(\ref{ope}).  A complete and independent operator
basis can be constructed by recursively applying the two-body quark
operator identities of Secs.~\ref{sec:classify} and~\ref{sec:derive}.
The anticommutator of two one-body operators occurs in the irreducible
$SU(2)\times SU(F)$ representations
given in the second column of Table~\ref{tab:opleft}. For
example, the anticommutator of $J^i$ with $J^j$ is a flavor singlet which
transforms in the symmetric tensor product of two spin ones, i.e. it is
either a $(0,0)$ or a $(2,0)$.
A similar analysis yields all the other entries in the second column of
Table~\ref{tab:opleft}.  Some of these operators can be eliminated
using the operator identities listed
in Table~\ref{tab:su2fiden}.  There are a total of 15 identities which
can be used to eliminate 15 operator products in Table~\ref{tab:opleft},
leaving only the representations listed in the third column of the table.
There is a simple structure to the operator products which
remain. Consider the operator products $\{T^a,T^b\}$, $\{T^a,G^{ib}\}$,
and $\{G^{ia},G^{jb}\}$ which each have two adjoint indices. These indices
can be contracted using $\delta^{ab}$ to give a flavor singlet, or with
$d^{abc}$ or $f^{abc}$ to give flavor adjoints. All these contractions
are eliminated by the identities.  The spin indices in $\{G^{ia},G^{jb}\}$ can
be contracted with $\delta^{ij}$ to give spin zero, or with $\epsilon^{ijk}$
to give spin one. These contractions are also eliminated using the identities.
In addition, the $(1,{\bar a a})$ in $\{T^a,G^{ib}\}$ and $(2,{\bar a a})$ in
$\{G^{ia},G^{jb}\}$ can be removed. All other products (including all operator
products involving $J$) remain. To summarize, the reduction of operator
products
is given by the

\medskip

\noindent{\bf Operator Reduction Rule:\ }
All operator products in
which two flavor indices are contracted using $\delta^{ab}$, $d^{abc}$ or
$f^{abc}$, or two spin indices on $G$'s are contracted using
$\delta^{ij}$ or  $\epsilon^{ijk}$ can be eliminated.\footnote{Operators such
as
$f^{acg}d^{bch} \left\{T^g,G^{ih}\right\}$ (which contains $i(\bar s a -
\bar a s)$) are different from $\left\{T^a,G^{ib}\right\}$ (which contains
$\bar s a +
\bar a s$), and are not removed, since the two indices on
$\left\{T^g,G^{ih}\right\}$
are not contracted using a $f$ or $d$-symbol. Many combinations
in which two adjoint indices $g$ and $h$ are contracted with $f^{acg}d^{bch}$,
$f^{acg}f^{bch}$, or $d^{acg}d^{bch}$ can be eliminated using
eqs.~(\ref{Axv})--(\ref{Axxii}).}
In addition, the
$(1,{\bar a a})$ in $\{T^a,G^{ib}\}$ and the $(2,{\bar a a})$ in
$\{G^{ia},G^{jb}\}$
can be eliminated.

\medskip

\noindent The last two exceptional cases are not important for examples of
physical interest, since there is no ${\bar a a}$ representation for $F=2$ or
$3$.
Note that it is possible to choose a different set of independent operators
using the
two-body identities.
The operator reduction rule we have chosen is appealing because it has a nice
physical interpretation, which is discussed in Section~\ref{sec:axial}.

The two-flavor case is special, since the $d$-symbol vanishes for $SU(2)$, so
there
are some additional simplifications. There is a symmetry between spin and
isospin for
the two-flavor case.  The operator reduction rule for two flavors becomes:
All operators in which two spin or isospin indices are contracted with a
$\delta$ or
$\epsilon$-symbol can be eliminated, with the exception of $J^2$.  Note that
the
inclusion of $J^2$, but not $I^2$, in the set of independent operators does not
break
the symmetry between spin and isospin, because of the identity $I^2 = J^2$.

In Sections~\ref{sec:axial} through~\ref{sec:nonlep}, the operator reduction
rule is
used to construct
$1/\N$ expansions for various static properties of baryons.  These include
baryon axial vector currents, masses, magnetic moments and hyperon non-leptonic
decay amplitudes.

\section{Operator Analysis For Completely Broken $\bf SU(3)$
Symmetry}\label{sec:broken}

The $1/\N$ expansion also provides information about the spin-flavor structure
of
baryons to all orders in $SU(3)$ symmetry breaking if the operator analysis
is performed for completely broken $SU(3)$ flavor symmetry.  In this section,
the quark operator identities and the classification of independent $n$-body
operators is analyzed for $SU(2) \times U(1)$ flavor symmetry.  For this
analysis, it is necessary to decompose the one-body operators
$J^i$, $T^a$ and $G^{ia}$ which transform under $SU(3)$ flavor symmetry
into operators with definite isospin and strangeness.  We define new one-body
operators
\begin{eqnarray}
&&I^a = T^a\ \  (a=1,2,3),\nonumber\\
&&G^{ia} = G^{ia}\ \  (a=1,2,3), \nonumber\\
&&t^\alpha = s^\dagger q^\alpha\ \ ,\nonumber\\
&&Y^{i\alpha} = s^\dagger J^i q^\alpha\ \ ,\\
&&N_s = s^\dagger s, \nonumber\\
&&J_s^i = s^\dagger J^i s,\nonumber
\end{eqnarray}
where $I^a$ and $G^{ia}$ are isospin one operators, and $t^\alpha$ and
$Y^{i\alpha}$
are isospin-$1/2$ operators.  The spin and strangeness quantum numbers of these
operators are obvious from the above definitions.
The independent one-body operators for completely broken $SU(3)$ symmetry are
$J^i$, $I^a$, $t^\alpha$, $t^\dagger_\alpha$, $N_s$, $G^{ia}$, $Y^{i\alpha}$,
$Y^{\dagger i}_\alpha$, and $J_s^i$.  This set of operators replaces the
$SU(3)$
one-body operators $J^i$, $T^a$ and $G^{ia}$.  Note that $t^\alpha$ and
$Y^{i\alpha}$,
$\alpha=1,2$, correspond to $T^a$ and $G^{ia}$ for $a=4-i5,6-i7$, respectively.
 The
strange quark number operator $N_s$ and the strange quark spin operator $J_s^i$
originate from
$T^8$ and $G^{i8}$,
\begin{eqnarray}\label{t8}
&&T^8= {1 \over {2\sqrt{3}}}\left( \N - 3 N_s \right),\nonumber\\
&&G^{i8} = {1 \over {2\sqrt{3}}}\left( J^i - 3 J^i_s \right).
\end{eqnarray}
The operator Lie algebra of Table~\ref{tab:su2fcomm} is
unaffected by $SU(3)$ breaking, so the commutation relations of the one-body
operators for $SU(2) \times U(1)$ flavor symmetry can be obtained directly
from the $SU(3)$ flavor commutation relations using the above identifications.
The commutation relations for $SU(2) \times U(1)$ flavor symmetry
are listed in Table~\ref{tab:brokencomm}.  Note that the spin-flavor algebra
contains an $SU(4)$ spin-flavor subgroup with generators $J_{ud}^i$, $I^a$
and $G^{ia}$, where $J_{ud}^i$, the spin operator for the $u$ and $d$ quarks,
is
linearly related to the spin operators $J^i$ and $J_s^i$,
\begin{equation}
J_{ud}^i = u^\dagger J^i u + d^\dagger J^i d = J^i - J_s^i.
\end{equation}
The commutation relations in
Table~\ref{tab:brokencomm} are written in terms of $J_{ud}^i$ and $J_s^i$,
rather
than $J^i$ and $J_s^i$ so that the $SU(4)$ spin-flavor symmetry is manifest.
The full spin-flavor symmetry of the algebra is $SU(4) \times SU(2) \times
U(1)$,
where the $SU(2)$ factor is strange quark spin and the $U(1)$ factor is the
number of strange quarks.

The two-body operator identities for $SU(4) \times SU(2) \times U(1)$
spin-flavor symmetry can be obtained by decomposing the $SU(6)$ identities in
Table~\ref{tab:su6iden} into irreducible representations with
definite spin $J$, isospin $I$, and strange quark number $S$\footnote{Note that
$S$ is
defined as strange quark number $N_s$, not strangeness.  Since the strangeness
of an
$s$-quark is $-1$, $S$ is the negative of strangeness.}. The resulting
identities are given in Tables~\ref{tab:broken0} and~\ref{tab:brokennot0}.  The
identities are denoted by their
$SU(2) \times SU(2)
\times U(1)$ quantum numbers
$(J,I)_S$.  Table~\ref{tab:broken0} contains the $S=0$ identities and
Table~\ref{tab:brokennot0} contains the $S=1$ and $S=2$ identities.  (The
$S=-1$ and
$S=-2$ identities are the conjugates of the identities in
Table~\ref{tab:brokennot0}.)
The identities are written most easily in terms of the spin
operators $J^i_{ud}$ and $J_s^i$.

The anticommutators of two one-body operators occur in the irreducible
$SU(2)\times SU(2) \times U(1)$ representations
given in the second column of Table~\ref{tab:brokenleft}.  The tables
of identities contain a total of 33 operator identities which can be
used to eliminate 33 different representations appearing in the second column
of
Table~\ref{tab:brokenleft}.  The operator products which remain appear
in the third column of the table.  There are several simplifications
which occur as a result of operator reduction.
All $Y Y^\dagger$, $Y t^\dagger$, $t Y^\dagger$ and $t t^\dagger$
anticommutators can be eliminated using the operator identities. This implies
that independent $n$-body operators with $\Delta S = 1$ ($\Delta S = -1$)
contain only one factor of $t$ or $Y$ ($t^\dagger$ or $Y^\dagger$).  It
also implies that $\Delta S=0$ operators can be simplified so that they do not
contain $t$, $t^\dagger$, $Y$ or $Y^\dagger$.
All operator combinations in which two isovector indices are
contracted with $\delta^{ab}$ or $\epsilon^{abc}$, or in which an
isovector and isospinor index are contracted with
$\left({\tau^a \over 2}\right)^\alpha_\beta$,
can be eliminated. In  addition, the product of two
$G$'s or two
$Y$'s or a
$G$ and $Y$ in which  the spin indices are contracted with a $\delta^{ij}$ or
$\epsilon^{ijk}$ can be  eliminated.  A few other operators also can be
eliminated.
To summarize, the reduction of operator products
for the spin $\otimes$ flavor group $SU(2) \times SU(2) \times U(1)$ obeys the

\medskip

\noindent{\bf Operator Reduction Rule II:\ }
\begin{enumerate}
\item All products of the form $t^\alpha t^\dagger_\beta$, $t^\alpha
Y^{\dagger i}_\beta$, $Y^{i\alpha} t^\dagger_\beta$ and $Y^{i\alpha}
Y^{\dagger j}_\beta$ can be eliminated.
\item All operator products in
which two isovector indices are contracted using $\delta^{ab}$ or
$\epsilon^{abc}$, or an isovector and an isospinor index are contracted
with $\left({\tau^a \over 2}\right)^\alpha_\beta$, can be eliminated.
\item All operator products $G^{ia} G^{jb}$, $Y^{i\alpha} G^{jb}$, and
$Y^{i\alpha} Y^{j\beta}$ (and their conjugates)
in which two spin indices are contracted using
$\delta^{ij}$ or $\epsilon^{ijk}$ can be eliminated.
\item The operators $\{ J_s^i, J_s^i \}$, $\{ G^{ia}, J_{ud}^i \}$,
$\{ Y^{i\alpha}, J_s^i \}$ and $i\epsilon^{ijk} \{Y^{i\alpha}, J_s^j\}$ (and
their
conjugates) can be eliminated.
\end{enumerate}.

Note that there are operators with two isospinor indices contracted with
$\epsilon_{\alpha\beta}$ which can ${\it not}$ be eliminated, namely
$\epsilon_{\alpha\beta}\{t^\alpha, t^\beta\}$ and
$i\epsilon_{\alpha\beta}\{t^\alpha, Y^{i\beta}\}$.

Again, it is possible to choose a different set of independent operators using
the
two-body identities.  The operator reduction rule chosen here is one of the
simplest.

\section{Axial Currents and Meson Couplings}\label{sec:axial}

The results of the previous sections will be used to obtain $1/\N$ expansions
for the
baryon axial currents and meson couplings, masses, magnetic moments and hyperon
non-leptonic decay amplitudes.  The axial currents and meson couplings are
considered in
this section.  We begin by deriving the $1/\N$ expansion for the baryon axial
vector
current in the $SU(F)$ flavor symmetry limit.  Expansions for the axial
currents to first order in $SU(3)$ breaking and for $SU(2) \times U(1)$ flavor
symmetry
are derived in subsequent subsections.

Only the space components of the axial current have non-zero matrix elements at
zero
recoil, so the axial vector current $A^{ia}$ transforms as $(1,adj)$ under
$SU(2) \times SU(F)$.  The $1/\N$ expansion for $A^{ia}$ derived in this
section
can be applied to any other baryon operator which transforms as a
$(1, adj)$, such as the magnetic moment operator.

The $n$-body quark operators in the $1/\N$ expansion of $A^{ia}$
reduce to operators containing only a single factor of $G^{ia}$ or $T^a$
by the operator reduction rule, which eliminates all operator products
with contracted flavor indices.  The only one-body operator is
$G^{ia}$. There are two allowed two-body operators, which can be written as
\begin{eqnarray}\label{VIIIii}
{\cal O}_2^{ia} &=& \epsilon^{ijk}\left\{ J^j, G^{ka} \right\}
= i \left[ J^2, G^{ia} \right] ,\\
{\cal D}_2^{ia} &=& J^i T^a .\label{VIIIiii}
\end{eqnarray}
There are two three-body operators, which can be written as
\begin{eqnarray}
{\cal O}_3^{ia} &=& \left\{ J^2, G^{ia} \right\} - \frac12
\left\{ J^i, \left\{ J^j, G^{ja} \right\} \right\} ,\label{VIIIiv}\\
{\cal D}_3^{ia} &=& \left\{ J^i, \left\{ J^j, G^{ja} \right\} \right\} .
\label{VIIIv}
\end{eqnarray}
All remaining $n$-body operators can be obtained recursively by applying
anticommutators of $J^2$ to the above operators.
For $n \ge 2$, the independent $(n+2)$-body operators are given by
\begin{eqnarray}
{\cal O}_{n+2}^{ia} &=& \left\{ J^2, {\cal O}_n^{ia} \right\},\label{VIIIvi}\\
{\cal D}_{n+2}^{ia} &=& \left\{ J^2, {\cal D}_n^{ia} \right\}. \label{VIIIvii}
\end{eqnarray}
The set of operators $G^{ia}$, ${\cal O}_n^{ia}$
and ${\cal D}_n^{ia}$, $n=2,3,\ldots,\N$, form a complete set of linearly
independent
spin-one adjoints. The operators ${\cal D}_n^{ia}$ are diagonal operators, in
the
sense that they have non-zero matrix elements only between states with the same
spin.
The operators ${\cal O}_n^{ia}$ are purely off-diagonal, in the sense that they
only
connect states with different spin.  The operator $G^{ia}$ is neither diagonal
or off-diagonal.  The operators $G^{ia}$, ${\cal D}_n^{ia}$ and ${\cal
O}_{2m+1}^{ia}$,
$m=1,2,\ldots$, are odd under time reversal, whereas the operators ${\cal
O}_{2m}^{ia}$,
$m=1,2,\ldots$ are even under time reversal.  Since $A^{ia}$ is $T$-odd, the
operators
${\cal O}_{2m}^{ia}$ do not occur in the $1/\N$ expansion of the axial vector
current.

The $1/\N$ expansion for $A^{ia}$ is
\begin{equation}\label{VIIIviii}
A^{ia} = a_1 G^{ia} + \sum_{n=2,3}^{\N} b_n {1 \over\N^{n-1}}{\cal D}_n^{ia} +
\sum_{n=3,5}^{\N} c_n {1\over\N^{n-1}}{\cal O}_n^{ia},
\end{equation}
where the coefficients $a_1$, $b_n$ and $c_n$ have expansions in powers of
$1/\N$ and are order unity at leading order in the $1/\N$ expansion.

The expansion for $A^{ia}$ eq.~(\ref{VIIIviii}) has a simple physical
interpretation
in terms of quark line diagrams (see Fig. 4).  An insertion of the axial
current
operator on a quark line $r$ gives $J^i_r T^a_r$ acting on that quark line.
(The
subscript $r$ implies that the operator acts only on the $r^{\rm th}$ quark.)
Summation
over all quark lines yields the operator
\[
\sum_r J^i_r T^a_r = G^{ia}.
\]
Spin-independent gluon exchange renormalizes the operator $G^{ia}$.
Spin-dependent gluon exchange produces
additional factors of $J$ acting on different quark lines.
The most
general operator structure is a flavor matrix
$T^a_r$ on some quark line $r$, and a product of $J$'s on one or more different
quark
lines $s_1,\ldots s_\ell$, summed over all possible choices for
$r,s_1,\ldots,s_\ell$.
Products of $J_r$ on a
single quark line can be reduced to at most one factor of $J_r$ since
\[
J_r^i J_r^j = \frac 14 \delta^{ij} + i\frac12\epsilon^{ijk} J_r^k
\]
for the spin-1/2 operator $J_r^i$. If $r$ is not equal to any of the $s_k$'s,
the
operator produced after  summing over all possible quark combinations is
\[
\sum_{r,s_1,\ldots,s_\ell} T^a_r J_{s_1} \ldots J_{s_\ell} = T^a J\ldots J,
\]
where the indices on the $J$'s are combined to form a spin-one operator.
If $r$ is equal to one of the $s_k$'s, the
operator produced after summing over all possible quark combinations is
\[
\sum_{r,s_1,\ldots,s_\ell} T^a_r J_r J_{s_1} \ldots J_{s_\ell} = G^{ia} J\ldots
J,
\]
where the indices on $G^{ia}$ and the $J$'s are combined to form a spin-one
operator.
The operator expansion eq.~(\ref{VIIIviii}) has this form. The above
diagrammatic
argument is similar to the one given in ref.~\cite{lm}.

The expansion for $A^{ia}$ eq.~(\ref{VIIIviii}) can be reduced to a finite
operator series using the fact that each factor of $J$ comes with a $1/\N$
suppression.
The truncation of the operator series is different for the two flavor and three
flavor cases.

\subsection{Meson Couplings in the Flavor Symmetry Limit}

\subsubsection{Two Flavors and the $\bf I=J$ Rule}\label{sec:twoflavor}

In the two-flavor case, the isospin $I$ is of order one for baryons
with spin $J$ of order one, since $I=J$. Thus every factor of $I$ and $J$
in eq.~(\ref{VIIIviii}) brings a $1/\N$ suppression. The matrix elements of the
operator $G^{ia}$ are of order $\N$, as are the matrix elements of
${\cal O}_m^{ia}$ and ${\cal D}_m^{ia}$, $m$ odd. The matrix elements of
${\cal D}_2^{ia}= J^i I^a$,
and ${\cal D}_m^{ia}$, $m$ even, are of order one.  Thus, the operator
expansion
eq.~(\ref{VIIIviii}) can be truncated after the one-body operator $G^{ia}$,
\begin{equation}\label{VIIIix}
A^{ia} = a_1\ G^{ia} \left[ 1 +{\cal O}\left({1\over\N^2}\right) \right].
\end{equation}
Eq.~(\ref{VIIIviii}) implies that there are no $1/\N$ corrections to ratios
of pion-baryon couplings for two quark flavors\cite{dm}. It also implies
that there are no $1/\N$ corrections to ratios of pion-baryon couplings
amongst baryons in a given strangeness sector for three quark
flavors\cite{djm}.

In addition, one can prove that pion couplings to baryons are purely
$p$-wave in the large $\N$ limit, which is an example of
the $I=J$ rule of Mattis and collaborators~\cite{mattis}.
For arbitrary $\N$, the baryons form a tower of states with $J$ ranging
from $1/2$ to $\N/2$. Neglecting baryon recoil (since the baryon mass is
of order $\N$), the only pseudoscalar meson coupling between the spin-1/2
and spin-3/2 states is a $p$-wave ($\ell=1$) coupling. However, higher spin
baryons can have couplings in other angular momentum channels, such as
$\ell=3,5,\ldots$, where $\ell$ must be odd by parity. A meson
coupling in the $\ell^{\rm th}$ partial wave is given by an operator
$A^{(i_1,\ldots, i_\ell) a}$ which is completely symmetric and traceless in
the $\ell$ spin indices. The operator reduction rule implies
that at leading order in $\N$, this operator is proportional to
$G^{i_1a} J^{i_2} J^{i_3}\ldots J^{i_\ell}/\N^{\ell-1}$, completely symmetrized
in
the spin indices, with a correction of relative order $1/\N^2$.
Since matrix elements of $G^{ia}$ are of order $\N$, and
matrix elements of $J^i$ are of order one, we conclude that the $\ell=1$
coupling
is of order $\N$ (which is just eq.~(\ref{VIIIix})), the $\ell=3$ coupling is
of order
$1/\N$, the $\ell=5$ coupling is of order $1/\N^3$, etc. In the large
$\N$ limit, all of the higher partial waves vanish, and the pion coupling to
baryons is purely $p$-wave, isospin one, i.e. it has $I=J$.
Mattis et al.\ \cite{mattis} originally derived the $I=J$ rule in the Skyrme
model.
The $I=J$ rule is true in large $\N$ QCD because $G^{ia}/\N$ which has $I=J=1$
is of order one, whereas the matrix elements of $I/\N$ and $J/\N$, which have
$|I-J|=1$, are of order $1/\N$. We also get the additional result
that the higher partial wave pion-couplings are of order
$1/\N^{\ell-2}$ in the $1/\N$ expansion. In general, pion couplings which
violate
the $I=J$ rule are
suppressed by a factor of $1/\N^{|I-J|}$ relative to the $I=J$ coupling, which
is of
order $\N$. A similar result also holds for other meson-baryon couplings. For
example,
$p$-wave kaon couplings (which have $J=1$ and $I=1/2$) are of order $\sqrt\N$,
and $p$-wave $\eta$ couplings (which have $J=1$ and $I=0$) are of order
one~\cite{djm};
the spin-one $\rho$ coupling and spin-zero $\omega$ couplings are of order
$\N$; and
the spin-zero $\rho$ coupling and spin-one $\omega$ coupling are of order
one~\cite{mattis,cgo}.

\subsubsection{Three Flavors}

The analysis of the $1/\N$ expansion of the axial current for three or
more flavors is essentially the same.  We employ the familiar language
of $SU(3)$ symmetry here for concreteness.

For three flavors, matrix elements of the flavor generators $T^a$ and matrix
elements of the spin-flavor generators $G^{ia}$ are not the same order in the
$1/\N$
expansion everywhere in the flavor weight diagram.  Consider, for example, the
weight
diagram of the spin-$1/2$ baryons (Fig.~2).  Baryons
with strangeness of order $\N^0$ (near the top of the weight diagram)
have matrix elements of $T^a$,
$a=1,2,3$ and $G^{i8}$ of order one; matrix elements of $T^a$ and $G^{ia}$,
$a=4,5,6,7$, of order $\sqrt{\N}$; and matrix elements of $T^8$ and $G^{ia}$,
$a=1,2,3$ of  order $\N$.  In other regions of the weight diagram, matrix
elements of different linear combinations of the $T$'s and $G$'s are of order
$\N$,
$\sqrt{\N}$ and one.  This non-trivial $\N$-dependence of the matrix elements
of $T^a$ and $G^{ia}$ makes the analysis of the $1/\N$ expansion for three (or
more) flavors more complicated than that for two  flavors, because matrix
elements of
the flavor generators $T^a$ are not suppressed relative to $G^{ia}$. In fact,
matrix elements of $T^a$ can be a factor of $\N$ larger than matrix elements of
$G^{ia}$ (for some values of the index $a$) in some regions of the flavor
weight
diagrams.

The $1/\N$ expansion of the axial vector current is tractable for three or more
flavors because of the operator reduction rule and because matrix elements of
the spin $J$ are suppressed for our choice of limit.
The general form of
the $1/\N$ expansion for the axial current eq.~(\ref{ope}) contains terms with
arbitrary powers of $T^a/\N$, $G^{ia}/\N$ and $J^i/\N$.
Factors of $T^a/\N$ and $G^{ia}/\N$ are of order one somewhere in the
weight diagram, but $J^i/N$ is of order $1/\N$ everywhere.  Thus, operators
with
arbitrary powers of $T^a/\N$ and $G^{ia}/\N$ are all equally important and must
be
retained.  The operator reduction rule, however, allows us to reduce this set
of operators to the subset with one factor of $T^a$ or $G^{ia}$,
eq.~(\ref{VIIIviii}).
The operator expansion eq.~(\ref{VIIIviii}) can be truncated after the first
two
terms,
\begin{equation}\label{VIIIx}
A^{ia} = a_1\ G^{ia} + b_2\ {1\over\N}{\cal D}_2^{ia} = a_1\ G^{ia} + b_2 {J^i
T^a\over\N},
\end{equation}
since all other terms in eq.~(\ref{VIIIviii}) are suppressed by at least
$1/\N^2$ relative to the terms which have been retained {\it everywhere} in
the weight diagram.
{}From the operator definitions eqs.~(\ref{VIIIii})--(\ref{VIIIvii}), it is
easy to see
that: (i) The operators ${{\cal O}_{n}^{ia}/\N^{n-1}}$, $n$ odd, are of order
$1/\N^{n-1}$  relative to $G^{ia}$ everywhere in the weight diagram;
(ii) The operators ${{\cal D}_{n}^{ia}/\N^{n-1}}$, $n$ odd, are suppressed by
$1/N^{n-1}$  relative to $G^{ia}$ everywhere in the weight diagram; and
(iii) The operators ${{\cal D}_{n}^{ia}/\N^{n-1}}$, $n$ even, are suppressed
by $1/N^{n-2}$ relative to ${\cal D}_2^{ia}/\N$ everywhere in the weight
diagram.
Although the second term in eq.~(\ref{VIIIx}) contains an explicit factor of
$1/\N$,
this term is not necessarily suppressed relative to the first term, and must
be retained for a valid truncation.

In eq.~(9.3) of ref.~\cite{djm}, we defined the $SU(3)$-invariant
meson couplings $\cal M$ and $\cal N$, and proved that
\begin{equation}\label{VIIIxi}
{{\cal N}\over{\cal M}} = {1\over2} + {\alpha\over\N} + {\cal
O}\left({1\over\N^2}\right).
\end{equation}
Eq.~(\ref{VIIIx}) implies
\begin{equation}\label{VIIIxii}
\alpha = - {3 \over 2}\left( 1 + {b_2 \over a_1} \right) ,
\end{equation}
so that $b_2/a_1$ determines the  parameter $\alpha$. Eq.~(\ref{VIIIx}) is the
quark
representation analog of eq.~(9.3) of ref.~\cite{djm} for the meson couplings
in
the $SU(3)$ limit.

We are interested in the baryon couplings for the physical case of
$\N=3$. In the $SU(3)$ limit, the octet baryon couplings are described by the
parameters $D$ and $F$, the decuplet to octet transition couplings are
described by ${\cal C}$, and the decuplet couplings are described by
${\cal H}$\cite{baryonxpt}. These parameters can be determined by taking matrix
elements of eq.~(\ref{VIIIx}) in quark model baryon states,
\begin{eqnarray}\label{dfch}
D &=& {1 \over 2} a_1, \nonumber\\
F&=& {1 \over 3} a_1 + {1 \over 6} b_2, \nonumber\\
{\cal C} &=& -a_1,\\
{\cal H} &=& -{3 \over 2} \left( a_1 + b_2  \right), \nonumber
\end{eqnarray}
where we have set $\N=3$.
Note that the diagonal operator ${\cal D}_2$, which is the product of spin and
flavor
generators, only affects the diagonal couplings $F$ and ${\cal H}$. The
parameter $b_2$
produces deviations from the $SU(6)$ prediction. One can eliminate the unknown
parameters
to get the two relations,
\begin{equation}\label{ch}
{\cal C}= - 2D,\qquad {\cal H} = 3 D - 9 F.
\end{equation}
The first relation is an $SU(6)$ relation.
The $F/D$ ratio is given by
\begin{equation}\label{fdeqn}
{F\over D}= \frac23 + \frac{b_2}{3a_1}.
\end{equation}
One cannot determine the accuracy in $1/\N$ of $F/D$ from eq.~(\ref{fdeqn}),
since
this equation is only valid at $\N=3$.

One can also analyze the meson couplings for arbitrary $\N$, and study the
limit as
$\N\rightarrow 3$.
The $1/\N$ dependence of meson-baryon couplings depends on the
location of the baryon in the $SU(3)$ weight diagram. If one considers baryons
with
strangeness of order
$\N^0$, the second  term in eq.~(\ref{VIIIx}) is of order $1/\N^2$ relative to
the first
term for $a=1,2,3$, ($\pi$  couplings); of relative order $1/\N$ for
$a=4,5,6,7$, ($K$ couplings); and of the same order for $a=8$ ($\eta$
couplings). For baryons with strangeness of order $\N^0$, the unknown ratio
$b_2/a_1$
affects pion couplings at order $1/\N^2$, kaon couplings at order $1/\N$, and
$\eta$
couplings at order one~\cite{djm}.
If one defines  $F/D$ using ratios of pion couplings
between baryons of strangeness $\N^0$, one concludes that $F/D=2/3 +{\cal
O}(1/\N^2)$.
Similarly a determination using $K$ and $\eta$ couplings gives $F/D=2/3 +{\cal
O}(1/\N)$ and $F/D=2/3 +{\cal O}(1)$, respectively. The ratio of $\pi$
couplings was
used in ref.~\cite{djm} since it gives the most accurate determination of
$F/D$.
 For a more detailed discussion of the
meson-couplings in the $SU(3)$ limit and the $F/D$ ratio, see Sections~IX
and~XI of
ref.~\cite{djm}.

At next order in the $1/\N$ expansion, the meson couplings have the form
\begin{equation}\label{mesonsubleading}
A^{ia} = a_1\ G^{ia} + b_2 {J^i T^a\over\N}+b_3 {{\cal D}_3^{ia}\over\N^2} +
c_3 {{\cal O}_3^{ia}\over\N^2}.
\end{equation}
Since there are four linearly independent operators in
eq.~(\ref{mesonsubleading}),
the four parameters $D$, $F$, ${\cal C}$ and ${\cal H}$,
are all independent at this order in $1/\N$ expansion, and there is no
prediction.

\subsection{Meson Couplings with Perturbative $\bf SU(3)$
Breaking}\label{sec:pert}

In this section, we compute the $1/\N$ expansion for a $(1,8)$
operator (such as the axial current or meson coupling) to first order
in $SU(3)$ breaking. Flavor $SU(3)$ breaking in QCD is due to the
light quark masses and transforms as a flavor octet. We will neglect isospin
breaking,
and work to linear order in the $SU(3)$ symmetry breaking perturbation
$\epsilon {\cal H}^8$.

The $SU(3)$ symmetry-breaking correction to the axial current is
computed to linear order in $SU(3)$ symmetry breaking from
the tensor product of the axial current which transforms as $(1,8)$, and the
perturbation $(0,8)$.  This tensor product contains a $(1,0)$, $(1,8)$, $(1,8)$
$(1,10 + \overline{10})$, and $(1,27)$.  The form of these operators is
determined by a
straightforward application of the operator reduction rule.

The series ${\cal O}_n$ for the $(1,0)$ operators starts with the one-body
operator
${\cal O}_1=J^i$. Consecutive terms in the series are generated by ${\cal
O}_{n+2}=
\left\{J^2,{\cal O}_n\right\}$. The higher order terms in the expansion of the
meson
couplings are all suppressed by at least $1/\N^2$ relative to the leading
operator
$J^i$, which gives the symmetry breaking contribution
\begin{equation}\label{pertsinglet}
\delta A^{ia}_1 \propto \delta^{a8} J^i \ .
\end{equation}

The $(1,8)$ operator has the same form
as eq.~(\ref{VIIIx}). The $(1,8)$ $SU(3)$-breaking correction to the axial
currents is
of the form
\begin{equation}\label{pertoctet}
\delta A^{ia}_8 \propto d^{ab8}\left[ c_1\ G^{ib} +
c_2 {J^i T^b\over\N}\right]\ .
\end{equation}
A similar term with the $d$-symbol replaced by an $f$-symbol is forbidden by
time
reversal. Neglected operators are suppressed by $1/\N^2$ relative to the
operators
we have retained.

There is a second $(1,8)$ operator of the form
\begin{equation}\label{fepsop}
\delta A^{ia}_8 \propto  f^{ab8} \epsilon^{ijk} \{ J^j, G^{kc} \} \ .
\end{equation}
Neglected operators contain additional factors of $J^2/\N^2$, and are
suppressed by
at least $1/\N^2$ relative to eq.~(\ref{fepsop}).
The operator eq.~(\ref{fepsop}) can be written as
\begin{equation}
\delta A^{ia}_8 \propto \left[J^2,\left[T^8,G^{ia}\right]\right],
\end{equation}
which shows that it contributes only to amplitudes which change both $J$ and
strangeness
($T^8$).

The operator reduction rule implies that the $(1,10+\overline{10})$
($(1,{\bar a s}+{\bar s a})$) representation is given by  (see
Table~\ref{tab:opleft} and
Appendix~\ref{app:suq})
\begin{equation}\label{eqassa}
\{G^{ia},T^b\}-\{G^{ib},T^a\}-{2\over3} f^{abc} f^{cgh}
\{G^{ig},T^h\}.
\end{equation}
Additional operators have at least two more factors of $J$, and are suppressed
by
$1/\N^2$ relative to the operator in eq.~(\ref{eqassa}). Note that
\begin{equation}\label{doublecomm}
\left[J^2,\left[T^b,G^{ia}\right]\right] =
\left[T^2,\left[T^b,G^{ia}\right]\right] =
f^{abc} f^{cgh} \{G^{ig},T^h\},
\end{equation}
where the first equality follows from eq.~(\ref{VIIx}), which implies that
$J^2-T^2$
equals a constant. Substituting eq.~(\ref{doublecomm}) into eq.~(\ref{eqassa})
gives
\begin{equation}\label{pertassa}
\delta A^{ia}_{10+\overline{10}} \propto
\{G^{ia},T^8\}-\{G^{i8},T^a\}-
{2\over3}\left[J^2,\left[T^8,G^{ia}\right]\right]\ .
\end{equation}

The $(1,27)$ symmetry-breaking
correction can be built out of the spin-zero $27$ in
$\{T^a,T^b\}$; the spin-one $27$ in $\{G^{ia},T^b\}+\{G^{ib},T^a\}$; and
the spin-two $27$ in $\{G^{ia},G^{ib}\}+\{G^{ia},G^{ib}\}$ (see
Table~\ref{tab:opleft}). The spin-zero
and spin-two $27$'s must be combined with additional factors of $J/\N$ to form
a
$(1,27)$, so to leading order in $1/\N$, only the spin-one $27$ needs to be
retained.
We do not need to subtract off the singlet and octet parts of
$\{G^{ia},T^b\}+\{G^{ib},T^a\}$ because
they can be  absorbed into the symmetry-breaking singlet and octet operators
(\ref{pertsinglet}) and~(\ref{pertoctet}) which have already been included.
Thus, the symmetry-breaking $(1,27)$ correction is proportional to
\begin{equation}\label{pert27}
\delta A^{ia}_{27} \propto
\{G^{ia},T^8\}+\{G^{i8},T^a\} \ .
\end{equation}
The neglected higher order $27$ operators are only suppressed by $1/\N$
relative to the
operator we have retained.

The leading order $1/\N$ expansion for the axial current to first order in
$SU(3)$ symmetry-breaking is given by the lowest order $SU(3)$ symmetry result
eq.~(\ref{VIIIx}) and the sum of the perturbations
eqs.~(\ref{pertsinglet})--(\ref{pert27}).  The perturbations involving $T^8$
and
$G^{i8}$ can be simplified using eq.~(\ref{t8}).  The final expression for the
axial current to linear order in symmetry breaking is
\begin{eqnarray}\label{axpluspert}
A^{ia} + \delta A^{ia} &=& \left(a + \epsilon c_1 d^{ab8}\right) G^{ib} +
\left(b + \epsilon c_2 d^{ab8} \right) {J^i T^b\over\N}\nonumber\\
&& + \epsilon c_3 {\left\{G^{ia},N_s\right\}\over \N}
+\epsilon c_4{\left\{J_s^i,T^a\right\}\over \N} \\
&& + \epsilon{c_5\over 3\N}
\left[J^2,\left[N_s,G^{ia}\right]\right] +\epsilon c_6 \delta^{a8}
J^i,\nonumber
\end{eqnarray}
where the parameter $\epsilon$ emphasizes which terms result from symmetry
breaking,
and the parameters $a$ and $b$ contain terms of order $\epsilon$ from the
substitutions
for $T^8$ and $G^{i8}$. The double commutator term in eq.~(\ref{axpluspert})
only
contributes to processes which change both spin and strangeness.

The coefficient $c_6$ in eq.~(\ref{axpluspert}) can be constrained due to the
following
considerations.  We focus on baryons containing no strange quarks.
The matrix elements of the isovector axial current ($\bar q \gamma^\mu\gamma_5
\tau^a q$)
are of order $\N$ in the strangeness zero sector, since the expansion for the
isovector
axial current involves the one-body operator $G^{ia}$, which has matrix
elements of order
$\N$. The flavor singlet axial current ($\bar q \gamma^\mu\gamma_5 q$) has
matrix
elements of order one in the strangeness zero sector, since matrix elements of
the
one-body operator $J^i$ are of order one.  The expansion of the strange axial
current
($\bar s \gamma^\mu\gamma_5 s$) involves the one-body operator
$J^i$, so naively one expects the matrix elements of the strange axial current
also to
be of order one. However, in the strangeness-zero sector, $\bar s
\gamma^\mu\gamma_5
s$ can only couple to the baryon through a closed $s$-quark loop, which is
accompanied by
a $1/\N$ suppression factor.  Thus, matrix elements of the strange axial
current
are of order $1/\N$, not order one, for strangeness-zero baryons.
$SU(3)$ breaking effects in the strangeness-zero
sector involve closed $s$-quark loops, and so $SU(3)$ breaking
has a $1/\N$ suppression factor.
Diagrams involving a closed loop, such as those in Fig.~(\ref{fig:sloop}), lead
to
quark mass dependence in axial current matrix elements.  Graphs, such as in
Fig.~(\ref{fig:sloop}a), in which the axial current is not inserted in the
closed loop
depend only on $\Tr M$, where $M$ is the quark mass matrix, and yield matrix
elements
which are quark mass-dependent, but which do not violate $SU(3)$. (An example
of this
kind is the pion-nucleon sigma term.) $SU(3)$ violation arises from diagrams
such
as Fig.~(\ref{fig:sloop}b) in which the axial current operator is inserted in
the closed
quark loop. This diagram can only produce $SU(3)$ violation in the baryon axial
currents
at order $1/\N$.  Thus, $SU(3)$ violation in the axial currents must be order
$1/\N$
for strangeness-zero baryons.

Imposing this constraint on the expansion eq.~(\ref{axpluspert}) fixes the
coefficient
$c_6$.  In the strangeness zero sector, eq.~(\ref{axpluspert}) reduces to
\begin{equation}
\left(a + \epsilon c_1 d^{ab8}\right) G^{ib} +
\left(b + \epsilon c_2 d^{ab8} \right) {J^i T^b\over\N}+\epsilon c_6
\delta^{a8} J^i.\nonumber
\end{equation}
Requiring that there is no $SU(3)$ symmetry breaking between the $\pi$ and
$\eta$ couplings at order one gives the constraint
\begin{equation}\label{constraint}
3 c_6 = c_1 + c_2,
\end{equation}
which reduces the number of parameters in eq.~(\ref{axpluspert}) to six
parameters.
Eqs.~(\ref{axpluspert}) and (\ref{constraint}) yield the following expressions
for the
pion, kaon and $\eta$ couplings.

The pion couplings are given by
\begin{eqnarray}\label{pionpert}
\pi^{ia} &=& \tilde a G^{ia} +
\tilde b{J^i T^a\over\N}\nonumber\\
&& + \epsilon c_3 {\left\{G^{ia},N_s\right\}\over \N}
+\epsilon c_4{\left\{J_s^i,T^a\right\}\over \N},
\end{eqnarray}
where
\[
\tilde a = a + \frac1{\sqrt3}\epsilon c_1,\ \ \
\tilde b = b + \frac1{\sqrt3}\epsilon c_2.
\]
If one considers baryons with strangeness of order $\N^0$, the matrix elements
of $G^{ia}$ are of order
$\N$, whereas the matrix elements of $J^i T^a$ and $J_s^i T^a$ are of order
unity. To
the order we are working, it is consistent to retain only
\begin{equation}\label{pionpertii}
\pi^{ia} = \tilde a G^{ia}
+ \epsilon c_3 {\left\{G^{ia},N_s\right\}\over \N} .
\end{equation}
Eq.~(\ref{pionpertii}) determines
the pion couplings of baryons with a given
strangeness to relative order $1/\N^2$, and it implies that pion couplings have
a linear dependence on strangeness at relative order $1/\N$~\cite{djm}.

The kaon couplings are given by
\begin{eqnarray}\label{kaonpert}
K^{ia} &=& \left(\tilde a -\frac{\sqrt3}{2} \epsilon  c_1\right) G^{ia} +
\left(\tilde b -\frac{\sqrt3}{2} \epsilon c_2 \right) {J^i
T^a\over\N}\nonumber\\
&& + \epsilon c_3 {\left\{G^{ia},N_s\right\}\over \N}
+\epsilon c_4{\left\{J_s^i,T^a\right\}\over \N} \\
&& + \epsilon{c_5\over 3\N}
\left[J^2,\left[N_s,G^{ia}\right]\right],\nonumber
\end{eqnarray}
where $\left[N_s,G^{ia}\right]=\pm G^{ia}$ depending on whether one is looking
at
terms which annihilate $K^+,K^0$ or $K^-,\bar K^0$. The double commutator term
only
contributes to processes which change $J^2$.

The $\eta$ couplings are given by
\begin{eqnarray}
\eta^i &=& \frac1{2\sqrt3}\left(\tilde a+\tilde b\right) J^i +
\left(-\frac{\sqrt3}{2}\tilde a + \epsilon c_1+ \frac1{\sqrt{3}} \epsilon
c_4\right)  J_s^i \nonumber
\\ && + \left( -\frac{\sqrt3}{2}\tilde b + \epsilon c_2+\frac1{\sqrt{3}}
\epsilon
c_3\right)
\frac{N_s}{\N} J^i \nonumber\\
&&\qquad -\sqrt3 \epsilon \left(c_3+c_4\right) \frac{N_s}{\N} J^i_s.
\label{etapert}
\end{eqnarray}

A detailed comparison of these equations with the experimental data is given in
ref.~\cite{ddjm}. They provide a good description of the data on decuplet
$\rightarrow$ octet decays (such as $\Delta\rightarrow N\pi$), and baryon
semileptonic decays.

\subsection{Meson Couplings Without $SU(3)$ Symmetry}

The $1/\N$ expansion can also be used to obtain meson couplings without
assuming
$SU(3)$ symmetry. For the $1/\N$ expansion to be valid, one must work with
states
which differ in strangeness by order unity. We will work near the top of the
$SU(3)$ weight diagram, where the number of strange quarks in the baryon is of
order one, and take the large-$\N$ limit with the number of strange
quarks of the baryon held fixed.  This form of the
$1/\N$ expansion was discussed in detail in ref.~\cite{djm}.  We rederive the
results
obtained there using the quark representation, and then compare with the
perturbative
symmetry breaking results obtained in the previous subsection.

\subsubsection{Pion Couplings}

The isovector axial current and $p$-wave pion couplings have $(J,I)_S$ quantum
numbers
$(1,1)_0$. The second operator reduction rule implies that all $t$,
$t^\dagger$, $Y$,
and $Y^\dagger$ operators can be eliminated using the identities, so the pion
coupling
can be written as a function of $G^{ia}$, $I^a$, $J^i$ (or $J_{ud}^i$), $J_s^i$
and
$N_s$. Operators with contracted isospin indices can be eliminated, so the
operators
which remain have either one $G^{ia}$ and no $I$'s or one $I^a$ and no $G$'s.
There are five operator series with the correct time-reversal properties.  They
are
given by the operators
\begin{eqnarray}
G^{ia}\ ,\nonumber\\
{1 \over \N}{J_{ud}^i I^a}\ ,\nonumber\\
{1 \over \N}{J_{s}^i I^a}\ ,\label{pionbroken}\\
{1 \over \N^2}{ \left\{ J_{ud}^i, \left\{G^{ka},J_s^k\right\}\right\}}
\ ,\nonumber\\
{1 \over \N^2}{ \left\{ J_{s}^i, \left\{G^{ka},J_s^k\right\}\right\}}
\ , \nonumber
\end{eqnarray}
times polynomials
\begin{equation}
{\cal P}\left( {N_s \over \N}, {J_{ud}^2 \over \N^2}, {{J_{ud}\cdot J_s} \over
\N^2}
\right)  \nonumber
\end{equation}
in the arguments $N_s/\N$,
$J_{ud}^2/\N^2$, and $J_{ud}\cdot J_s/\N^2$.
Each operator series involves a different polynomial.
Once the operator
series has been determined, it is more convenient to rewrite the polynomials as
functions of $N_s/\N$, $J^2/\N^2$ and $I^2/\N^2$, using $I^2=J_{ud}^2$ and $J^2
= (J_{ud}
+ J_s)^2$.

For baryons with strangeness of order unity, the matrix elements of $G^{ia}$
are order
$\N$, and the matrix elements of $I$, $J$ (or $J_{ud}$), and $J_s$ are order
one, so
the dominant operator series is the series generated by $G^{ia}$.  The four
other
operators in eq.~(\ref{pionbroken}) are suppressed by a factor of $1/\N^2$
relative to
$G^{ia}$.  The first operator series contains the operators $G^{ia}$ and $N_s
G^{ia}/\N$ up to terms of relative order $1/\N^2$ compared to the leading
operator $G^{ia}$.  Thus, eq.~(\ref{pionbroken}) produces the same result as
the
perturbative breaking formula eq.~(\ref{pionpertii}) to relative order
$1/\N^2$.  The
derivation of this formula using only $SU(2) \times U(1)$ flavor symmetry
implies that
the equal spacing rule for pion-baryon couplings is valid to all orders in
$SU(3)$
symmetry breaking \cite{djm}.

\subsubsection{Kaon Couplings}

The kaon couplings transform as $(1,1/2)_1$ and contain either one $t$
and no $Y$, or one $Y$ and no $t$. There are six basic operator series
which contribute; they are generated by the operators
\begin{eqnarray}\label{kaonbroken}
Y^{i\alpha}\ ,\nonumber\\
{1 \over \N} \left\{t^\alpha, J_{ud}^i\right\}\ ,\nonumber\\
{1 \over \N} \left\{t^\alpha, J_{s}^i \right\}\ ,\nonumber\\
{1 \over \N} i\epsilon^{ijk}\left\{Y^{j\alpha},J_{ud}^k\right\}\ , \\
{1 \over \N^2} \left\{ J_{ud}^i, \left\{Y^{k\alpha},J_{ud}^k\right\}\right\}\
,\nonumber\\
{1 \over \N^2} \left\{ J_{s}^i, \left\{Y^{k\alpha},J_{ud}^k\right\}\right\}\
,\nonumber
\end{eqnarray}
times polynomials of $N_s/\N$, $I^2/\N^2$ and $J^2/\N^2$.

For baryons with strangeness of order unity, the matrix elements of
$Y^{i\alpha}$ and
$t^\alpha$ are order $\sqrt \N$.  Thus, the leading order operator for the kaon
couplings is $Y^{i\alpha}$.  There are four additional operators which
contribute at
relative order $1/\N$.  They are $\{N_s, Y^{i\alpha} \}/ \N$ and the three
operators
proportional to $1/\N$ in eq.~(\ref{kaonbroken}).

The perturbative $SU(3)$ breaking expansion eq.~(\ref{kaonpert}) has the same
structure outlined above. The commutation relations in
Table~\ref{tab:brokencomm} imply that the double commutator term in
eq.~(\ref{kaonpert}) is equal to
$i\epsilon^{ijk}\left\{Y^{j\alpha},J^k\right\}$.  This operator can be
rewritten in
terms of $J_{ud}$ and $J_s$.  The piece involving $J_s$ reduces to a linear
combination of the other operators in eq.~(\ref{kaonpert}) by the operator
identities.
Thus, eq.~(\ref{kaonpert}) contains the same five operators as the completely
broken
analysis to relative order $1/\N^2$.  The perturbative breaking formula
determines four
of the five kaon coefficients in terms of the coefficients for the $\pi$ and
$\eta$
couplings.  This relationship is
lost for completely broken
$SU(3)$ symmetry.

\subsubsection{Eta Couplings}

The isoscalar axial current transforms as $(1,0)_0$.
The second operator reduction rule implies that the
general expansion is generated by the operators
\begin{eqnarray}\label{etabroken}
J_{ud}^i\  ,\nonumber\\
J_{s}^i\  ,
\end{eqnarray}
times polynomials of $N_s/\N$, $I^2/\N^2$ and $J^2/\N^2$.
Eq.~(\ref{etapert}) gives the same expansion up to terms of order $1/\N^2$ as
eq.~(\ref{etabroken}). There is no relationship between the $\pi$ and $\eta$
coefficients
in eq.~(\ref{pionpertii}) and eq.~(\ref{etapert}) for  perturbatively broken or
for completely broken $SU(3)$ symmetry.

\section{Baryon Masses}\label{sec:masses}

In this section, we study the baryon masses in the flavor $SU(3)$ limit,
to first order in $SU(3)$ breaking, and for $SU(2) \times U(1)$ flavor
symmetry.

\subsection{Baryon Masses in the Flavor Symmetry Limit}

The $1/\N$ expansion for the baryon masses in the $SU(F)$
flavor symmetry limit can be obtained using
using the operator reduction rule.  The general form of the quark operator
expansion of the baryon Hamiltonian is given by eq.~(\ref{ope}). The
Hamiltonian
is a spin and flavor singlet.  The expansion involves the zero-body operator
$\openone$
and polynomials in the one-body operators $J^i$, $T^a$ and $G^{ia}$ which
transform
as the $(0,0)$ representation of $SU(2) \times SU(F)$.
To obtain a flavor singlet, all flavor indices on the $T$'s and $G$'s must be
contracted using $SU(F)$ tensors with adjoint indices, which can be written as
products of the $SU(F)$ invariant tensors $\delta^{ab}$, $d^{abc}$ and
$f^{abc}$.
All such objects can be removed by the operator reduction rule. Thus, the
Hamiltonian
can be written purely as an expansion in $J^i$, and, by rotation invariance, it
can
only be a function of $J^2$.  Thus, the baryon mass operator is given by
\begin{equation}\label{masssym}
M = \N\ {\cal P}\left({J^2\over\N^2}\right)
\end{equation}
where $\cal P$ is a polynomial. This result reproduces the form of the
baryon mass expansion obtained previously\cite{j,djm,cgo,lm}.

\subsection{Baryon Masses with Perturbative $SU(3)$ Breaking}

Flavor $SU(3)$ symmetry is broken because the light quarks have different
masses.
The perturbation transforms as the $(0, adj)$ irreducible representation
of $SU(2) \times SU(F)$. The dominant $SU(3)$ breaking transforms as
the eighth component of a flavor octet. Isospin
breaking effects are much smaller, and will be neglected here.

The quark operator expansion for a $(0, adj)$ QCD operator is of the
form given in eq.~(\ref{ope}).  The operator reduction rule implies that
only $n$-body operators with a single factor of either $T^a$ or $G^{ia}$ need
to be retained.  There is only one one-body operator,
\begin{equation}\label{adjonebody}
{\cal O}^a_1 = T^a,
\end{equation}
and there is only one two-body operator,
\begin{equation}\label{adjtwobody}
{\cal O}^a_2 = \left\{ J^i, G^{ia} \right\},
\end{equation}
allowed by the operator reduction rule.
In general, there is only one independent $n$-body operator for each $n$.
All of these operators can be generated recursively from operators~${\cal
O}^a_1$
and~${\cal O}^a_2$ by anticommuting with $J^2$,
\begin{equation}
{\cal O}^a_{n+2} = \left\{J^2, {\cal O}^a_n \right\}\ .
\end{equation}
The set of operators ${\cal O}^a_n$, $n=1,2,\ldots,\N$, forms a complete
set of linearly independent spin-zero adjoints.
Thus, the flavor symmetry breaking component of the Hamiltonian has
the expansion
\begin{equation}\label{IXv}
{\cal H}^a=\sum_{n=1}^{\N} b_n {1\over \N^{n-1}} {\cal O}^a_n,
\end{equation}
where $b_n$ are unknown coefficients. Since $J$ is of order one, the
contribution of
${\cal O}_{n+2}$ to the baryon mass in eq.~(\ref{IXv}) is suppressed by
$1/\N^2$
relative to that of ${\cal O}_n$.  Thus, the expansion of the symmetry breaking
perturbation can be truncated after the first two terms, up to corrections of
relative
order $1/\N^2$.
The expansion for the baryon masses, including $SU(3)$ breaking perturbatively
to linear order, is
\begin{equation}
M = a_0 \N\ + a_2 {J^2\over\N}\label{epsmasses}
+ \epsilon b_1 T^8
+ \epsilon {1 \over \N} b_2 \left\{ J^i, G^{i8} \right\}
\end{equation}
up to terms of order $1 / \N^2$.
An explicit factor of $\epsilon$ appears
in front of the last two terms in eq.~(\ref{epsmasses}) to emphasize
which terms arise from symmetry breaking.  Note that $\epsilon$ should not
be regarded as an additional parameter, since $b_1$ and $b_2$ are unknowns.
The general expansion for the masses to linear order in symmetry breaking has
the form
\begin{eqnarray}
M &&= \N\ {\cal P}_0\left({J^2\over\N^2}\right)
+ \epsilon{\cal P}_1\left( {J^2\over\N^2} \right) T^8 \nonumber\\
&&+ \epsilon {1 \over \N} {\cal P}_2\left( {J^2\over\N^2} \right)
\left\{ J^i, G^{i8} \right\}\ ,
\end{eqnarray}
where ${\cal P}_i$ are arbitrary polynomials in their argument.

Eq.~(\ref{epsmasses}) can be rewritten using the substitutions~(\ref{t8}) for
$T^8$ and $G^{i8}$,
\begin{equation}\label{epsmasstwo}
M = a_0 \N\ + a_2 {J^2\over\N}
+ \epsilon b_1 N_s
+ \epsilon {1 \over \N} b_2 \left\{ J^i, J^i_s \right\},
\end{equation}
where pieces of the last two terms in eq.~(\ref{epsmasstwo}) have been
reabsorbed
into the first two terms.  Note that
\begin{equation}
\left\{ J^i, J^i_s \right\} = J^2 + J_s^2 - I^2
\end{equation}
since $J^i-J_s^i = J_{ud}^i$ and $J_{ud}^2 = I^2$.
The masses of the
eight isomultiplets of the spin-$1/2$ octet and spin-$3/2$ decuplet
baryons are written in terms of four unknown parameters in
eq.~(\ref{epsmasses}),
so there are four mass relations which are satisfied to linear order in
symmetry
breaking and to order $1/\N^2$ in the $1/\N$ expansion,
\begin{eqnarray}
&&{1 \over 3}\left( \Sigma + 2 \Sigma^* \right) - \Lambda = {2 \over 3} \left(
\Delta - N \right),\label{massone} \\
&&\Sigma^* - \Sigma = \Xi^* - \Xi, \label{masstwo}\\
&&{3 \over 4}\Lambda + {1 \over 4}\Sigma - {1 \over 2}\left( N + \Xi \right) =
0,
\label{massthree}\\
&&\left(\Sigma^* - \Delta \right) = \left(\Xi^* - \Sigma^* \right)
,\label{massfour}\\
&&\left( \Xi^* - \Sigma^*  \right)= \left( \Omega - \Xi^* \right)
,\label{massfive}
\end{eqnarray}
where only four of the above five relations are linearly independent.  The
first
two relations are spin-flavor relations.  The third
relation is the Gell-Mann--Okubo formula for the baryon octet, and the last two
relations are the equal spacing rule of the decuplet. The Gell-Mann--Okubo
formula
and the equal spacing rule are consequences of $SU(3)$ symmetry alone, but the
other
two relations connect mass splittings in the octet and decuplet, and depend on
the
spin-flavor structure of the mass splittings.

\subsection{Baryon Masses with Completely Broken $SU(3)$ Symmetry}

The analysis of the baryon masses can be performed using only
$SU(2) \times U(1)$ flavor symmetry.  Such an analysis yields baryon
mass relations which are valid to all orders in $SU(3)$ breaking even
for large nonperturbative $SU(3)$ breaking.  The completely broken
$SU(3)$ analysis constructs a $1/\N$ expansion of all quark operators
transforming as spin, isospin and strangeness singlets, $(0,0)_0$.
Application of the second operator reduction rule yields
the mass expansion,
\begin{equation}\label{brokenmass}
M = a_0 \N\ + a_1 N_s + a_{21} {J^2\over\N}
+ a_{22} {I^2\over\N} + a_{23} {N_s^2 \over \N} ,
\end{equation}
which is valid up to relative order $1 / \N^3$.
Eq.~(\ref{brokenmass})
yields three linearly independent mass relations.  Relations~(\ref{massone})
and~(\ref{masstwo}) are still valid.  The last three relations are replaced by
\begin{eqnarray}
&&{3 \over 4}\Lambda + {1 \over 4}\Sigma - {1 \over 2}\left( N + \Xi \right) =
-{1 \over 4} \left( \Omega - \Xi^* - \Sigma^* + \Delta\right),
\nonumber\\   
&&{1 \over 2}\left(\Sigma^* - \Delta \right) -
\left(\Xi^* - \Sigma^*
\right) + {1 \over 2}\left(\Omega - \Xi^* \right)=0 . \label{masseight}
\end{eqnarray}
The first relation relates breaking of the Gell-Mann--Okubo formula
to breaking of one linear combination of the two equal spacing rule relations.
The second relation is the other linear combination of the two equal spacing
rule
relations.  The above results were derived previously in ref.~\cite{djm}, and
the
reader is referred to this work for additional discussion.  Comparison of
eq.~(\ref{brokenmass}) and the perturbative formula with linear symmetry
breaking
eq.~(\ref{epsmasstwo}) shows that the completely broken case replaces the
operator
$\left\{ J^i, J_s^i \right\}$ by two independent operators,
$I^2$ and $N_s^2$.

The general form of the mass expansion for completely broken flavor $SU(3)$
symmetry
to all orders in the $1/\N$ expansion is
\begin{equation}\label{genmassbroken}
M = \N\
{\cal P}\left({N_s \over \N}, {J^2\over\N^2}, {I^2\over\N^2} \right)\ ,
\end{equation}
where ${\cal P}$ is an arbitrary polynomial in its arguments.

\section{Magnetic Moments}\label{sec:magnetic}

The baryon magnetic moments were analyzed in ref.~\cite{jm} using the $1/\N$
expansion in
the Skyrme representation. In this section, we compare those results with the
analysis
in the quark representation.

The baryon magnetic moments transform as $(1,8)$ under $SU(2)\times SU(3)$, so
the
expansions derived in Section~\ref{sec:axial} for the axial couplings can be
applied to
the magnetic moments as long as one remembers that the coefficients of the
operators in
the $1/\N$ expansion can have different values for the magnetic moments than
for the
axial couplings.  The magnetic moments are proportional to the quark charge
matrix
${\cal Q} = {\rm diag} ( 2/3, -1/3, -1/3)$, so they can be separated into
isovector and
isoscalar components.  The analysis of Section~\ref{sec:axial} showed that the
$1/\N$
expansions of the isovector and isoscalar axial currents are unrelated for
perturbatively broken $SU(3)$ symmetry, and that the same expansions are
obtained for
perturbatively broken and completely broken $SU(3)$ symmetry.  Thus, the
analysis
of the magnetic moments can be restricted to the cases of exact $SU(3)$
symmetry and
completely broken $SU(3)$ symmetry.

The analysis of the magnetic moments in the $SU(3)$ symmetry limit is
completely
analogous to the analysis of the meson couplings in the flavor symmetry limit,
and
will not be repeated here.  The magnetic moments for $\N=3$ are parametrized by
four
$SU(3)$ invariants, $\mu_D$,
$\mu_F$, $\mu_{\cal C}$ and $\mu_{\cal H}$ \cite{cptmagmom}.  These parameters
satisfy
equations which are the analogues of eqs.~(\ref{dfch}--\ref{fdeqn}) for the
meson
couplings $D$, $F$, ${\cal C}$, and ${\cal H}$.

The isovector and isoscalar magnetic moments for completely broken $SU(3)$
symmetry
are given by the operator expansions eqs.~(\ref{pionbroken}) and
eq.~(\ref{etabroken}).
Retaining terms to order $1/\N^2$ in these expansions yields
\begin{eqnarray}
\mu^{i3} &=& c_1 G^{i3} + c_2\frac{N_s}{\N} G^{i3},\label{mageqn}\\
\mu^{i8} &=& d_1 J^i + d_2 J_s^i + d_3 \frac{N_s}{\N} J^i +
d_4 \frac{N_s}{\N} J_s^i,\nonumber
\end{eqnarray}
which is precisely the same expansion used in ref.~\cite{jm}, with the quark
operator
$G^{ia}$ of the quark representation replacing the analogous Skyrme
representation
operator $\N X^{ia}$.

The eight isoscalar magnetic moment relations $S1$--$S8$ of ref.~\cite{jm} hold
without change, since the matrix elements of $J^i$ and $J^i_s$ are identical in
the
Skyrme and quark representations.  All eight relations are valid in the
non-relativistic quark model.  Relations
$S1$--$S6$ are valid to order
$1/\N^2$. Only one of these relations, $S1$, is measured experimentally; it
holds to
$4\pm5\%$. Relation $S7$,
\begin{equation}
5(p+n)-(\Xi^0+\Xi^-)=4(\Sigma^{+}+\Sigma^-) \ , \nonumber
\end{equation}
is true in large-$\N$ QCD at leading order, but is violated at order $1/\N$. It
works
experimentally to $22\pm4\%$. Relation $S8$,
\begin{equation}
(p+n)-3\Lambda=\frac12(\Sigma^{+}+\Sigma^-)-(\Xi^0+\Xi^-)\ , \nonumber
\end{equation}
is true in large-$\N$ QCD at leading order. It is also
valid at order $1/\N$ if one imposes $SU(3)$ symmetry in the
$1/\N$ terms, without imposing $SU(3)$ symmetry in the leading order terms.
(This
restriction corresponds to neglecting the order $\epsilon$ terms in $J^i
N_s/\N$ and
$J^i_s N_s/\N$ in eq.~(\ref{etapert}), but retaining them in $J^i$ and $J^i_s$,
which
eliminates the $J^i N_s/\N$ operator from the expansion.) Thus, the violation
of this relation is suppressed by $SU(3)$ symmetry breaking/$\N$.
Relation $S8$ holds experimentally to $7\pm1\%$.
Thus, the $1/\N$ expansion provides some
understanding of why one quark model relation works better than the others.
This
example also shows that the $1/\N$ expansion is not the same as the
non-relativistic
quark model, even in the quark representation.

Ten isovector relations were derived in ref.~\cite{jm}. Relations $V1$--$V7$
hold without change in the quark representation to relative order $1/\N^2$.
Relations
$V8$--$V10$ had slightly different forms in the non-relativistic quark model,
and in the
$1/\N$ expansion using the Skyrme representation. The isovector relations
derived using
eq.~(\ref{mageqn}) of the quark representation are identical to those in the
quark model,
i.e.\ one obtains relations $V8_2$, $V9_2$ and
$V10_4$.  Relations $V8_2$ and $V9_2$ are valid to relative order $1/\N^2$
using
the quark representation.  Relation $V10_4$ is valid at leading order in the
$1/\N$
expansion and at first subleading order in the $1/\N$ expansion in the $SU(3)$
limit
using the quark representation, and is the analogue of relations $V10_2$ and
$V10_3$ using the Skyrme representation. The difference between the Skyrme and
quark
representation forms of the isovector relations is due to a difference of order
$1/\N^2$ in the matrix elements of the two different representations. Since we
have
neglected terms of order $1/\N^2$, the difference between the results is not
significant
to this order. In particular, one of the conclusions of ref.~\cite{jm}, that
the $1/\N$
expansion gives a better prediction than the quark model for the
$\Delta^+\rightarrow
p\gamma$ transition moment, is incorrect; the difference in the two predictions
is an
order $1/\N^2$ effect.

The relation $S/V_1$ between the isoscalar and isovector magnetic
moments is valid in both the Skyrme and quark representations in the $SU(3)$
limit
to two orders in $1/\N$.  Note that the matrix element of $G^{ia}$ in the
proton
is $(N_c+2)/12$ for our definition of $G^{ia}$.

Finally, note that eq.~(\ref{kaonpert}) can be used to obtain the weak
magnetism
form factors in terms of the baryon magnetic moments.

\section{Hyperon Non-Leptonic Decays}\label{sec:nonlep}

Hyperon non-leptonic decays are $\Delta S=-1$ non-leptonic weak
interaction processes\footnote{Recall that $S$ is defined to be strange quark
number,
not strangeness, in this work.}.  The weak Hamiltonian is proportional to
\begin{equation}
\left( \bar s \gamma^\mu P_L u \right) \left( \bar u \gamma_\mu P_L d \right)\
\end{equation}
at the weak scale.
At leading order in $1/\N$, factorization is exact, so that the weak
Hamiltonian can be written as the product of currents. As a result, large $\N$
considerations do not seem to lead to the $\Delta I=1/2$ rule for $K$ decays,
one of the
most striking features of non-leptonic weak interactions. It has been suggested
that a
naive application of
$1/\N$ counting is incorrect, however, because of large logarithms of the form
$\ln
M_{\rm W}/\Lambda_{\rm QCD}$ from renormalization group scaling (see
refs.~\cite{buras,cfg} for discussion of this issue). We do not consider the
issue
of whether the $\Delta I = 1/2$ rule can be derived in large $\N$ in this work.
 We
will assume octet dominance and
$\Delta I=1/2$ enhancement in the following analysis of the hyperon
non-leptonic decay
amplitudes in the $1/\N$ expansion.  This assumption appears to be valid
experimentally.

The general form of the decay amplitude for spin-1/2 baryons is~\cite{pdg}
\begin{equation}\label{hnldi}
{\cal M} = G_F m^2_{\pi^+} \bar u_{B_f}\left(A-B\gamma_5\right) u_{B_i},
\end{equation}
where $A$ and $B$ are parity violating $s$-wave and parity conserving $p$-wave
decay amplitudes. The decay rates and asymmetry parameters are given in terms
of
the amplitudes $s$ and $p$ which are related to $A$ and $B$ by
\begin{equation}
s=A,\quad p = B \frac{\left|{\bf p}_f\right|}{E_f+M_f},
\end{equation}
where ${\bf p}_f$, $E_f$ and $M_f$ are the three-momentum, energy and mass of
the
final baryon. The dimensionless amplitudes $s$ are given in
Table~\ref{tab:hnlds}, where we quote the experimental values
extracted in ref.~\cite{ej:hnld}.

The assumption of octet dominance or $\Delta I = 1/2$ enhancement leads to
three isospin relations amongst the seven
decay amplitudes for the spin-$1/2$ octet baryons,
\begin{eqnarray}\label{hnlisospin}
&&{\cal A}\left(\Sigma^+\rightarrow n \pi^+\right)
\!=\!\sqrt2{\cal A}\left(\Sigma^+\rightarrow p \pi^0\right)
\!+\!{\cal A}\left(\Sigma^-\rightarrow n \pi^-\right),\nonumber\\
&&{\cal A}\left(\Lambda^0\rightarrow p \pi^-\right)=-
\sqrt2\, {\cal A}\left(\Lambda^0\rightarrow n \pi^0\right),\\
&&{\cal A}\left(\Xi^-\rightarrow \Lambda^0 \pi^-\right)=
-\sqrt2\,{\cal A}\left(\Xi^0\rightarrow \Lambda^0 \pi^0\right),\nonumber
\end{eqnarray}
where the relations eq.~(\ref{hnlisospin}) apply to both the $s$- and $p$-wave
amplitudes.  There are two isospin relations amongst the five $\Omega^-$
$p$-wave
amplitudes,
\begin{eqnarray}
&&{\cal A}\left(\Omega^-\rightarrow \Xi^0 \pi^-\right)=
\sqrt2\,{\cal A}\left(\Omega^-\rightarrow \Xi^- \pi^0\right),\nonumber\\
&&{\cal A}\left(\Omega^-\rightarrow \Xi^{*0} \pi^-\right)=
-\sqrt2\,{\cal A}\left(\Omega^-\rightarrow \Xi^{*-} \pi^0\right).
\end{eqnarray}
These isospin relations are evident in the experimental data.

\subsection{$S$-Wave Decay Amplitudes}

The
$s$-wave hyperon non-leptonic decay amplitude does not vanish at zero momentum,
and can
be obtained using a soft pion theorem,
\begin{equation}
{\cal A}(B_i\rightarrow B_f + \pi^a) = \frac{i}{f_\pi} \left\langle B_f\right|
\left[Q_5^a,{\cal H}_W\right]\left|B_i\right\rangle,
\end{equation}
where ${\cal H}_W$ is the weak Hamiltonian, $Q_5^a$ is the axial charge, and
$f_\pi$
is the pion decay constant. Since the weak Hamiltonian transforms as $(8,1)$
under
chiral $SU(3)_L\times SU(3)_R$, $\left[Q_5^a,{\cal H}_W\right]=
\left[Q^a,{\cal H}_W\right]$, where the vector charge $Q^a = I^a$.  Thus, the
$s$-wave
non-leptonic weak decay amplitudes, which are obtained from matrix elements of
$\left[Q^a,{\cal H}_W\right]$, involve the $1/\N$ expansion for the weak
Hamiltonian.

Assuming octet dominance, the weak Hamiltonian
transforms as the $(6+i7)$ component of a $(0,8)$ representation of
$SU(2)\times
SU(3)$. The $1/\N$ expansion for a $(0,adj)$ operator in the $SU(3)$ symmetry
limit was
derived in Section~\ref{sec:masses}. There are two operators series, given by
the operators $T^a$ and $\left\{J^i, G^{ia}\right\}/\N$ times polynomials in
$J^2/\N^2$.  Thus, the weak Hamiltonian has the expansion,
\begin{equation}\label{hweak}
{\cal H}_W = b_1 T^{6+i7} + b_2 { {\left\{J^i, G^{i(6+i7)}\right\} } \over \N
},
\end{equation}
up to corrections of relative order $1/\N^2$.
For baryons with strangeness of order $\N^0$,
$T^{6+i7}$ is of order $\sqrt{\N}$ and $\left\{J^i, G^{i(6+i7)}\right\}/\N$ is
of order
$1/\sqrt{\N}$, so the second term in the expansion eq.~(\ref{hweak}) is
suppressed by
a factor of $1/\N$ relative to the first term.  Thus, to leading order in
the $1/\N$ expansion, the $s$-wave non-leptonic decay amplitudes are given by
the matrix
elements of $[ I^a, T^{6+i7} ]$, and are purely $F$-coupling. At order $1/\N$,
the
second term in eq.~(\ref{hweak}) produces a $1/\N$-suppressed $D$-coupling. It
is known
that an $SU(3)$ symmetric fit works well for the $s$-wave amplitudes, with the
$D$
coupling smaller than the $F$ coupling by about a factor of three~\cite{amhg}.

At leading order in $1/\N$, the $s$-wave amplitudes are pure $F$, and there are
three
relations amongst the four independent amplitudes,
\begin{eqnarray}
&&{\cal A}\left(\Sigma^+\rightarrow n \pi^+\right)=0,\nonumber\\
&&{\cal A}\left(\Xi^-\rightarrow \Lambda^0 \pi^-\right)=
-{\cal A}\left(\Lambda^0\rightarrow p \pi^-\right)\\
&&{\cal A}\left(\Lambda^0\rightarrow p \pi^-\right)=\sqrt{\frac32}{\cal
A}\left(\Sigma^-\rightarrow n \pi^-\right).\nonumber
\end{eqnarray}
These three relations are valid up to a correction of relative order $1/\N$.
The one-parameter fit to the $s$-wave amplitudes is given in the third column
of
Table~\ref{tab:hnlds}.  The fit agrees with the experimental data at the $30\%$
level,
which is consistent with the level expected for a $1/\N$ correction.

When the subleading $1/\N$ correction in eq.~(\ref{hweak}) is included, there
are two
relations valid to relative order $1/\N^2$ amongst the four independent
$s$-wave
amplitudes.  These relations are ${\cal A}\left(\Sigma^+\rightarrow n
\pi^+\right)=0$,
and the Lee-Sugawara relation,
\begin{eqnarray}
&&\sqrt{\frac32}\,{\cal A}\left(\Sigma^-\rightarrow n \pi^-\right)+
{\cal A}\left(\Lambda^0\rightarrow p \pi^-\right)\nonumber\\
&&\qquad\qquad +2\, {\cal A}\left(\Xi^-\rightarrow \Lambda^0 \pi^-\right)=0.
\end{eqnarray}
This two-parameter fit to the $s$-wave amplitudes is given in the fourth column
of
Table~\ref{tab:hnlds}.  The fit agrees with the experimental data at the $10\%$
level,
which is the level expected for $1/\N^2$ corrections.

The analysis of the $s$-wave amplitudes can be repeated adding perturbative
$SU(3)$
symmetry breaking or using only $SU(2) \times U(1)$ flavor symmetry.  The
completely
broken $SU(3)$ flavor symmetry case yields the same operator expansion as for
perturbative symmetry breaking to linear order.  The completely broken analysis
is
supplied here, since it gives results which are valid to all orders in $SU(3)$
symmetry
breaking.  The weak Hamiltonian for completely broken $SU(3)$ symmetry has the
$1/\N$ expansion,
\begin{equation}\label{hweakbroken}
{\cal H}_W = c_1 t^{\dagger}_2
+ c_2 { {\left\{J_{ud}^i,Y^{\dagger\,i}_2 \right\} }\over\N }
+ c_3 { { \{ N_s, t^{\dagger}_2 \} } \over \N} ,
\end{equation}
up to corrections of relative order $1/\N^2$.  Note that $t^\dagger_2 =
T^{6+i7}$, so
the leading operator in eq.~(\ref{hweakbroken}) is the same as the leading
operator in
the
$SU(3)$ symmetric expansion.  At relative order $1/\N$, the single operator in
the
$SU(3)$ symmetric expansion is replaced by two operators.  These two operators
are
contained in $\{ J^i, G^{i(6+i7)} \} = \{ J_{ud}^i + J_s^i, Y^{\dagger\,i}_2
\}$, since
$\{ J_s^i, Y^{\dagger\, i}_2 \}$ reduces to a linear combination of
$t^\dagger_2$ and $\{ N_s,
t^\dagger_2 \}$ by the operator identities.  There are three parameters in the
completely broken $SU(3)$ case, so there is one relation amongst the four
independent
$s$-wave amplitudes, ${\cal A}\left(\Sigma^+\rightarrow n \pi^+\right)=0$. This
amplitude can be non-zero only due to corrections to the soft-pion limit.

Finally, note the subdominant $(0, 27)$ component of the weak Hamiltonian,
which
contains a $\Delta I = 3/2$ piece, is given by the two-body operator
$\{t^\dagger_1, I^+ \}$ at leading order in the $1/\N$ expansion, so it is
suppressed by one factor of $1/\N$ relative to the leading $(0,8)$ one-body
operator
$t^\dagger_\alpha$.

\subsection{$P$-Wave Decay Amplitudes}

The $p$-wave decay amplitude vanishes at zero-momentum, and so soft-pion
theorems
can not be used to simplify the calculation. The weak Hamiltonian transforms as
$(0,8)$, and the pion coupling transforms as part of an $SU(3)$ octet. Thus the
$p$-wave hyperon non-leptonic amplitude transforms as $(1,8\otimes
8)=(1,1)+(1,8)+(1,8)+(1,10+\overline{10})+ (1,27)$. The
operators contributing to the $p$-wave amplitudes were classified previously in
the
analysis of meson couplings with perturbative $SU(3)$ breaking in
Sec.~\ref{sec:axial}.  The form of the $1/\N$ expansion for the $p$-wave
amplitudes
in the $SU(3)$ symmetry limit is the same as the expansion for the meson
couplings
with perturbative $SU(3)$ symmetry breaking, except that the weak Hamiltonian
transforms
as $T^{6+i7}$, whereas the symmetry breaking Hamiltonian transforms as $T^8$.
Thus,
the expression derived in Sec.~\ref{sec:axial} must be rotated to the $6+i7$
direction.  The result for the $p$-wave amplitudes can be written as
\begin{eqnarray}\label{pwaveeq}
P^{ia} &=& d^{a\,(6+i7)\,c} \left( c_1 G^{ic} +
c_2 {J^i T^c\over\N} \right) \nonumber\\
&& + c_3 {\left\{G^{ia}, T^{6+i7} \right\}\over \N}
+c_4{\left\{T^a, G^{i\,(6+i7)} \right\}\over \N} \\
&& + {c_5\over \N}
\left[J^2,\left[T^{6+i7},G^{ia}\right]\right] \nonumber\\
&&+c_6 \delta^{a\,(6+i7)} J^i,\nonumber
\end{eqnarray}
where $a$ denotes the flavor of the pion (or kaon).  The term proportional to
$c_6$
does not contribute to any of the observed $p$-wave decay amplitudes.  The
double
commutator term requires the initial and final baryons to have different spin,
so it
does not contribute to any of the octet baryon decay amplitudes, but does
contribute to the $p$-wave $\Omega^-$ decays to octet baryons.
The analysis of $p$-wave non-leptonic decay amplitudes in the chiral quark
model~\cite{amhg} resembles the $1/\N$ expansion (\ref{pwaveeq}).  The chiral
quark
model also predicts the $p$-wave amplitudes in terms of five one-body and
two-body operators.

There is a significant contribution to the $p$-wave decay amplitudes from pole
graphs, which are sensitive to $SU(3)$ breaking. This introduces additional
calculable operators to those given in eq.~(\ref{pwaveeq}). The analysis of the
$p$-wave amplitudes
including pole graphs is complicated, and will be given elsewhere~\cite{ddjm}.

\section{The Skyrme Representation}\label{sec:skyrme}

The large-$\N$ consistency conditions derived in ref.~\cite{dm,j,djm}
can be analyzed by constructing irreducible representations of the contracted
spin-flavor algebra using the
theory of induced representations.  This construction naturally leads to a
description of large-$\N$ baryons in terms of the Skyrme
representation\footnote{We  will
refer to the Skyrme representation rather than the Skyrme model to emphasize
that we
are using the Skyrme model to provide an operator basis for the
$1/\N$ expansion of baryons in QCD, not in the Skyrme model.}.
The analysis of large-$\N$ baryons using the Skyrme representation is discussed
in detail
in ref.~\cite{djm}, and will not be repeated here. In this section, we
elucidate some
connections between the quark and Skyrme representations of large-$\N$ baryons.

In the Skyrme representation, the space
components of the axial currents are proportional to $\N$ times
\begin{equation}
X^{ia} = 2\ \Tr\ A T^i A^{-1} T^a,\label{sk:1}
\end{equation}
where $A$, the Skyrmion collective coordinate, is an element of
$SU(F)$ in the $F$-flavor case. The spin operators $J^i$
generate the right transformations of $A$,
\begin{equation}
A \rightarrow A U^{-1},\label{sk:2}
\end{equation}
where $U$ is an element of a $SU(2)$ subgroup of $SU(F)$,
and the flavor operators $T^a$ generate left transformations
\begin{equation}
A \rightarrow U A,\label{sk:3}
\end{equation}
where $U$ is an element of $SU(F)$.  The Skyrme representation
gives an exact realization of the contracted spin-flavor algebra since $X^{ia}$
is a c-number which satisfies the commutation relation $[ X^{ia}, X^{jb}]=0$ to
all
orders in the $1/\N$ expansion.

The Skyrme and quark representations are equivalent in the large-$\N$ limit,
but differ at subleading orders in the $1/\N$ expansion.  The equivalence is
most transparent for the two-flavor case; for three or more flavors, there are
additional subtleties.

The operator structure of the Skyrme representation is particularly simple for
the case
of two light flavors. The Skyrme representation operators $J^i$, $I^a$ and
$X^{ia}$ have
well-defined, $\N$-independent matrix elements, whereas the quark
representation
operator $G^{ia}$ has matrix elements of order $\N$.  The equivalence of the
Skyrme
and quark representations follows from the identification
\begin{equation}
X^{ia} = \lim_{\N\rightarrow\infty}\ -{4\over {\N + 2}}\, G^{ia} + {\cal
O}\left({1 \over \N^2}\right).\label{sk:5}
\end{equation}
The Skyrme representation identities derived in ref.~\cite{djm} are reproduced
by taking
the limit eq.~(\ref{sk:5}) of the quark representation identities in
Table~\ref{tab:su4iden}, and dropping subleading terms in $1/\N$,
\begin{eqnarray}
X^{ia} X^{ia} &=&3, \nonumber \\
X^{ia} J^i &=& - I^a, \nonumber \\
X^{ia} I^a &=& - J^i, \nonumber \\
\epsilon^{ijk} \epsilon^{abc} X^{ia} X^{jb} &=& 2 X^{ic}, \nonumber \\
I^2 &=& J^2, \label{sk:4} \\
X^{ia} X^{ib} &=& \delta^{ab},\nonumber \\
\epsilon^{ijk} \left\{J^j,X^{ka}\right\} &=&
\epsilon^{abc}\left\{I^b,X^{ic}\right\}, \nonumber\\
X^{ia} X^{ja} &=& \delta^{ij}. \nonumber
\end{eqnarray}
Thus, the operator structure of the Skyrme and quark representations is
identical
at leading order in the $1/\N$ expansion.  Either operator basis can be used to
expand
QCD operators in a $1/\N$ expansion.  The two expansions will have different
coefficients at subleading order in $1/\N$, but will give identical
predictions for physical quantities.  The operator structure of the Skyrme
representation is simpler than that of the quark representation for two
flavors,
since the Skyrme representation operators and operator identities do not depend
on
$\N$.  It is important to note, however, that the baryon spectrum in the Skyrme
representation contains more states than in the quark representation.
The baryon spectrum in the quark
representation is a tower of states with $(J,I) = (1/2,1/2),\ (3/2,3/2),\ldots,
(\N/2,\N/2)$. The spectrum in the Skyrme representation is an infinite tower of
states
$(J,I) = (1/2,1/2),\  (3/2,3/2),\ldots$. The extra states in the Skyrme model
are
sometimes  regarded as ``spurious'' states from the point of view of the quark
model. They have the quantum numbers of hadrons containing $\N$ quarks
plus some $\bar q q$ pairs.  The existence of extra states in the Skyrme
representation
does not affect the conclusion that the quark and Skyrme representations yield
equivalent operator bases since any operator of finite spin (such as the mass
operator
with spin  zero, or the axial current with spin one) does not couple states at
the
bottom of the tower to these additional states with spin of order
$\N$.

For two flavors, the quark representation operator $G^{ia}$ can be written
explicitly in terms of the Skyrme
representation operator $X^{ia}$ to all orders in $1/\N$. The matrix elements
of $X^{ia}$ between baryon states are known~\cite{anw}. The matrix elements of
$G^{ia}$ between baryon states can be computed using the method of
ref.~\cite{karlpaton}.
Baryons with $I_3=1/2$ are made of $(\N+1)/2$ $u$-quarks combined into a state
with
spin $J_u$, and $(\N-1)/2$ $d$-quarks combined into a state with spin $J_d$,
where
$J_u$ and $J_d$ are combined to form a state with total spin $J$. The matrix
elements of
$J_u$ and $J_d$ can be computed using standard methods~\cite{edmonds} to obtain
the matrix elements of $G^{ia}$. Writing the matrix elements of $G^{ia}$ in
terms of $X^{ia}$, one obtains
\begin{eqnarray}\label{gxexpand}
G^{ia} &=& -\frac{\N+2}{4}\ \sqrt{1-\hat z}\ X^{ia} \\
&&\quad + \frac1{\N+2} \left(\frac{1}{1+\sqrt{1-
\hat z}}\right) J^i I^a,\nonumber
\end{eqnarray}
where $\hat z$ is the operator
\begin{equation}
\hat z =  \frac2{\left(\N+2\right)^2} Ad_+ J^2,
\end{equation}
and $Ad_+ J^2$ is the operator defined by
\begin{equation}
\left(Ad_+ J^2\right) {\cal O} \equiv \left\{J^2,{\cal O} \right\}.
\end{equation}
Eq.~(\ref{gxexpand}) is valid for matrix elements of $G^{ia}$ between states
with
$J\le \N/2$. Matrix elements of $G^{ia}$ in which at least one state has $J >
\N/2$
vanish.

The comparison of the Skyrme and quark representations is more interesting when
the
number of flavors is greater than two. We will concentrate on the three-flavor
case
in this discussion, since all the subtleties already occur for this case.  The
baryon
spectrum of the Skyrme representation contains additional states which couple
to
baryon states with low spin.  The baryon spectrum is determined by quantization
of
the collective coordinate $A$ in eq.~(\ref{sk:1}).
The collective coordinate $A$ must be  quantized subject to the constraint that
the
body-centered hypercharge is $\N/3$ (i.e. $T^8$ is $\N/\sqrt{12}$) \cite{guad},
as
dictated by the Wess-Zumino term. It is important to note that this constraint
depends
on $\N$. Many errors in the Skyrme model literature arise from quantizing the
Skyrmion
with the hypercharge set to its value for $\N=3$, $Y=1$, while
expanding in $1/\N$.  Because of this constraint, the
spectrum of the  Skyrme model depends explicitly on $\N$, in contrast to the
two-flavor
case.  The spectrum of the Skyrme representation contains the same tower of
states
as the quark representation.  These representations, which are given in
Table~\ref{tab:su2f->suf}, consist of Young tableaux with
$\N$ boxes for spin and flavor.  In addition, the Skyrme representation
contains
states which consist of Young tableaux with $\N+3$ boxes, $\N+6$ boxes, etc.
\cite{am}.
These ``spurious'' states can have low spin, such as
$1/2$, $3/2$, etc., and they can couple to the standard spin-$1/2$ and
spin-$3/2$
baryons via operators of finite spin, such as the axial currents. These
additional
states have the quantum numbers of a state composed of $\N$ quarks and $\bar
qq$ pairs.
In large-$\N$ QCD, pair creation of an additional $\bar q q$ pair is suppressed
by a
factor of $1/\sqrt{\N}$, since pair creation plus annihilation produces a
closed fermion
loop, which is down by $1/\N$. It is straightforward to check by explicit
computation of Clebsch-Gordan coefficients for arbitrary $\N$ that the matrix
elements of
operators between the ``normal'' and ``spurious'' states
have this $1/\N$ suppression. For example, the amplitude for a spin-1/2 baryon
with $\N$ boxes to couple via the axial current to a baryon with $\N+3$ boxes
(i.e. a
baryon with one extra $\bar q q$ pair) is suppressed by $1/\sqrt{N}$.  Thus,
the
additional states of the Skyrme representation affect the couplings of the
$\N$-quark
baryon states only at subleading order.

The Skyrme operators $J^i$, $T^a$ and $X^{ia}$ can be used to obtain a
$1/\N$ expansion for three flavors.
Since the matrix elements of the Skyrme operator $X^{ia}$ now have a
$\N$-dependence, the relation eq.~(\ref{sk:5}) between the quark and
Skyrme operators $G^{ia}$ and $X^{ia}$ is no longer valid, and the Skyrme model
operator
identities are not given by taking the $\N\rightarrow\infty$ limit of the
quark model identities.  The $\N$-dependence of the matrix elements of the
operators $T^a$ and $X^{ia}$ is different in different regions of the weight
diagram.  This non-trivial $\N$-dependence of operator matrix elements is
what made the
analysis in ref.~\cite{djm} of the $SU(3)$ flavor symmetry limit
complicated.  In ref.~\cite{djm}, the coupling of baryons to octet mesons
was given in terms of two invariant amplitudes $\cal M$ and $\cal N$, with
\begin{equation}
{\cal N \over M} = {1\over 2} + {\alpha\over\N} + {\cal
O}\left({1\over\N^2}\right),\label{sk:6}
\end{equation}
where $\alpha$ is an undetermined parameter.  In the quark representation, one
can show that the operator $G^{ia}$ implies that
\begin{equation}
{\cal N \over M} = {1\over2}{\N-1\over\N+2} = {1 \over 2}-{3 \over
{2\N}}+\ldots,
\end{equation}
so that $\alpha=-3/2$ in the non-relativistic quark model.  Similarly,
the ratio $\cal N/M$ can be calculated for arbitrary $\N$ for the Skyrme model
operator
$X^{ia}$,
\begin{equation}\label{sk:7}
{\cal N \over M} = {1\over2}\,{\left(\N-1\right)\left(\N+9\right)\over
\N^2+8\N+9} = {1 \over 2} + {\cal O}\left( {1 \over \N^2} \right),
\end{equation}
so that $\alpha=0$ in the Skyrme model.  In the $1/\N$ expansion of the meson
couplings, the operator $J^i T^a /\N$ changes the prediction for the
parameter $\alpha$ in either the quark or Skyrme representation away from these
quark
and Skyrme model values.  The coefficient of $J^i T^a/\N$ must be different in
the
$1/\N$ expansions in the quark and Skyrme representations to produce a given
value of
$\alpha$ since $G^{ia}$ and $X^{ia}$ give different contributions to $\alpha$.
Using
eq.~(\ref{VIIIxii}), one finds the non-trivial relation
\begin{equation}\label{sk:8}
X^{ia} = -{4\over{\N+2}}\left[ G^{ia} - {J^i T^a\over\N}\right]+ {\cal
O}\left({1\over\N^2}\right),
\end{equation}
between the Skyrme and quark model operators,\footnote{The lefthand side of
Eq.~(\ref{sk:8}) is the Skyrme model $X^{ia}$ projected onto the set of states
which
exist in the quark model. There are extra states in the Skyrme model. The full
Skyrme
model operator $X^{ia}$ can not be written as an expansion in terms of quark
operators
since it has matrix elements between the quark model states and the extra
Skyrme model states.}
where the overall coefficient in front of the parentheses is determined by
requiring
that the Skyrme model identity
$X^{ia} T^a = -J^i$ is satisfied to this order.  Eq.~(\ref{sk:8}) implies that
some of
the operator identities in the Skyrme and quark model are different, even at
leading
order in $\N$.  For instance, the Skyrme model identity
$X^{ia} T^a = - J^i$ is valid in the case of two or three flavors, but the
analogous quark model identity $\left\{T^a,G^{ia}\right\}=(\N +F)(1 - 1/F) J^i$
has a
different coefficient of proportionality for two and three flavors, even at
leading order
in $1/\N$.  This occurs because the $J^i T^a/\N$ term in eq.~(\ref{sk:8}) for
$a=8$ is
unsuppressed relative to $G^{ia}$, and it produces a leading order change in
the
$X^{ia} T^a$ identity.
Using eq.~(\ref{sk:8}) and eq.~(\ref{VIIx}), one finds that
\begin{eqnarray}
X^{ia} T^a &=& -{4\over{\N+2}}\left[ G^{ia} T^a  -{1 \over \N} J^i T^2
\right] + \ldots\nonumber\\
&=&  -{4\over{\N+2}}\left[ {1 \over 2}\left(\N +F\right)\left(1 - {1 \over
F}\right)\right.
\nonumber\\
&&\qquad\left.-{1 \over {4F}}\left( \N + 2F \right)\left( F - 2 \right) \right]
J^i +
\ldots \\
&=&-J^i,\nonumber
\end{eqnarray}
for any number of flavors $F$.

We will not work out the operator identities in the Skyrme basis. Some of
these identities are given in ref.~\cite{djm}.

\section{Conclusions}\label{sec:conc}

The $1/\N$ expansion allows one to compute properties of baryons in a
systematic
expansion of QCD. There are only a few terms in the expansion at any given
order
in $1/\N$ once
redundant operators are eliminated using the operator reduction rule. Results
which hold
to two orders in $1/\N$ typically work at the 10--15\% level.  For $\N=3$, the
$1/\N$
expansion usually can not be carried beyond second order because the number of
independent operators becomes too large to make any non-trivial predictions.
The
$1/\N$ expansion explains which features of the quark and Skyrme models follow
directly
from the spin-flavor symmetry structure of QCD.

\acknowledgements

This work was supported in part by the Department of Energy under grant
DOE-FG03-90ER40546 and by the National Science Foundation under grant NSF
PHY90-21984.
E.J. was supported in part by NYI award PHY-9457911 from the National Science
Foundation.  A.M. was supported in part by PYI award PHY-8958081 from the
National
Science Foundation.

\appendix

\section{$\bf SU(Q)$ Group Theory}\label{app:suq}

This appendix reviews some $SU(Q)$ group theory which is
needed for the derivation of the quark operator identities in
Secs.~\ref{sec:classify} and~\ref{sec:derive}.
The results in this appendix can be applied to the spin-flavor group
by setting $Q=2F$, and to the flavor group by setting $Q=F$.

Irreducible representations of $SU(Q)$ are denoted in a number of different
ways;
these include their dimensions, Young tableaux, tensors, and Dynkin labels.
The fundamental representation
of $SU(Q)$ can be represented by its dimension $Q$, the Young tableau $\sqr$,
the tensor with
one upper index $T^\alpha$, or the Dynkin label $[1,0,0,\ldots,0]$,
where the number of entries in the Dynkin label is $(Q-1)$, the rank of
$SU(Q)$.
The complex conjugate representation of the fundamental also is dimension $Q$,
and
is often denoted by $\overline Q$.  The Young tableau of $\overline Q$ is
\begin{equation}\nonumber
\nboxconj
\end{equation}
(sometimes written $\overline {\sqr}$), the completely antisymmetric product
of $(Q-1)$ fundamental representations.  $\overline Q$ also can be represented
as the tensor with one lower index, $T_\beta$, or by the Dynkin label
$[0,0,0,\ldots,0,1]$.

The tensor product of the fundamental representation and its complex
conjugate contains the singlet and the adjoint representations of $SU(Q)$.
The adjoint
representation can be represented by its dimension $(Q^2-1)$, the Young tableau
\begin{equation}\nonumber
\nboxadj \ ,
\end{equation}
the traceless tensor $T^\alpha_\beta$ with one upper and one lower index,
or the Dynkin label $[1,0,0,\ldots,0,1]$. The adjoint
representation is a real representation. A tensor
with one upper index $\alpha$ and one lower index $\beta$ in the
fundamental representation can be converted into a tensor with a single
index $A$ in the adjoint representation using the $SU(Q)$
group generators in the fundamental representation, $\Lambda^A$,
\begin{equation}\label{Ai}
T^\alpha_\beta \rightarrow T^A=\left(\Lambda^A\right)^\beta_\alpha
T^\alpha_\beta.
\end{equation}

The commutator of two $SU(Q)$ generators is defined by structure constants
$f^{ABC}$ which are completely antisymmetric and real,
\begin{equation}\label{Aii}
\left[\Lambda^A,\Lambda^B\right] = i f^{ABC} \Lambda^C.
\end{equation}
The anticommutator of two $SU(Q)$
generators {\it in the fundamental representation} is
\begin{equation}\label{Aiii}
\left\{\Lambda^A,\Lambda^B\right\} = {1\over Q} \delta^{AB} +  d^{ABC}
\Lambda^C,
\end{equation}
which defines the $d$-symbol, a real and completely symmetric tensor with three
adjoint indices. The $f$- and $d$-symbols for $SU(Q)$
can be written in terms of traces of
$SU(Q)$ generators in the fundamental representation,
\begin{eqnarray}
f^{ABC}  &=& - 2 i\ \Tr \Lambda^A\left[\Lambda^B,\Lambda^C\right]
,\nonumber\\
d^{ABC}  &=&  2\ \Tr \Lambda^A\left\{\Lambda^B,\Lambda^C\right\} .
\label{Aiv}
\end{eqnarray}

A number of identities for contractions of $f$- and $d$-symbols are used
in the derivation of the quark operator identities.  These are:
\begin{eqnarray}
d^{AAB} &=& 0 ,\nonumber\\
d^{ABC} d^{ABD} &=& \left( Q - {4 \over Q} \right) \ \delta^{CD}, \nonumber\\
f^{ABC} f^{ABD} &=& Q  \ \delta^{CD}, \label{Av}\\
f^{ABC} f^{ADE} d^{BDF} &=& {Q \over 2} \ d^{CEF}, \nonumber\\
d^{ABC} d^{ADE} d^{BDF} &=& \left({Q \over 2}-{6 \over Q}\right) \
d^{CEF}, \nonumber \\
d^{ABC} d^{ADE} f^{BDF} &=& \left({Q \over 2}-{2 \over Q} \right)\
f^{CEF} .\nonumber
\end{eqnarray}
These identities can be proved using eq.~(\ref{Aiv}) and the trace identities
\begin{eqnarray}
\Tr\ \Lambda^A X \Lambda^A Y &=& {1\over 2} \Tr\ X\ \ \Tr Y - {1\over 2Q}
\Tr\ XY , \nonumber\\
\Tr\ \Lambda^A X\ \ \Tr\ \Lambda^A Y &=& {1\over 2} \Tr\ X Y - {1\over 2Q}
\Tr\ X \Tr\ Y,\label{Avi}
\end{eqnarray}
which follow from the Fierz identity
\begin{equation}
\left(\Lambda^A\right)^\alpha_\beta \left(\Lambda^A\right)^\gamma_\delta=
{1\over 2}\, \delta^\alpha_\delta \delta^\gamma_\beta - {1\over 2Q}\,
\delta^\alpha_\beta \delta^\gamma_\delta .
\end{equation}

The tensor product of two adjoint representations can be divided into the
symmetric product $\left(adj\times adj\right)_S$ and the antisymmetric
product $\left(adj\times adj\right)_A$. The irreducible representations
in the symmetric and antisymmetric products of two
adjoints are given in Tables~\ref{tab:adj2s} and~\ref{tab:adj2a}, respectively.
These decompositions can be written as
\begin{eqnarray}
\left(adj\otimes adj\right)_S &=& 1 + adj + {\bar a a} + {\bar s s}
,\nonumber\\
\left(adj\otimes adj\right)_A &=& adj + {\bar a s} + {\bar s a} , \label{Avii}
\end{eqnarray}
where the Dynkin labels of the irreducible representations in eq.~(\ref{Avii})
are listed in Tables~\ref{tab:adj2s} and~\ref{tab:adj2a}.
The designation of the irreducible representations
by $adj$, ${\bar a a}$, ${\bar s s}$, ${\bar a s}$ and ${\bar s a}$
is used throughout this paper.  This designation is not conventional.
The dimensions, Casimirs, Dynkin labels and Young tableaux for these
representations
are listed in Table~\ref{tab:suqreps}.

The decomposition of the tensor product of two adjoints into the
irreducible representations in eq.~(\ref{Avii}) is straightforward
using tensor methods, if
the adjoint representation is given as a tensor $T^\alpha_\beta$ with
indices in the fundamental representation.  The
derivation of the quark operator identities requires the decomposition
when the adjoint representation is given as a tensor $T^A$ with an index
in the adjoint representation.

Consider first the decomposition of
$\left(adj\times adj\right)_A$. Given two adjoints $T_1^A$
and $T_2^A$, one can define the antisymmetric tensor
\begin{equation}\label{Aviii}
{\cal X}^{AB}_- = T^A_1 T^B_2 - T^B_1 T^A_2.
\end{equation}
The tensor ${\cal X}^{AB}_-$, which is antisymmetric in the two adjoint
indices,
can be decomposed into $adj+{\bar a s}+{\bar s a}$. The adjoint is obtained by
contracting with the $f$-symbol,
\begin{equation}\label{Aix}
{\cal X}_{-,adj}^C= f^{ABC} {\cal X}_-^{AB}.
\end{equation}
The ${\bar a s}+{\bar s a}$ representations can be obtained by subtracting  the
adjoint from ${\cal X}_-^{AB}$
\begin{equation}\label{Ax}
{\cal X}_{{\bar a s}+{\bar s a}}^{AB} = {\cal X}_-^{AB} - {1\over Q}\, f^{ABC}
f^{CGH} {\cal X}_-^{GH}.
\end{equation}
The coefficient of the last term has been chosen so that
\begin{equation}
f^{ABC}\ {\cal X}_{{\bar a s}+{\bar s a}}^{AB} =0.
\end{equation}
The adjoint representation is real, so ${\bar a s}$ and ${\bar s a}$ are
conjugate representations.

The decomposition of the symmetric product of two adjoints is more
involved. Define the symmetric tensor
\begin{equation}\label{Axi}
{\cal X}^{AB}_+ = T^A_1 T^B_2 + T^B_1 T^A_2.
\end{equation}
The tensor ${\cal X}^{AB}_+$, which is symmetric in the two adjoint indices,
can be decomposed into $1+adj+{\bar a a}+{\bar s s}$. The singlet is
obtained by contracting the two adjoint indices, whereas the
adjoint representation in ${\cal X}_+^{AB}$ can be projected out using the
$d$-symbol,
\begin{eqnarray}
{\cal X}_{1} &=& {\cal X}_+^{AA}, \nonumber\\
{\cal X}_{+,adj}^C &=& d^{ABC} {\cal X}_+^{AB}. \label{Axii}
\end{eqnarray}
The sum of the ${\bar a a}$ and ${\bar s s}$ representations is given by
subtracting off the singlet and adjoint components of ${\cal X}_+^{AB}$,
\begin{eqnarray}\label{Axiii}
{\cal X}_{{\bar a a}+{\bar s s}}^{AB} &=& {\cal X}_+^{AB} - {1\over Q^2-1}\,
\delta^{AB}\ {\cal X}_+^{CC}
\\ &&\qquad -{Q\over Q^2-4}\, d^{ABC} d^{CGH}\ {\cal X}_+^{GH}.\nonumber
\end{eqnarray}

Eq.~(\ref{Axiii}) can be separated into the individual ${\bar a a}$ and
${\bar s s}$ representations
using a trick. Any tensor $X^{AB}$ with two adjoint indices
transforms under infinitesimal $SU(Q)$ transformations by the generators
in the adjoint representation (which are the structure constants)
acting on the two indices separately
\begin{equation}
\delta^C {\cal X}^{AB} = -i f^{CAD} {\cal X}^{DB} - i f^{CBD} {\cal X}^{AD}.
\end{equation}
Thus, the Casimir operator ${C_2}$ acting on ${\cal X}^{AB}$ is
\begin{eqnarray}
\left({C_2} {\cal X}\right)^{AB}& =& - \Bigl(f^{CAE} f^{CEG} \delta^{BH} +
f^{CBE} f^{CEH} \delta^{AG} \nonumber \\
&&\quad  + 2 f^{CAG} f^{CBH}\Bigr) {\cal X}^{GH}.
\end{eqnarray}
The Casimir operator has the values (see Table~\ref{tab:suqreps})
\begin{equation}\label{Axiv}
{C_2} = \cases{ 2 Q & ${\bar a s}$, \cr
                   2 Q & ${\bar s a}$ ,\cr
                   2(Q-1) & ${\bar a a}$ ,\cr
                   2(Q+1) & ${\bar s s}$ ,\cr}
\end{equation}
in the representations of interest. Using eqs.~(\ref{Av},\ref{Axiv}), we find
that
\begin{equation}\label{Axv}
f^{ACG} f^{BCH} {\cal X}^{GH}  = \cases{ 0 & ${\bar a s}$ ,\cr
                                  0 & ${\bar s a}$ ,\cr
                                  {\cal X}^{AB} & ${\bar a a}$ ,\cr
                                  -{\cal X}^{AB} & ${\bar s s}$ .\cr}
\end{equation}
The operator $f^{ACG} f^{BCH}$ can be used to split ${\cal X}_{{\bar a a}+{\bar
s s}}^{AB}$ into
${\bar a a}$ and ${\bar s s}$ representations, using the projection operators
\begin{eqnarray}
\left(P_{{\bar a a}} {\cal X}\right)^{AB} &=& {1\over2}\left(\delta^{AG}
\delta^{BH} +
f^{ACG} f^{BCH}\right) {\cal X}^{GH}, \nonumber\\
\left(P_{{\bar s s}} {\cal X}\right)^{AB} &=& {1\over2}\left(\delta^{AG}
\delta^{BH} -
f^{ACG} f^{BCH}\right) {\cal X}^{GH} .\label{Axvi}
\end{eqnarray}

Other useful identities are
\begin{equation}\label{Axxi}
d^{ACG} d^{BCH} {\cal X}^{GH}  = \cases{ 0 & ${\bar a s}$ ,\cr
                                  0 & ${\bar s a}$ ,\cr
                                  -\left(1+2/Q\right){\cal X}^{AB} & ${\bar a
a}$ ,\cr
                                   \left(1-2/Q\right){\cal X}^{AB} & ${\bar s
s}$ ,\cr}
\end{equation}
\begin{equation}\label{Axxii}
f^{ACG} d^{BCH} {\cal X}^{GH}  = \cases{ -i{\cal X}^{AB}  & ${\bar a s}$ ,\cr
                                  i{\cal X}^{AB} & ${\bar s a}$ ,\cr
                                  -i{\cal X}^{AB} & ${\bar a a}$ ,\cr
                                   i{\cal X}^{AB} & ${\bar s s}$ ,\cr}
\end{equation}
so that $f^{ACG} d^{BCH}$ can be used to split ${\cal X}_{{\bar a s}+{\bar s
a}}^{AB}$
into $\bar a s$ and $\bar s a$ representations.

\section{$\bf SU(2F)\rightarrow SU(2)\times SU(F)$}\label{app:su2f}

To decompose the two-body quark operator identities into irreducible
representations of $SU(2)\times SU(F)$, we need the decomposition
of the $SU(2F)$ irreducible representations $adj$, ${\bar a a}$ and ${\bar s
s}$
into $SU(2) \times SU(F)$ representations.  A straightforward
computation yields
\begin{eqnarray}
&adj &\rightarrow (0,adj)+(1,0)+(1,adj) , \label{Bi} \\
&{\bar a a} &\rightarrow (0,0) + (0,adj) + (0,{\bar a a}) + (0,{\bar s s})
\nonumber\\
&&+ (1,adj) +  (1,adj) + (1,{\bar a a}) + (1,{\bar a s}+{\bar s a}) \nonumber
\\
&& + (2,0) + (2,adj) + (2,{\bar a a}),\nonumber\\
&{\bar s s} &\rightarrow (0,0) + (0,adj) + (0,{\bar a a}) +
(0,{\bar s s}) \nonumber \\
&&+ (1,adj)+ (1,adj) + (1,{\bar s s}) + (1,{\bar a s}+{\bar s a}) \nonumber\\
&& + (2,0) + (2,adj) + (2,{\bar s s}). \nonumber
\end{eqnarray}
For the special cases of $SU(4) \rightarrow SU(2) \times SU(2)$ and
$SU(6) \rightarrow SU(2) \times SU(3)$, eq.~(\ref{Bi}) reduces to
\begin{eqnarray}
&15 &\rightarrow (0,1)+(1,0)+(1,1),\nonumber \\
&20 &\rightarrow (0,0) + (1,1) + (2,0) + (0,2),\label{Bii}  \\
&84 &\rightarrow (0,0) + (1,1) + (1,1) + (0,2) + (2,0) \nonumber\\
&& + (1,2) + (2,1) + (2,2),\nonumber
\end{eqnarray}
and
\begin{eqnarray}
&35 &\rightarrow (0,8)+(1,0)+(1,8),\nonumber \\
&189 &\rightarrow (0,0) + (0,8) + (0,27) + (1,8) + (1,8) \nonumber\\
&&\qquad + (1,10+\overline {10})
+ (2,0) + (2,8) , \label{Biii}  \\
&405 &\rightarrow (0,0) + (0,8) + (0,27) + (1,8) +(1,8)
\nonumber \\
&&+(1,27)+(1,10+\overline{10})
+ (2,0) + (2,8) + (2,27), \nonumber
\end{eqnarray}
respectively.

The decomposition of the $SU(2F)$ $d$-symbol $d^{ABC}$ for
$SU(2F) \rightarrow SU(2) \times SU(F)$ is required.  The adjoint
of $SU(2F)$ decomposes into the representations
$(1,0)$, $(0,adj)$ and $(1,adj)$ of $SU(2)\times SU(F)$, so an
adjoint index $A$ is replaced by
the $SU(2)\times SU(F)$ indices $i$, $a$ and $ia$, respectively.
The $SU(2F)$ $d$-symbol can be written as
\begin{equation}\label{Bv}
d^{ABC} = \cases{ 0 & $\{A,B,C\}=\{i,j,k\}$\cr
                  0 & $\{A,B,C\}=\{i,j,a\}$\cr
                  0 & $\{A,B,C\}=\{i,a,b\}$\cr
                  {1\over\sqrt 2} d^{abc} & $\{A,B,C\}=\{a,b,c\}$\cr
                  0 & $\{A,B,C\}=\{ia,j,k\}$\cr
                 {1\over \sqrt F} \delta^{ij} \delta^{ab} &
$\{A,B,C\}=\{ia,j,b\}$\cr
                  0 & $\{A,B,C\}=\{ia,b,c\}$\cr
                  0 & $\{A,B,C\}=\{ia,jb,k \}$\cr
                  {1\over\sqrt2}\delta^{ij}d^{abc} & $\{A,B,C\}=\{ia,jb,c\}$\cr
                  -{1\over\sqrt2}\epsilon^{ijk}f^{abc} &
$\{A,B,C\}=\{ia,jb,kc\}$\cr
                  }
\end{equation}
in terms of the $SU(F)$ $d$- and $f$-symbols.

\onecolumn 

\widetext

\begin{table}[htbp]
\caption{$SU(2) \otimes SU(F)$ decomposition of the $SU(2F)$
representation $\nbox$ of the ground state baryons.  All Young tableaux contain
$\N$ boxes.}
\label{tab:su2f->suf}
\smallskip
\centerline{\vbox{ \tabskip=0pt \offinterlineskip
\def\tablerule{\noalign{\hrule}}
\def\space{height 2pt&\omit&&\omit&\cr}
\halign{
\vrule #&\strut\hfil\ $ # $\ \hfil&&
\vrule #&\strut\hfil\ $ # $\ \hfil\cr
\tablerule\space
& SU(2) && SU(F) &\cr
\space\tablerule\space\space\space
& \raise8pt\hbox{$1 \over 2$}
&&
\nboxF
&\cr
\space\tablerule\space\space\space
& \raise8pt\hbox{$3 \over 2$}
&&
\nboxE
&\cr
\space\tablerule\space
& \cdot
&&
\cdot
&\cr
& \cdot
&&
\cdot
&\cr
& \cdot
&&
\cdot
&\cr
\space\tablerule\space\space\space
& \raise8pt\hbox{${\N-2} \over 2$}
&&
\nboxA
&\cr
\space\tablerule\space\space
& \raise2pt\hbox{$\N \over 2$}
&&
\nbox
&\cr
\space\tablerule
}}}
\end{table}

\begin{table}[htbp]
\caption{$SU(2F)$ Commutation Relations}
\bigskip
\label{tab:su2fcomm}
\centerline{\vbox{ \tabskip=0pt \offinterlineskip
\halign{
\strut\quad $ # $\quad\hfil&\strut\quad $ # $\quad \hfil\cr
\multispan2\hfil $\left[J^i,T^a\right]=0,$ \hfil \cr
\noalign{\medskip}
\left[J^i,J^j\right]=i\epsilon^{ijk} J^k,
&\left[T^a,T^b\right]=i f^{abc} T^c,\cr
\noalign{\medskip}
\left[J^i,G^{ja}\right]=i\epsilon^{ijk} G^{ka},
&\left[T^a,G^{ib}\right]=i f^{abc} G^{ic},\cr
\noalign{\medskip}
\multispan2\hfil$\left[G^{ia},G^{jb}\right] = {i\over 4}\delta^{ij} f^{abc} T^c
+ {i\over
2F} \delta^{ab} \epsilon^{ijk} J^k + {i\over 2} \epsilon^{ijk} d^{abc}
G^{kc}.$ \hfill\cr
}}}
\end{table}

\begin{table}[htbp]
\caption{$\left( adj \otimes adj \right)_S $}
\smallskip
\label{tab:adj2s}
\centerline{\vbox{ \tabskip=0pt \offinterlineskip
\def\tablerule{\noalign{\hrule}}
\def\space{height 2pt&\omit&&\omit&&\omit&&\omit&\cr}
\halign{
\vrule #&\strut\hfil\ $ # $\ \hfil&&
\vrule #&\strut\hfil\ $ # $\ \hfil\cr
\tablerule\space
&  && SU(Q) && SU(6) && SU(4) &\cr
\space\tablerule\space
& \left( adj \otimes adj \right)_S
&&
\left( \ \left[1, 0, 0, 0, \ldots, 0, 0, 1 \right]^2 \
\right)_S &&
\left( \ \left[1, 0, 0, 0, 1 \right]^2 \ \right)_S
&&
\left( \ \left[1, 0, 1 \right]^2 \ \right)_S
&\cr
\space\tablerule\space
& 1
&& \left[0, 0, 0, 0, \ldots, 0, 0, 0 \right]
&& \left[0, 0, 0, 0, 0 \right]
&& \left[0, 0, 0 \right]
&\cr
& adj
&& \left[1, 0, 0, 0, \ldots, 0, 0, 1 \right]
&& \left[1, 0, 0, 0, 1 \right]
&& \left[1, 0, 1 \right]
&\cr
& {\bar a a}
&& \left[0, 1, 0, 0, \ldots, 0, 1, 0 \right]
&& \left[0, 1, 0, 1, 0 \right]
&& \left[0, 2, 0 \right]
&\cr
& {\bar s s}
&& \left[2, 0, 0, 0, \ldots, 0, 0, 2 \right]
&& \left[2, 0, 0, 0, 2 \right]
&& \left[2, 0, 2 \right]
&\cr
\space\tablerule
}}}
The $SU(6)$ and $SU(4)$ representations are 1+35+189+405
and 1+15+20+84, respectively.
\end{table}

\begin{table}[htbp]
\caption{$\left( adj \otimes adj \otimes adj \right)_S $}
\smallskip
\label{tab:adj3s}
\centerline{\vbox{ \tabskip=0pt \offinterlineskip
\def\tablerule{\noalign{\hrule}}
\def\space{height 2pt&\omit&&\omit&&\omit&\cr}
\halign{
\vrule #&\strut\hfil\ $ # $\ \hfil&&
\vrule #&\strut\hfil\ $ # $\ \hfil\cr
\tablerule\space
& SU(Q) && SU(6) && SU(4) &&\omit\cr
\space\tablerule\space
& \left( \ \left[1, 0, 0, 0, \ldots, 0, 0, 1\right]^{3} \
\right)_S
&& \left( \ \left[1, 0, 0, 0, 1\right]^{3} \ \right)_S
&& \left( \ \left[1, 0, 1\right]^{3} \ \right)_S
&&\omit\cr
\space\tablerule\space
& \left[0, 0, 0, 0, \ldots, 0, 0, 0 \right]
&& \left[0, 0, 0, 0, 0 \right]
&& \left[0, 0, 0 \right]
&&\omit\cr
& \left[1, 0, 0, 0, \ldots, 0, 0, 1 \right]
&& \left[1, 0, 0, 0, 1 \right]
&& \left[1, 0, 1 \right]
&&\omit\cr
& \left[1, 0, 0, 0, \ldots, 0, 0, 1 \right]
&& \left[1, 0, 0, 0, 1 \right]
&& \left[1, 0, 1 \right]
&&\omit\cr
& \left[0, 0, 1, 0, \ldots, 1, 0, 0 \right]
&& \left[0, 0, 2, 0, 0 \right]
&& \omit
&&\omit\cr
& \left[0, 1, 0, 0, \ldots, 0, 1, 0 \right]
&& \left[0, 1, 0, 1, 0 \right]
&& \omit
&&\omit\cr
& \left[2, 0, 0, 0, \ldots, 0, 1, 0 \right]
&& \left[2, 0, 0, 1, 0 \right]
&& \left[2, 1, 0 \right]
&&\omit\cr
& \left[0, 1, 0, 0, \ldots, 0, 0, 2 \right]
&& \left[0, 1, 0, 0, 2 \right]
&& \left[0, 1, 2 \right]
&&\omit\cr
& \left[2, 0, 0, 0, \ldots, 0, 0, 2 \right]
&& \left[2, 0, 0, 0, 2 \right]
&& \left[2, 0, 2 \right]
&&\omit\cr
& \left[1, 1, 0, 0, \ldots, 0, 1, 1 \right]
&& \left[1, 1, 0, 1, 1 \right]
&& \left[1, 2, 1 \right]
&&\omit\cr
& \left[3, 0, 0, 0, \ldots, 0, 0, 3 \right]
&& \left[3, 0, 0, 0, 3 \right]
&& \left[3, 0, 3 \right]
&&\omit\cr
\space\tablerule
}}}
\end{table}

\begin{table}[htbp]
\caption{$\left( {\bar a a} \otimes adj \right) $}
\smallskip
\label{tab:aaadj}
\centerline{\vbox{ \tabskip=0pt \offinterlineskip
\def\tablerule{\noalign{\hrule}}
\def\space{height 2pt&\omit&&\omit&&\omit&\cr}
\halign{
\vrule #&\strut\hfil\ $ # $\ \hfil&&
\vrule #&\strut\hfil\ $ # $\ \hfil\cr
\tablerule\space
& SU(Q) && SU(6) && SU(4) &&\omit\cr
\space\tablerule\space
&
\left( \ \left[0, 1, 0, 0, \ldots, 0, 1, 0\right]
\otimes adj \ \right)
&&
\left( \ \left[0, 1, 0, 1, 0\right] \otimes adj \
\right)   &&
\left( \ \left[0, 2, 0\right] \otimes adj \
\right)   &&\omit\cr
\space\tablerule\space
& \left[1, 0, 0, 0, \ldots, 0, 0, 1 \right]
&& \left[1, 0, 0, 0, 1 \right]
&& \left[1, 0, 1 \right]
&&\omit\cr
& \left[0, 1, 0, 0, \ldots, 0, 1, 0 \right]
&& \left[0, 1, 0, 1, 0 \right]
&& \omit
&&\omit\cr
& \left[0, 1, 0, 0, \ldots, 0, 1, 0 \right]
&& \left[0, 1, 0, 1, 0 \right]
&& \omit
&&\omit\cr
& \left[0, 0, 1, 0, \ldots, 1, 0, 0 \right]
&& \left[0, 0, 2, 0, 0 \right]
&& \left[0, 2, 0 \right]
&&\omit\cr
& \left[2, 0, 0, 0, \ldots, 0, 1, 0 \right]
&& \left[2, 0, 0, 1, 0 \right]
&& \left[2, 1, 0 \right]
&&\omit\cr
& \left[0, 1, 0, 0, \ldots, 0, 0, 2 \right]
&& \left[0, 1, 0, 0, 2 \right]
&& \left[0, 1, 2 \right]
&&\omit\cr
& \left[1, 1, 0, 0, \ldots, 1, 0, 0 \right]
&& \left[1, 1, 1, 0, 0 \right]
&& \omit
&&\omit\cr
& \left[0, 0, 1, 0, \ldots, 0, 1, 1 \right]
&& \left[0, 0, 1, 1, 1 \right]
&& \omit
&&\omit\cr
& \left[1, 1, 0, 0, \ldots, 0, 1, 1 \right]
&& \left[1, 1, 0, 1, 1 \right]
&& \left[1, 2, 1 \right]
&&\omit\cr
\space\tablerule
}}}
\end{table}

\begin{table}[htbp]
\caption{$SU(2F)$ Identities: The second column gives the transformation
properties of the identities under $SU(2)\times SU(F)$.}
\smallskip
\label{tab:su2fiden}
\centerline{\vbox{ \tabskip=0pt \offinterlineskip
\def\tablerule{\noalign{\hrule}}
\def\space{height 4pt&\omit&&\omit&\cr}
\halign{
\vrule #&\strut\hfil\ $ # $\ \hfil&\vrule #&\strut\hfil\ $ # $\ \hfil&\vrule
#\cr
\tablerule\space
&2\ \left\{J^i,J^i\right\} + F\ \left\{T^a,T^a\right\} + 4 F\ \left\{G^{ia},
G^{ia}\right\} = N \left(N+2F\right) \left(2F-1\right)&&(0,0)&\cr
\space\tablerule\space
&d^{abc}\ \left\{G^{ia}, G^{ib}\right\} + {2\over F}\ \left\{J^i, G^{ic}
\right\} + {1\over4}\ d^{abc}\ \left\{T^a, T^b\right\} = \left(N+F\right)
\left(1-{1\over F}\right)\ T^c && (0,adj)&\cr
\space
&\left\{T^a,G^{ia}\right\} = \left(N+F\right) \left(1-{1\over F}\right)\
J^i
&&(1,0)&\cr
\space
&{1\over F}\ \left\{J^k,T^c\right\} +  d^{abc}\ \left\{T^a,G^{kb} \right\}
-\epsilon^{ijk} f^{abc} \left\{G^{ia}, G^{jb}\right\} = 2\left(N+F\right)
\left(1-{1\over F} \right)\ G^{kc} && (1,adj)&\cr
\space\tablerule\space
&4F\left(2-F\right)\ \left\{G^{ia},G^{ia}\right\} + 3 F^2\ \left\{T^a,
T^a\right\} + 4\left(1-F^2\right)\ \left\{J^i,J^i\right\}=0&& (0,0)&\cr
\space
&\left(4-F\right)d^{abc}\ \left\{G^{ia}, G^{ib}\right\} + {3\over 4} F
\ d^{abc}\ \left\{T^a, T^b\right\} - 2\left(F-{4\over F}\right)\
\left\{J^i,G^{ia}
\right\} = 0 && (0,adj) &\cr
\space
&4\ \left\{G^{ia},G^{ib}\right\} = -3\ \left\{T^a,T^b\right\}\qquad ({\bar a
a})
&& (0,{\bar a a})&\cr
\space
&4\ \left\{G^{ia},G^{ib}\right\} = \left\{T^a,T^b\right\}\qquad ({\bar s s})
&& (0,{\bar s s})&\cr
\space
&\epsilon^{ijk}\ \left\{ J^i,G^{jc}\right\} = f^{abc} \ \left\{T^a,G^{kb}
\right\}&&(1,adj)&\cr
\space
&d^{abc}\ \left\{T^a,G^{kb}\right\} = \left(1-{2\over F}\right) \left(
\left\{J^k,T^c\right\} -  \epsilon^{ijk} f^{abc}\ \left\{G^{ia}, G^{jb}
\right\}\right)&&(1,adj)&\cr
\space
&\epsilon^{ijk}\ \left\{G^{ia},G^{jb}\right\} = f^{acg} d^{bch}\
\left\{T^g,G^{kh}\right\}\qquad ({\bar a s}+{\bar s a})&&(1,{\bar a s}+{\bar s
a})&\cr
\space
&\left\{T^a,G^{ib}\right\} =0\qquad ({\bar a a})&&(1,{\bar a a})&\cr
\space
&\left\{G^{ia}, G^{ja}\right\} = {1\over2}\left(1-{1 \over F}\right)\
\left\{J^i, J^j \right\}
\qquad (J=2) && (2,0)&\cr
\space
&d^{abc}\ \left\{G^{ia}, G^{jb}\right\} = \left(1-{2\over F}\right) \
\left\{J^i,G^{jc}\right\}\qquad (J=2) && (2,adj)&\cr
\space
&\left\{G^{ia},G^{jb}\right\} = 0\qquad (J=2,{\bar a a})&& (2,{\bar a a})&\cr
\space\tablerule
}}}
\end{table}

\begin{table}[htbp]
\caption{$SU(4)$ Identities: The second column gives the transformation
properties of the identities under $SU(2)\times SU(2)$.}
\smallskip
\label{tab:su4iden}
\centerline{\vbox{ \tabskip=0pt \offinterlineskip
\def\tablerule{\noalign{\hrule}}
\def\space{height 4pt&\omit&&\omit&\cr}
\halign{
\vrule #&\strut\hfil\ $ # $\ \hfil&\vrule #&\strut\hfil\ $ # $\
\hfil&\vrule #\cr
\tablerule\space
&\left\{J^i,J^i\right\} + \left\{I^a,I^a\right\} +
4\ \left\{G^{ia},G^{ia}\right\} ={3 \over2} N \left(N+4\right)
&&(0,0)&\cr
\space\tablerule\space
&2 \left\{J^i,G^{ia}\right\} = \left(N+2\right)\ I^a
&&(0,1)&\cr
\space
&2 \left\{I^a,G^{ia}\right\} = \left(N+2\right)\ J^i
&&(1,0)&\cr
\space
&{1\over 2}\ \left\{J^k,I^c\right\}
-\epsilon^{ijk} \epsilon^{abc} \left\{G^{ia}, G^{jb}\right\} =
\left(N+2\right)\ G^{kc}
&&(1,1)&\cr
\space\tablerule\space
&\left\{I^a,
I^a\right\} - \left\{J^i,J^i\right\}=0 &&(0,0)&\cr
\space
&4\ \left\{G^{ia},G^{ib}\right\} = \left\{I^a,I^b\right\}\qquad (I=2)
&&(0,2)&\cr
\space
&\epsilon^{ijk}\ \left\{ J^i,G^{jc}\right\} = \epsilon^{abc} \
\left\{I^a,G^{kb}
\right\}&&(1,1)&\cr
\space
&4\ \left\{G^{ia}, G^{ja}\right\} = \left\{J^i, J^j
\right\}\qquad (J=2)&&(2,0)&\cr
\space\tablerule
}}}
\end{table}

\begin{table}[htbp]
\caption{$SU(6)$ Identities: The second column gives the transformation
properties of the identities under $SU(2)\times SU(3)$.}
\smallskip
\label{tab:su6iden}
\centerline{\vbox{ \tabskip=0pt \offinterlineskip
\def\tablerule{\noalign{\hrule}}
\def\space{height 4pt&\omit&&\omit&\cr}
\halign{
\vrule #&\strut\hfil\ $ # $\ \hfil&\vrule #&\strut\hfil\ $ # $\
\hfil&\vrule #\cr
\tablerule\space
&2\ \left\{J^i,J^i\right\} + 3\ \left\{T^a,T^a\right\} + 12\
\left\{G^{ia},G^{ia}\right\} = 5 N \left(N+6\right)&&(0,0)&\cr
\space\tablerule\space
&d^{abc}\ \left\{G^{ia}, G^{ib}\right\} + {2\over 3}\ \left\{J^i,G^{ic}
\right\} + {1\over4}\ d^{abc}\ \left\{T^a, T^b\right\} = {2\over 3}
\left(N+3\right)\ T^c  && (0,adj) &\cr
\space
&\left\{T^a,G^{ia}\right\} = {2\over3}\left(N+3\right)\ J^i && (1,0)&\cr
\space&{1\over 3}\ \left\{J^k,T^c\right\} +  d^{abc}\ \left\{T^a,G^{kb}\right\}
-\epsilon^{ijk} f^{abc} \left\{G^{ia}, G^{jb}\right\} =  {4\over3}
\left(N+3\right)\ G^{kc} && (1,adj)&\cr
\space\tablerule\space
&-12\ \left\{G^{ia},G^{ia}\right\} + 27\ \left\{T^a,
T^a\right\} - 32\ \left\{J^i,J^i\right\}=0 && (0,0)&\cr
\space
&d^{abc}\ \left\{G^{ia}, G^{ib}\right\} + {9\over 4} \ d^{abc}\ \left\{
T^a, T^b\right\} - {10\over3}\ \left\{J^i,G^{ic}\right\} = 0 && (0,adj)&\cr
\space
&4\ \left\{G^{ia},G^{ib}\right\} = \left\{T^a,T^b\right\}\qquad ({\bar s s})
&& (0,{\bar s s})&\cr
\space
&\epsilon^{ijk}\ \left\{ J^i,G^{jc}\right\} = f^{abc} \ \left\{T^a,G^{kb}
\right\}&& (1,adj)&\cr
\space
&3\ d^{abc}\ \left\{T^a,G^{kb}\right\} = \left\{J^k,T^c\right\} -
\epsilon^{ijk} f^{abc}\ \left\{G^{ia}, G^{jb}\right\}&& (1,adj)&\cr
\space
&\epsilon^{ijk}\ \left\{G^{ia},G^{jb}\right\} = f^{acg} d^{bch}\ \left\{
T^g,G^{kh}\right\}\qquad ({\bar a s}+{\bar s a}) && (1,{\bar a s}+{\bar s
a})&\cr
\space
&3\ \left\{G^{ia}, G^{ja}\right\} = \left\{J^i, J^j
\right\}\qquad (J=2) && (2,0)&\cr
\space
&3\ d^{abc}\ \left\{G^{ia}, G^{jb}\right\} =
\left\{J^i,G^{jc}\right\}\qquad (J=2) && (2,adj)&\cr
\space\tablerule
}}}
\end{table}

\begin{table}[htbp]
\caption{Operator reduction for $F$ flavors. The second column gives the
allowed $SU(2)\times SU(F)$ representations for the operators in
column one. The third column gives the combinations left after eliminating
all redundant linear combinations using the identities in
Table~\protect\ref{tab:su2fiden}.}
\smallskip
\label{tab:opleft}
\centerline{\vbox{ \tabskip=0pt \offinterlineskip
\def\tablerule{\noalign{\hrule}}
\def\space{height 4pt&\omit&&\omit&&\omit&\cr}
\halign{
\vrule #&\strut\hfil\ $ # $\ \hfil&\vrule #&&\strut\hfil\ $ # $\ \hfil&\vrule
#\cr
\tablerule\space
&\left\{J^i,J^j\right\}&&(0,0)\ (2,0) && (0,0)\ (2,0)&\cr
\space
\space
&\left\{G^{ia},G^{jb}\right\}&&(0,0)\ (0,adj)\ (0,{\bar a a})\ (0,{\bar s s})\
(1,adj)&& &\cr
& && (1,{\bar a s}+{\bar s a})\ (2,0)\ (2,adj)\ (2,{\bar a a})\ (2,{\bar s
s})&& (2,{\bar s s})&\cr
\space
\space
&\left\{T^{a},T^{b}\right\}&&(0,0)\ (0,adj)\ (0,{\bar a a})\ (0,{\bar s
s})&&(0,{\bar a a})\
(0,{\bar s s}) &\cr
\space
\space
&\left\{J^i,T^{a}\right\}&&(1,adj)&&(1,adj) &\cr
\space
\space
&\left\{J^i,G^{ja}\right\}&&(0,adj)\ (1,adj)\ (2,adj)&&(0,adj)\ (1,adj)\
(2,adj)&\cr
\space
\space
&\left\{T^a,G^{ib}\right\}&&(1,0)\ (1,adj)\ (1,adj)\ (1,{\bar a s}+{\bar s
a})&&(1,{\bar a s}+{\bar s a}) &\cr
& &&(1,{\bar a a})\ (1,{\bar s s})&& (1,{\bar s s})&\cr
\space\tablerule
}}}
\end{table}

\begin{table}[htbp]
\caption{$SU(4)\times SU(2)\times U(1)$ Commutation Relations}
\smallskip
\label{tab:brokencomm}
\centerline{\vbox{ \tabskip=0pt \offinterlineskip
\def\space{\noalign{\medskip}}
\begin{eqnarray*}
&\begin{array}{lll}
\space
\left[J^i_{ud},J^j_{s}\right]=0,&\left[J^i_{ud},I^a\right]=0,&
\left[J^i_{s},I^a\right]=0, \\
\space
\left[J^i_{s},G^{ja}\right]=0,&\left[Y^{i\alpha},Y^{j\beta}\right]=0,&
\left[t^\alpha,t^\beta\right]=0, \\
\space
\ \left[J^i_{ud},J^j_{ud}\right]=i\epsilon^{ijk} J^k_{ud},\ &
\ \left[J^i_{s},J^j_{s}\right]=i\epsilon^{ijk} J^k_{s},\ &
\ \left[I^a,I^b\right]=i \epsilon^{abc} I^c,\ \\
\space
\ \left[J^i_{ud},G^{ja}\right]=i\epsilon^{ijk} G^{ka},\ &
\ \left[I^a,G^{ib}\right]=i \epsilon^{abc} G^{ic},\ &
\ \left[J_{s}^i,t^\alpha\right]= Y^{i\alpha},\ \\
\space
\ \left[J_{ud}^i,t^\alpha\right]= - Y^{i\alpha},\ &
\ \left[Y^{i\alpha},t^\beta\right]=0,\ &
\ \left[I^a,t^\alpha\right]=-\left({\tau^a \over 2}\right)^\alpha_\beta
t^\beta,\ \\
\space
\ \left[N_s,I^a\right]= 0,\ &
\ \left[N_s,J_{ud}^i\right]=0,\ &
\ \left[N_s,J_s^i\right]=0,\ \\
\space
\ \left[N_s,G^{ia}\right]= 0,\ &
\ \left[N_s,t^\alpha\right]=t^\alpha,\ &
\ \left[N_s,Y^{i\alpha}\right]=Y^{i\alpha},\ \\
\space
\end{array}&\\
&\begin{array}{ll}
\ \left[G^{ia},t^\alpha\right]=-\left({\tau^a \over 2}\right)^\alpha_\beta
Y^{i\beta},\ &
\ \left[I^a,Y^{i\alpha}\right]=-\left({\tau^a \over 2}\right)^\alpha_\beta
Y^{i\beta},\
\\
\space
\ \left[J_{ud}^i,Y^{j\alpha}\right]={i\over 2}\epsilon^{ijk}
Y^{k\alpha} - {1\over 4}\delta^{ij} t^\alpha,\ &
\ \left[G^{ia},G^{jb}\right] = {i\over 4}\delta^{ij}
\epsilon^{abc} I^c +  {i\over 4} \delta^{ab} \epsilon^{ijk} J^k_{ud},\ \\
\space
\ \left[J_s^i,Y^{j\alpha}\right]={i\over 2}\epsilon^{ijk} Y^{k\alpha}
+{1\over 4}\delta^{ij} t^\alpha,\ &
\left[t^\alpha,t^\dagger_\beta\right]={1\over2}\delta^\alpha_\beta\left(
3 N_s - N\right) - 2 \left({\tau^a \over 2}\right)^\alpha_\beta I^a,\ \\
\space
\ \left[Y^{i\alpha},t^\dagger_\beta\right]=\delta^\alpha_\beta\left(
J_s^i-{1\over2} J_{ud}^i\right) - 2 \left({\tau^a \over 2}\right)^\alpha_\beta
G^{ia},\ &
\ \left[G^{ia},Y^{j\alpha}\right]=-\left({\tau^a \over
2}\right)^\alpha_\beta\left(
{1\over4}\delta^{ij}t^\beta - {i\over2}\epsilon^{ijk} Y^{k\beta}\right),\ \\
\space\end{array}&\\
&\begin{array}{c}
\ \left[Y^{i\alpha},Y^{\dagger j}_\beta\right]={1\over8}\delta^\alpha_\beta
\delta^{ij}\left( 3N_s - N\right) + {i\over2}
\delta^\alpha_\beta \epsilon^{ijk}\left(J_s^k + {1\over2} J_{ud}^k
\right) + \left({\tau^a \over 2}\right)^\alpha_\beta\left(-{1\over2}\delta^{ij}
I^a +
i \epsilon^{ijk} G^{ka}\right),\ \\
\space
\end{array}&
\end{eqnarray*}
}}
\end{table}

\begin{table}[htbp]
\caption{$SU(4)\times SU(2)\times U(1)$ Identities for $\Delta S=0$. The
second column gives the $(J,I)_S$ quantum numbers of the operator
identities.}
\smallskip
\label{tab:broken0}
\centerline{\vbox{ \tabskip=0pt \offinterlineskip
\def\tablerule{\noalign{\hrule}}
\def\space{height 4pt&\omit&&\omit&\cr}
\halign{
\vrule #&\strut\hfil\ $ # $\ \hfil&\vrule #&\strut\hfil\ $ # $\
\hfil&\vrule #\cr
\tablerule\space
&
\left\{G^{ia},G^{ia}\right\}=-{1\over2}\left\{J_{ud}^i,J_{ud}^i\right\}+
{3\over8}\left(N-N_s\right)\left(N-N_s+4\right) && (0,0)_0
&\cr\space
&
\left\{J_s^i,J_s^i\right\} = {1\over2} N_s\left(N_s+2\right) && (0,0)_0
&\cr\space
&
\left\{I^a,I^a\right\} = \left\{ J_{ud}^i, J_{ud}^i \right\} && (0,0)_0
&\cr\space
&
\left\{t^\alpha,t^\dagger_\alpha\right\} = 2\ \left\{J_{ud}^i,J_s^i
\right\} + N +  N_s + N N_s - N_s^2 && (0,0)_0
&\cr\space
&
\left\{Y^{i\alpha},Y^{\dagger i}_\alpha\right\}=-{1\over2}\
\left\{J_{ud}^i,J_s^i \right\} + {3\over4}\left(N +  N_s + N N_s - N_s^2
\right) && (0,0)_0
&\cr
\space\tablerule\space
&
\left\{G^{ia},J_{ud}^i\right\} = {1\over2}\left(N-N_s+2\right)\
I^a && (0,1)_0
&\cr\space
&
\left({\tau^a\over2}\right)^\alpha_\beta\ \left\{t^\beta,
t^\dagger_\alpha\right\} = \left(N_s+1\right) I^a  +
2\ \left\{ G^{ia},J_s^i\right\} && (0,1)_0
&\cr\space
&
\left\{Y^{i\beta},Y^{\dagger i}_\alpha\right\} = {3\over4}\
\left(N_s+1\right)I^a-{1\over2}\ \left\{ G^{ia},J_s^i\right\} && (0,1)_0
&\cr
\space\tablerule\space
&
\left\{ G^{ia},G^{ib}\right\}={1\over4}\ \left\{I^a,I^b\right\}
\qquad (I=2) && (0,2)_0
&\cr
\space\tablerule\space
&
\left\{G^{ia},I^a\right\} = {1\over2}\left(N-N_s+2\right)
\ J_{ud}^i && (1,0)_0
&\cr\space
&
\left\{Y^{k\alpha},t^\dagger_\alpha\right\} = i \epsilon^{ijk}\
\left\{J_{ud}^i,J_s^j\right\}+ \left(N_s+1\right)
J_{ud}^i + \left(N-N_s+2\right) J_s^i && (1,0)_0
&\cr\space
&
\left\{Y^{\dagger k}_\alpha,
t^\alpha\right\} = -i \epsilon^{ijk}\
\left\{J_{ud}^i,J_s^j\right\}+ \left(N_s+1\right)
J_{ud}^i + \left(N-N_s+2\right) J_s^i && (1,0)_0
&\cr\space
&
i\epsilon^{ijk}\ \left\{Y^{i\alpha},Y^{\dagger j}_\alpha\right\}=\left(
N_s+1\right) J_{ud}^k-\left(N-N_s+2\right) J_s^k  && (1,0)_0
&\cr
\space\tablerule\space
&
i\epsilon^{ijk}\ \left\{G^{ic},J_{ud}^j\right\} = i\epsilon^{abc}\left\{
G^{ka},I^b\right\} && (1,1)_0
&\cr\space
&
\epsilon^{ijk}\epsilon^{abc}\ \left\{G^{ia},G^{jb}\right\}={1\over2}\
\left\{J_{ud}^k,I^c\right\}-  \left(N-N_s+2\right) G^{kc}  && (1,1)_0
&\cr\space
&
\left({\tau^a\over2}\right)^\alpha_\beta\left\{Y^{k\beta}, t^\dagger_\alpha
\right\}= i\epsilon^{ijk} \left\{G^{ia},J_s^j\right\}
+{1\over2}\ \left\{J_s^i, I^a\right\}+ \left(N_s+1\right) G^{ia}&& (1,1)_0
&\cr\space
&
\left({\tau^a\over2}\right)^\alpha_\beta\left\{Y^{\dagger i}_\alpha,
t^\beta\right\}= -i\epsilon^{ijk} \left\{G^{ia},J_s^j\right\}
+{1\over2}\ \left\{J_s^i, I^a\right\}+ \left(N_s+1\right) G^{ia}&& (1,1)_0
&\cr\space
&
i \epsilon^{ijk}\left({\tau^a\over2}
\right)^\alpha_\beta\left\{ Y^{i\beta},Y^{\dagger j}_\alpha\right\} =
-{1\over2}\ \left\{J_s^k,I^a\right\} +\left(N_s+1\right)G^{ka}  && (1,1)_0
&\cr
\space\tablerule\space
&
\left\{ G^{ia},G^{ja}\right\}={1\over4}\ \left\{J_{ud}^i,J_{ud}^j\right\}
\qquad (J=2) && (2,0)_0
&\cr\space
&
\left\{ Y^{i\alpha},Y^{\dagger j}_\alpha\right\} = \left\{ J_{ud}^i,
J_s^j\right\}\qquad (J=2) && (2,0)_0
&\cr
\space\tablerule\space
&
\left\{ G^{ia},J_s^j\right\} = \left({\tau^a\over2}\right)^\alpha_\beta\
\left\{
Y^{i\beta},Y^{\dagger j}_\alpha\right\}\qquad (J=2) && (2,1)_0
&\cr
\space\tablerule
}}}
\end{table}

\begin{table}[htbp]
\caption{$SU(4)\times SU(2) \times U(1)$ Identities for $\Delta S\not=0$.
The second column gives the $(J,I)_{S}$ transformation properties of the
operator identities.}
\smallskip
\label{tab:brokennot0}
\centerline{\vbox{ \tabskip=0pt \offinterlineskip
\def\tablerule{\noalign{\hrule}}
\def\space{height 4pt&\omit&&\omit&\cr}
\halign{
\vrule #&\strut\hfil\ $ # $\ \hfil&\vrule #&\strut\hfil\ $ # $\
\hfil&\vrule #\cr
\tablerule\space
&
\left({\tau^a\over2}\right)^\alpha_\beta\ \left\{I^a,t^\beta\right\} = \left\{
Y^{i\alpha},J_{ud}^i\right\} && (0,1/2)_1
&\cr\space
&
\left\{Y^{i\alpha},J_s^i\right\} = {1\over4}\left\{N_s,t^\alpha\right\}+
{1\over2}t^\alpha
&& (0,1/2)_1
&\cr\space
&
\left({\tau^a\over2}\right)^\alpha_\beta\ \left\{Y^{i\beta},G^{ia}\right\} =
{3\over8}\
\left(N+2\right) t^\alpha - {3\over16}\left\{N_s,t^\alpha\right\}-
{1\over2}\ \left\{Y^{i\alpha},J_{ud}^i\right\} && (0,1/2)_1
&\cr
\space\tablerule\space
&
\left\{Y^{i\alpha},G^{ia}\right\} ={1\over4}\left\{t^\alpha,I^a\right\}\qquad
(I=3/2)&& (0,3/2)_1
&\cr
\space\tablerule\space
&
i \epsilon^{ijk} \left\{Y^{i\alpha},J_s^j\right\} =-{1\over2} \left\{
t^\alpha,J_s^k\right\}+{1\over2}\left\{N_s,Y^{k\alpha}\right\}+
Y^{k\alpha}&& (1,1/2)_1
&\cr\space
&
\left({\tau^a\over2}\right)^\alpha_\beta\ \left\{t^\beta,G^{ka}\right\}=
{1\over2}i
\epsilon^{ijk}\ \left\{Y^{i\alpha},J_{ud}^j\right\}+{1\over2} \left(N+2
\right) Y^{k\alpha}-{1\over4}\left\{N_s,Y^{k\alpha}\right\} && (1,1/2)_1
&\cr\space
&
i\epsilon^{ijk}\left({\tau^a\over2}\right)^\alpha_\beta\
\left\{Y^{i\beta},G^{ja}
\right\}={1\over4}\left\{t^\alpha,J_{ud}^k\right\}-{1\over4}i\epsilon^{ijk}
\ \left\{Y^{i\alpha},J_{ud}^j\right\}-{1\over2}\left(N+2\right) Y^{k\alpha}
+{1\over4}\left\{N_s,Y^{k\alpha}\right\}&& (1,1/2)_1
&\cr\space
&
\left({\tau^a\over2}\right)^\alpha_\beta\
\left\{Y^{k\beta},I^a\right\}={1\over4}
\left\{t^\alpha,J_{ud}^k\right\}-{1\over2} i \epsilon^{ijk}\ \left\{
Y^{i\alpha},J_{ud}^j\right\} && (1,1/2)_1
&\cr
\space\tablerule\space
&
i \epsilon^{ijk}\ \left\{Y^{i\alpha},G^{ja}\right\} = {1\over2}\left\{
t^\alpha,G^{ka}\right\}-{1\over2} \left\{Y^{k\alpha}, I^a\right\}
\qquad(I=3/2) && (1,3/2)_1
&\cr
\space\tablerule\space
&
\left({\tau^a\over2}\right)^\alpha_\beta\ \left\{Y^{i\beta},G^{ja}\right\} =
{1\over4}\left\{Y^{i\alpha},J^j_{ud}\right\}\qquad(J=2)&& (2,1/2)_1
&\cr
\space\tablerule\noalign{\vskip 1pt}\tablerule\space
&
\left\{ Y^{i\alpha},Y^{i\beta}\right\}={1\over4}\left\{ t^\alpha, t^\beta
\right\}\qquad(I=1)&& (0,1)_2
&\cr
\space\tablerule\space
&
\epsilon^{ijk} \epsilon_{\alpha\beta} \ \left\{Y^{i\alpha},Y^{j\beta}
\right\} = i \epsilon_{\alpha\beta} \left\{t^\alpha,Y^{k\beta}\right\}
&& (1,0)_2
&\cr
\space\tablerule
}}}
\end{table}

\begin{table}[htbp]
\caption{Operator reduction for $SU(2)\times U(1)$ flavor symmetry. The second
column
gives the  allowed $(J,I)_{S}$ quantum numbers for the operators in
column one. The third column gives the operators left after eliminating
all redundant linear combinations using the identities in
Tables~\protect\ref{tab:broken0} and \protect\ref{tab:brokennot0}. Operator
products not
shown, such as $\left\{J_{ud}^i,J_s^j\right\}$, do not have any linear
combinations which can be eliminated.}
\smallskip
\label{tab:brokenleft}
\centerline{\vbox{ \tabskip=0pt \offinterlineskip
\def\tablerule{\noalign{\hrule}}
\def\space{height 2pt&\omit&&\omit&&\omit&\cr}
\halign{
\vrule #&\strut\hfil\ $ # $\ \hfil&\vrule #&&\strut\hfil\ $ # $\ \hfil&\vrule
#\cr
\tablerule\space
&\left\{G^{ia},G^{jb}\right\}&&(0,0)_0\ (1,1)_0\ (0,2)_0\ (2,0)_0\
(2,2)_0&&(2,2)_0 &\cr
\space\space
&\left\{I^{a},I^{b}\right\}&&(0,0)_0\ (0,2)_0 && (0,2)_0&\cr
\space\space
&\left\{I^a,G^{ib}\right\}&&(1,0)_0\ (1,1)_0\ (1,2)_0&&(1,2)_0 &\cr
\space\space
&\left\{J_{ud}^i,G^{ja}\right\}&&(0,1)_0\ (1,1)_0\ (2,1)_0&&(1,1)_0\ (2,1)_0
&\cr
\space\space
&\left\{J_{s}^i,J_s^{j}\right\}&&(0,0)_0\ (2,0)_0&&(2,0)_0 &\cr
\space\space
&\left\{Y^{i\alpha},Y^{\dagger j}_\beta\right\}&&(0,0)_0\ (0,1)_0\
(1,0)_0\ (1,1)_0\ (2,0)_0\ (2,1)_0&& &\cr
\space\space
&\left\{Y^{i\alpha},t^{\dagger}_\beta\right\}&&(1,0)_0\ (1,1)_0&&  &\cr
\space\space
&\left\{t^{\alpha},Y^{\dagger i}_\beta\right\}&&(1,0)_0\ (1,1)_0&&  &\cr
\space\space
&\left\{t^{\alpha},t^{\dagger}_\beta\right\}&&(0,0)_0\ (0,1)_0&&  &\cr
\space\tablerule\space
&\left\{I^a,t^\alpha\right\}&&(0,1/2)_1\ (0,3/2)_1&&(0,3/2)_1  &\cr
\space\space
&\left\{J_{s}^i,Y^{j\alpha}\right\}&&(0,1/2)_1\ (1,1/2)_1\
(2,1/2)_1&&(2,1/2)_1 &\cr
\space\space
&\left\{I^a,Y^{i\alpha}\right\}&&(1,1/2)_1\ (1,3/2)_1&&(1,3/2)_1  &\cr
\space\space
&\left\{G^{ia},t^\alpha\right\}&&(1,1/2)_1\ (1,3/2)_1&&(1,3/2)_1  &\cr
\space\space
&\left\{G^{ia},Y^{j\alpha}\right\}&&(0,1/2)_1\ (0,3/2)_1\ (1,1/2)_1\
(1,3/2)_1\ (2,1/2)_1\ (2,3/2)_1&&(2,3/2)_1  &\cr
\space\tablerule\space
&\left\{Y^{i\alpha},Y^{j\beta}\right\}&&(0,1)_2\ (1,0)_2\ (2,1)_2&&(2,1)_2
&\cr
\space\tablerule
}}}
\end{table}

\begin{table}[htbp]
\caption{$SU(Q)$ Representations}
\smallskip
\label{tab:suqreps}
\centerline{\vbox{ \tabskip=0pt \offinterlineskip
\def\tablerule{\noalign{\hrule}}
\def\space{height 2pt&\omit&&\omit&&\omit&&\omit&&\omit&&\omit\cr}
\halign{
\vrule #&\strut\hfil\ $\raise25pt\hbox{$ # $}$\ \hfil&&
\vrule #&\strut\hfil\ $\raise25pt\hbox{$ # $}$\ \hfil\cr
\tablerule\space\space
& \omit \hfil\ Rep\ \hfil
&& \omit \hfil\ Dimension\ \hfil
&& \omit \hfil\ Casimir\ \hfil
&&\omit \hfil\ Dynkin\ Label\ \hfil
&& \omit \hfil\ Young\ Tableau\ \hfil
&&\omit\cr
\space\space\tablerule\space
& \omit \hfil 1 \hfil
&& \omit
&& \omit \hfil 0 \hfil
&& \omit \hfil\ $\left[0, 0, 0, 0, \ldots, 0, 0, 0 \right]$\ \hfil
&& \omit
&&\omit\cr
\space\tablerule\space\space
& \omit \hfil $\onebox$ \hfil
&& \omit \hfil $Q$ \hfil
&& \omit \hfil ${Q^2-1\over2Q}$ \hfil
&& \omit \hfil\ $\left[1, 0, 0, 0, \ldots, 0, 0, 0 \right]$\ \hfil
&& \omit \hfil $\onebox$ \hfil
&&\omit\cr
\space\tablerule\space\space\space
& \overline\onebox
&& Q
&& {Q^2-1\over2Q}
&& \left[0, 0, 0, 0, \ldots, 0, 0, 1 \right]
&& \omit \hfil $\nboxconj$ \hfil
&&\omit\cr
\space\tablerule\space\space\space
& adj
&& \left( Q^2 - 1 \right)
&& Q
&& \left[1, 0, 0, 0, \ldots, 0, 0, 1 \right]
&& \omit \hfil $\nboxadj$ \hfil
&&\omit\cr
\space\tablerule\space\space\space
& {\bar a s}
&& {1\over4}  \left( Q^2 - 1 \right) \left( Q^2-4 \right)
&& 2Q
&& \left[2, 0, 0, 0, \ldots, 0, 1, 0 \right]
&& \omit \hfil $\nboxas$ \hfil
&&\omit\cr
\space\tablerule\space\space\space
& {\bar s a}
&& {1\over4}  \left( Q^2 - 1 \right) \left( Q^2-4 \right)
&& 2Q
&& \left[0, 1, 0, 0, \ldots, 0, 0, 2 \right]
&& \omit \hfil $\nboxsa$ \hfil
&&\omit\cr
\space\tablerule\space\space\space
& {\bar a a}
&& {1\over4} Q^2 \left( Q + 1 \right) \left( Q - 3 \right)
&& 2\left(Q-1\right)
&& \left[0, 1, 0, 0, \ldots, 0, 1, 0 \right]
&& \omit \hfil $\nboxaa$ \hfil
&&\omit\cr
\space\tablerule\space\space\space
& {\bar s s}
&&{1\over4} Q^2 \left( Q - 1 \right) \left( Q + 3 \right)
&& 2\left(Q+1\right)
&& \left[2, 0, 0, 0, \ldots, 0, 0, 2 \right]
&& \omit \hfil $\nboxss$ \hfil
&&\omit\cr
\space\tablerule
}}}
\bigskip
\end{table}

\begin{table}[htbp]
\caption{$\left( adj \otimes adj \right)_A $}
\smallskip
\label{tab:adj2a}
\centerline{\vbox{ \tabskip=0pt \offinterlineskip
\def\tablerule{\noalign{\hrule}}
\def\space{height 2pt&\omit&&\omit&&\omit&&\omit&\cr}
\halign{
\vrule #&\strut\hfil\ $ # $\ \hfil&&
\vrule #&\strut\hfil\ $ # $\ \hfil\cr
\tablerule\space
&  && SU(Q) && SU(6) && SU(4) &\cr
\space\tablerule\space
& \left( adj \otimes adj \right)_A
&&
\left( \ \left[1, 0, 0, 0, \ldots, 0, 0, 1 \right]^2 \
\right)_A &&
\left( \ \left[1, 0, 0, 0, 1 \right]^2 \ \right)_A
&&
\left( \ \left[1, 0, 1 \right]^2 \ \right)_A
&\cr
\space\tablerule\space
& adj
&& \left[1, 0, 0, 0, \ldots, 0, 0, 1 \right]
&& \left[1, 0, 0, 0, 1 \right]
&& \left[1, 0, 1 \right]
&\cr
& {\bar a s}
&& \left[2, 0, 0, 0, \ldots, 0, 1, 0 \right]
&& \left[2, 0, 0, 1, 0 \right]
&& \left[2, 1, 0 \right]
&\cr
& {\bar s a}
&& \left[0, 1, 0, 0, \ldots, 0, 0, 2 \right]
&& \left[0, 1, 0, 0, 2 \right]
&& \left[0, 1, 2 \right]
&\cr
\space\tablerule
}}}
\end{table}

\twocolumn 

\begin{table}[htbp]
\caption{$S$-wave hyperon non-leptonic decay amplitudes $s$. Experimental
amplitudes
are given in the second column. The third column is the one-parameter $SU(3)$
symmetric fit keeping only the leading operator in the $1/\N$ expansion. The
fourth
column is the two-parameter $SU(3)$ symmetric fit including the leading
operator
and the subleading $1/\N$ correction. A soft-pion theorem has been used in the
calculation.}
\medskip
\label{tab:hnlds}
\centerline{\vbox{ \tabskip=0pt \offinterlineskip
\def\tablerule{\noalign{\hrule}}
\def\space{height 2pt&\omit&\omit&\omit&&\omit&\omit&\omit&&
\omit&&\omit&\cr}
\def\minus{\hphantom{-}}
\halign{
\vrule #&\strut\hfil\quad $ # $&$\rightarrow #$&\strut $ # $\quad  \hfil&
\vrule #&\strut\hfil\quad $ # $&$\pm #$&\strut $ # $\quad  \hfil&&
\vrule#&\strut\hfil\quad $ # $\quad \hfil\cr
\tablerule\space
&\multispan3\hfil Decay \hfil &&\multispan3 \hfil {\rm Expt} \hfil &&
1 && 1/\N &\cr
\space\tablerule\space
& \Sigma^+ && n\pi^+ && 0.06 && 0.01 && \minus 0.0 &&  \minus 0.0 &\cr
& \Sigma^+ && p\pi^0 && -1.43 && 0.05 && -1.00 &&  -1.35  &\cr
& \Sigma^- && n\pi^- && 1.88 && 0.01 && \minus 1.41  && \minus 1.90 &\cr
& \Lambda^0 && p\pi^- && 1.42 && 0.01 && \minus 1.73 && \minus 1.44 &\cr
& \Lambda^0 && n\pi^0 && -1.04 && 0.01 && -1.22  && -1.02  &\cr
& \Xi^- && \Lambda\pi^- && -1.98 && 0.01 && -1.73  && -1.88  &\cr
& \Xi^0 && \Lambda\pi^0 && 1.52 && 0.02 && \minus 1.22  && \minus 1.33 &\cr
\space\tablerule
}}}
\end{table}

\vfill\break\eject

\narrowtext


\begin{figure}
\caption{$SU(2F)$ representation for ground-state baryons.  The Young tableau
has $\N$ boxes.}
\label{fig:groundstate}
\end{figure}

\begin{figure}
\caption{Weight diagram for the $SU(3)$ flavor representation of
the spin-${1 \over 2}$ baryons.  The top of the weight diagram has
baryons with zero strange quarks. The long side of the weight diagram
contains ${1 \over 2}\left( \N + 1 \right)$ weights. The numbers denote
the multiplicity of the weights.}
\label{fig:weight1/2}
\end{figure}

\begin{figure}
\caption{Weight diagram for the $SU(3)$ flavor representation of
the spin-${3 \over 2}$ baryons.  The top of the weight diagram has
baryons with zero strange quarks. The long side of the weight diagram
contains ${1 \over 2}\left( \N - 1 \right)$ weights. The numbers denote
the multiplicity of the weights.}
\label{fig:weight3/2}
\end{figure}

\begin{figure}
\caption{Feynman diagrams depicting the insertion of a one-quark QCD
operator on the $\N$ quark lines of the baryon.  Graphs (b) contain
additional planar gluons, and are of the same order as (a).}
\label{fig:opmatrix}
\end{figure}

\begin{figure}
\caption{The hyperfine mass splittings within a baryon tower.
Splittings at the bottom of the tower are of order $1/\N$, whereas splittings
at the top of the tower are of order $1$. There are $(\N+1)/2$ energy
levels in the flavor symmetry limit.}
\label{fig:spectrum}
\end{figure}

\begin{figure}
\caption{Diagrams contributing to the quark mass dependence of the axial
current matrix
element. The axial current is denoted by $\otimes$, and the symmetry breaking
Hamiltonian by a solid square. Diagrams such as (a) do not break the flavor
symmetry.
Flavor symmetry breaking arises from diagrams such as (b).}
\label{fig:sloop}
\end{figure}

\vfill\break\eject

\centerline{$$\nbox$$}
\bigskip
\centerline{Figure 1}
\vskip1in

\def\onedot{\makebox(0,0){$\scriptstyle 1$}}
\def\twodot{\makebox(0,0){$\scriptstyle 2$}}
\def\threedot{\makebox(0,0){$\scriptstyle 3$}}
\def\fourdot{\makebox(0,0){$\scriptstyle 4$}}

\setlength{\unitlength}{3mm}

\centerline{\hbox{
\begin{picture}(20.79,18)(-10.395,-8)
\multiput(-1.155,10)(2.31,0){2}{\onedot}
\multiput(-2.31,8)(4.62,0){2}{\onedot}
\multiput(-3.465,6)(6.93,0){2}{\onedot}
\multiput(-4.62,4)(9.24,0){2}{\onedot}
\multiput(-5.775,2)(11.55,0){2}{\onedot}
\multiput(-6.93,0)(13.86,0){2}{\onedot}
\multiput(-8.085,-2)(16.17,0){2}{\onedot}
\multiput(-9.24,-4)(18.48,0){2}{\onedot}
\multiput(-10.395,-6)(20.79,0){2}{\onedot}
\multiput(-9.24,-8)(2.31,0){9}{\onedot}
\multiput(0,8)(2.31,0){1}{\twodot}
\multiput(-1.155,6)(2.31,0){2}{\twodot}
\multiput(-2.31,4)(2.31,0){3}{\twodot}
\multiput(-3.465,2)(2.31,0){4}{\twodot}
\multiput(-4.62,0)(2.31,0){5}{\twodot}
\multiput(-5.775,-2)(2.31,0){6}{\twodot}
\multiput(-6.93,-4)(2.31,0){7}{\twodot}
\multiput(-8.085,-6)(2.31,0){8}{\twodot}
\end{picture}
}}
\bigskip
\centerline{Figure 2}
\vskip1in

\centerline{\hbox{
\begin{picture}(20.79,18)(-8.085,-8)
\multiput(-1.155,10)(2.31,0){4}{\onedot}
\multiput(-2.31,8)(9.24,0){2}{\onedot}
\multiput(-3.465,6)(11.55,0){2}{\onedot}
\multiput(-4.62,4)(13.86,0){2}{\onedot}
\multiput(-5.775,2)(16.17,0){2}{\onedot}
\multiput(-6.93,0)(18.48,0){2}{\onedot}
\multiput(-8.085,-2)(20.79,0){2}{\onedot}
\multiput(-6.93,-4)(18.48,0){2}{\onedot}
\multiput(-5.775,-6)(16.17,0){2}{\onedot}
\multiput(-4.62,-8)(2.31,0){7}{\onedot}
\multiput(0,8)(2.31,0){3}{\twodot}
\multiput(-1.155,6)(6.93,0){2}{\twodot}
\multiput(-2.31,4)(9.24,0){2}{\twodot}
\multiput(-3.465,2)(11.55,0){2}{\twodot}
\multiput(-4.62,0)(13.86,0){2}{\twodot}
\multiput(-5.775,-2)(16.17,0){2}{\twodot}
\multiput(-4.62,-4)(13.86,0){2}{\twodot}
\multiput(-3.465,-6)(2.31,0){6}{\twodot}
\multiput(1.155,6)(2.31,0){2}{\threedot}
\multiput(0,4)(4.62,0){2}{\threedot}
\multiput(-1.155,2)(6.93,0){2}{\threedot}
\multiput(-2.31,0)(9.24,0){2}{\threedot}
\multiput(-3.465,-2)(11.55,0){2}{\threedot}
\multiput(-2.31,-4)(2.31,0){5}{\threedot}
\multiput(2.31,4)(2.31,0){1}{\fourdot}
\multiput(1.155,2)(2.31,0){2}{\fourdot}
\multiput(0,0)(2.31,0){3}{\fourdot}
\multiput(-1.155,-2)(2.31,0){4}{\fourdot}
\end{picture}
}}
\bigskip
\centerline{Figure 3}
\vskip1in
\insertfig{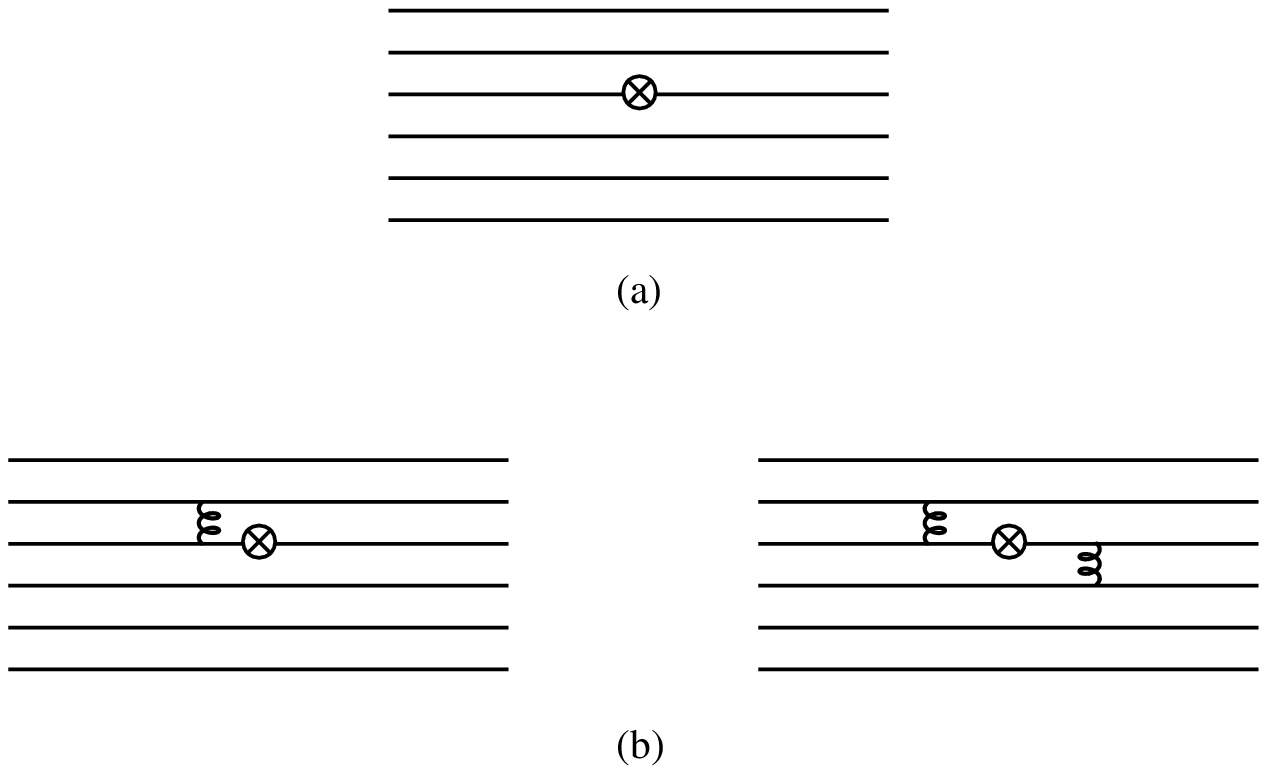}
\bigskip
\centerline{Figure 4}

\vskip 1in
\setlength{\unitlength}{6mm}
\centerline{\hbox{
\begin{picture}(10,7.725)(-1.9,-0.525)
\def\level{\line(1,0){5}}
\thicklines
\put(0,6.4){\level}
\put(0,4.9){\level}
\put(0,3.6){\level}
\put(0,2.5){\level}
\put(0,1.6){\level}
\put(0,0.9){\level}
\put(0,0.4){\level}
\put(0,0.1){\level}
\put(0,0){\level}
\thinlines
\put(5.5,6.4){\line(1,0){1}}
\put(5.5,4.9){\line(1,0){1}}
\put(5.5,0.125){\line(1,0){1}}
\put(5.5,-0.025){\line(1,0){1}}
\put(-1.9,0){\line(1,0){1}}
\put(-1.9,6.4){\line(1,0){1}}
\put(6,5.65){\makebox(0,0){$1$}}
\put(7.5,0){\makebox(0,0){$1/N_c$}}
\put(-1.65,3.05){$N_c$}
\put(-1.4,3.6){\vector(0,1){2.7}}
\put(-1.4,2.8){\vector(0,-1){2.7}}
\put(6,7.2){\vector(0,-1){0.8}}
\put(6,4.1){\vector(0,1){0.8}}
\put(6,-0.525){\vector(0,1){0.5}}
\put(6,0.625){\vector(0,-1){0.5}}
\end{picture}
}}
\bigskip
\centerline{Figure 5}
\vskip1in
\insertfig{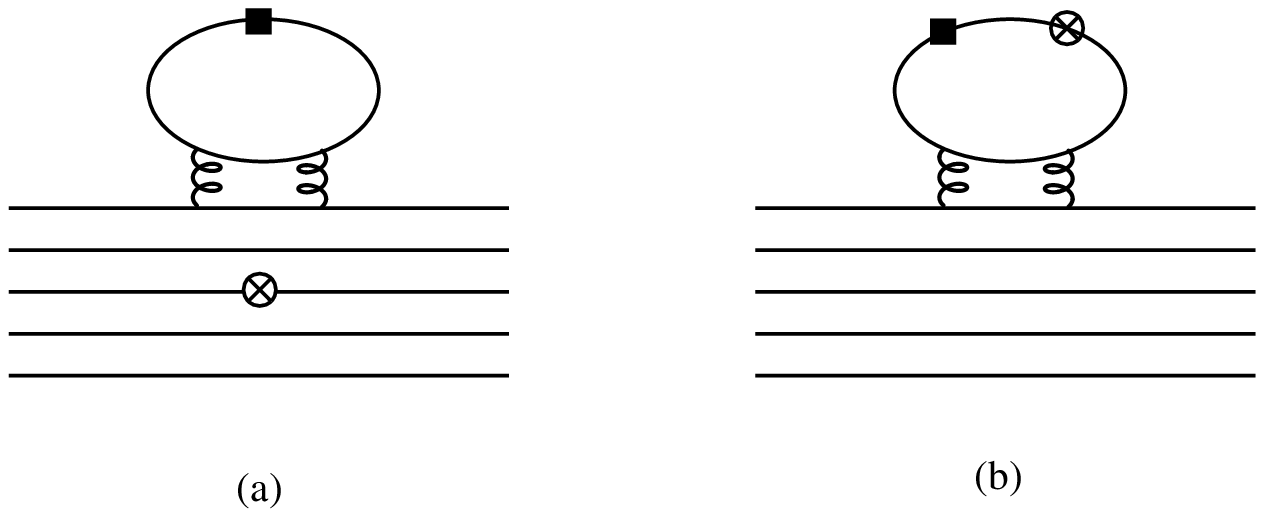}
\smallskip
\centerline{Figure 6}
\end{document}